\documentclass[12pt]{iopart}
\usepackage{amssymb}
\usepackage{epsfig,newlfont}
\usepackage{subfigure}
\usepackage[colorlinks, 
linkcolor=blue,
urlcolor=blue,
citecolor=blue,
bookmarksopen=false]{hyperref}
\usepackage{graphicx, xcolor}
\usepackage{caption}

\newcommand{\fsa}[1]{\left\langle{#1}\right\rangle}

\begin{document}

\title[Fast evaluation of prompt losses of energetic ions in stellarators]{A model for the fast evaluation of prompt losses of energetic ions in stellarators}

\author{J. L. Velasco$^1$, I. Calvo$^1$, S. Mulas$^1$, E. S\'anchez$^1$, F. I. Parra$^2$, \'A. Cappa$^1$ and the W7-X team}

\address{$^1$ Laboratorio Nacional de Fusi\'on, CIEMAT, Madrid, Spain}

\address{$^2$ Rudolf Peierls Centre for Theoretical Physics, University of Oxford, Oxford, United Kingdom}

\ead{joseluis.velasco@ciemat.es}

\vspace{10pt}
\begin{indented}
\item[]\today
\end{indented}

\begin{abstract}

A good understanding of the confinement of energetic ions in non-axisymmetric magnetic fields is key for the design of reactors based on the stellarator concept. In this work, we develop a model that, based on the radially-local bounce-averaged drift-kinetic equation, classifies orbits and succeeds in predicting configuration-dependent aspects of the prompt losses of energetic ions in stellarators. Such a model could in turn be employed in the optimization stage of the design of new devices.

\end{abstract}

\section{Introduction}

Good confinement of fusion-generated alpha particles is one of the basic properties of a fusion reactor. On the one hand, these alpha particles are expected to contribute to heat the plasma, which requires their confinement time to be sufficiently longer than the time that it takes them to thermalize by giving their energy to the bulk plasma. This is the so called \textit{slowing-down-time}. Specifically, the widely-employed figure of merit for energetic ion confinement is the accumulated fraction of alphas that are lost between their birth and one slowing-down time. %While no hard criterion can exist on what value of this quantity is low enough, a few percents is generally considered satisfactory, see e.g.~\cite{ku2008ariescs}.

However, a more restrictive criterion may be set by the heat loads on the walls, see e.g.~\cite{najmabadi2008ariescs}: alphas that are promptly lost, and that therefore retain most of their original energy, could damage the plasma facing components (the degree to which this happens will depend to some extent on the details of the magnetic configuration beyond the last closed flux-surface, as well as on the design of the facing components themselves~\cite{mau2008ariescs}). This issue turns the fraction of alphas that are promptly lost (i.e., that are lost in a time scale much shorter than the slowing-down time) into an additional quantity to be targeted in the design of a magnetic configuration.

In stellarators, neoclassical processes are the main concern with respect to energetic ion confinement. Particles trapped in the magnetic field of axisymmetric tokamaks, while moving back and forth along the field lines, experience radial excursions that produce banana-shaped orbits, but no net radial displacement takes place in the absence of collisions (their bounce-averaged radial drift, in the terminology that we will employ later, is exactly zero). Things are different in a generic stellarator (or in a tokamak with broken symmetry), where collisionless trapped orbits are not confined. The component of the magnetic drift that is tangential to the flux-surface causes a precession within the flux-surface that tends to keep the orbits close to the original surface~\cite{calvo2017sqrtnu} (for thermal ions of low collisionality, the $E\times B$ drift plays this role, but it is negligible for energetic ions). In a generic stellarator, however, a fraction of the particles have small tangential magnetic drift. These particles drift radially very fast, following a so-called \textit{superbanana} orbit. Along this work, we will denote as superbananas the orbits of particles whose bounce-averaged radial magnetic drift is much larger than their bounce-averaged tangential magnetic drift.

Although not always explicitly named, superbananas have been long known to cause significant energetic ion prompt losses, and are usually targeted in stellarator optimization. In the stellarator Wendelstein 7-X (W7-X), good confinement of the energetic ions relies on the diamagnetic effect at finite $\beta$, via an enhancement of their poloidal precession~\cite{wobig1999helias}. W7-X, like other quasi-isodynamic stellarators, seeks to satisfy the \textit{maximum-J} property: the contours of constant second adiabatic invariant $J$ are ideally aligned with the flux surfaces, and the function $J(s)$, being $s$ the flux-surface label, is made monotonously decreasing~\cite{mikhailov2002qipc,kolesnichenko2006er}. Since particles move at constant $J$, if these conditions are met for all velocities, no superbananas exist. In heliotron devices, inward-shifting of the magnetic axis is known to contribute to align the contours of $J$ with the flux-surfaces, and it has been experimentally demonstrated to improve energetic ion confinement~\cite{kaneko2002fastions}. In quasisymmetric devices, optimization with respect to energetic ion confinement has often been addressed indirectly, through reduction of the Fourier components of the magnetic field strength with helicity different than that corresponding to the direction of symmetry~\cite{ku2010qhs,henneberg2019fastions} (in a perfectly quasisymmetric stellarator, $J$ is constant on the flux-surface). Additionally, recent optimizations using the $\Gamma_{\mathrm{c}}$ proxy have yielded good energetic ion confinement in a quasihelically symmetric stellarator~\cite{bader2019fastions}. The  $\Gamma_{\mathrm{c}}$ proxy~\cite{nemov2008gammac} measures the ratio between the average radial component and the average tangential component of the magnetic drift: when it is 0, no superbanana orbits exist; it is 1 for a flux-surface where the ions cannot precess. Good correlation has been recently found between $\Gamma_c$ and the performance of several reactor-size stellarator configurations with respect to energetic ion confinement~\cite{bader2021gammac}. As a result of these optimization efforts, configurations belonging to the main stellarator concepts can be found that have negligible prompt losses, at least for energetic ions born at a certain radial region, see e.g.~\cite{subbotin2006qipc,bader2019fastions,henneberg2019fastions,masaoka2013fastions}. %\ongoing{However, even in these cases, there are energetic ions that escape the plasma on a longer time scale, caused by stochastic diffusion, see~\cite{beidler2001stochastic} and references therein. Ions that are shallowly trapped precess very fast within the flux-surface. During each poloidal turn, they undergo one or more transitions between \textit{locally trapped} (e.g. being trapped in one field-period) and \textit{locally passing} (e.g. with their periodic motion along the magnetic field line extending to a larger, but finite, number of field periods). Each of these transitions is accompanied by a small radial excursion. The accumulation of these excursions, that are uncorrelated in time, produces a radial diffusion. }

%\ongoing{Optimizing stellarators with respect to prompt losses should have, generally speaking, a positive effect on stochastic losses: in a stellarator in which the contours of $J$ coincide with the flux-surfaces, no stochastic losses take place. However, the region of phase space responsible for prompt and stochastic losses is generally different. If the alignment of $J-$ and $\Psi_t-$ contours is achieved mainly for deeply trapped particles at the cost of a worse alignment for barely trapped ions, then the reduction of prompt losses will come at the cost of worsening stochastic losses, see e.g. figure 3 of~\cite{bader2019fastions}. It is possible to target specifically the region of phase space that causes stochastic diffusion~\cite{tykhyy2007stochastic,drevlak2014fastions}. Indeed, in~\cite{drevlak2014fastions}, helias configurations with basically no stochastic losses are found (and poor correlation is observed between the change in the level of prompt losses and that of the stochastic losses).}

For an accurate evaluation of the confinement of energetic ions by a given magnetic configuration, Monte Carlo simulations are usually employed~\cite{isaev1999fastions,mikhailov2002qipc,ku2010qhs,masaoka2013fastions,drevlak2014fastions,henneberg2019fastions,bader2019fastions,cole2019xgc}. Guiding-center orbits are distributed similarly to fusion-born alpha particles and followed as they explore a reactor-size magnetic configuration at least for one slowing-down time. While most of these studies have focused on estimating the loss fraction, and on assessing that the optimization strategy has indeed improved the desired figure of merit, they have also confirmed several aspects of the picture drawn in the previous paragraphs. Two time scales are  observed in the time dependence of the loss fraction, corresponding to very different populations, see e.g.~\cite{masaoka2013fastions,drevlak2014fastions,faustin2016loss,cole2019xgc,bader2019fastions}: prompt losses take place mainly among relatively deeply trapped particles that are born on a~\textit{loss cone}, and they can be reduced by optimization as outlined in the previous paragraph; on a longer time scale, shallowly trapped ions escape. These slower losses are thought to be caused by stochastic diffusion~\cite{beidler2001stochastic}.

The goal of this work is to provide a more exhaustive characterization of the prompt losses of energetic ions in a stellarator configuration. To that end, we will develop a simple model that, based on the bounce-averaged drift kinetic equation, will classify bounce-averaged orbits into confined or unconfined. To the extent that this model is able to describe important aspects of these prompt losses, and if the dependence of these features on the magnetic configuration is  well captured, it could in turn be employed within the optimization loop in the design of new stellarator configurations.

For these studies, we will use modules of the neoclassical code {\ttfamily KNOSOS}~(KiNetic Orbit-averaging SOlver for Stellarators)\footnote{{\ttfamily KNOSOS} can be downloaded from \href{https://github.com/joseluisvelasco/KNOSOS}{https://github.com/joseluisvelasco/KNOSOS}.}, that treats rigorously the effect of the component of the magnetic drift that is tangent to magnetic surfaces~\cite{velasco2020knosos}. The set of magnetic equilibria for our tests has been selected from the configuration space of W7-X, and it includes a scan in the mirror term, the rotational transform, and $\beta$. They correspond to free-boundary VMEC~\cite{hirshman1983vmec} equilibria, which guarantees that the relevant effect of the coil ripple~\cite{nemov2014ripple} is taken into account. The selection of a relatively narrow stellarator configuration space will allow us to concentrate on fine parameters of the magnetic field, rather than on differences on zero-dimensional quantities such as e.g. the major radius. The rationale for this is that the former are the ones actually explored in an optimization process, while the latter are mainly set on beforehand based on considerations of a different kind. We will argue that our results should nevertheless be relevant for the characterization and optimization of energetic ion confinement in other types of stellarators. They can also be useful for the design of real tokamaks, in which neoclassical energetic ion losses, associated to the deviation from perfect axis-symmetry (caused, e.g., by coil ripple), have to be mitigated~\cite{tobita2003fi}. The predictions of the model will be compared to Monte Carlo collisionless simulations performed with the code ASCOT~\cite{akaslompolo2019ascot}.

The rest of the work is distributed as follows. Section~\ref{SEC_EQUATIONS} presents the notation, and the relevant equations and identities. Section~\ref{SEC_SUPERBANANAS} contains a discussion on superbananas from both the radially-global and radially-local perspective.  Our proposed model for prompt ion losses is presented in section~\ref{SEC_MODEL_PL} and validated against full orbit simulations in section~\ref{SEC_VAL_PL}. Section~\ref{SEC_DISCUSSION} is devoted to discussing the results and future steps.

%%%%%%%%%%%%%%%%%%%%%%%%%%%%%%%%%%%%%%%%%%%%%%%%%%%%%%%%%%%%%%%%%%%%%%%%%%%%%%%%%%%%%

\section{Equations}\label{SEC_EQUATIONS}

Let us first define  the coordinate system that we will use. Flux surfaces are labelled by the radial coordinate
\begin{equation}
s=\frac{\Psi_t}{\Psi_{LCMS}}\,,
\end{equation}
where $2\pi\Psi_t$ is the toroidal magnetic flux, and $\Psi_t=\Psi_{LCMS}$ at the last-closed flux-surface. Magnetic field lines on the surface are labelled by an angular coordinate
\begin{equation}
\alpha = \theta -\iota\zeta\,,
\end{equation}
where $\theta$ and $\zeta$ are poloidal and toroidal Boozer angles, respectively, and $\iota$ is the rotational transform. Finally, $l$ is the arc-length along the magnetic field line. In these spatial coordinates, the magnetic field $\mathbf{B}$ can be written as
\begin{equation}
\mathbf{B}=\Psi_{LCMS}\nabla s \times\nabla\alpha\,.
\end{equation}

As velocity coordinates, we choose the magnitude of the particle velocity
\begin{equation}
v = |\mathbf{v}|\,,
\end{equation}
the pitch-angle coordinate
\begin{equation}
\lambda=\frac{1}{B}\frac{v_\perp^2}{v^2}\,,
\end{equation}
and the sign of the parallel velocity
\begin{equation}
\sigma= \frac{v_\parallel}{|v_\parallel|}=\pm 1\,,
\end{equation}
with
\begin{eqnarray}
v_\parallel &= \mathbf{v}\cdot\frac{\mathbf{B}}{|\mathbf{B}|}=  \mathbf{v}\cdot\frac{\mathbf{B}}{B}\,,\nonumber  \\
v_\perp &= \sqrt{v^2-v_\parallel^2}\,.
\end{eqnarray}

\begin{figure}
\centering
\includegraphics[angle=0,width=0.45\columnwidth]{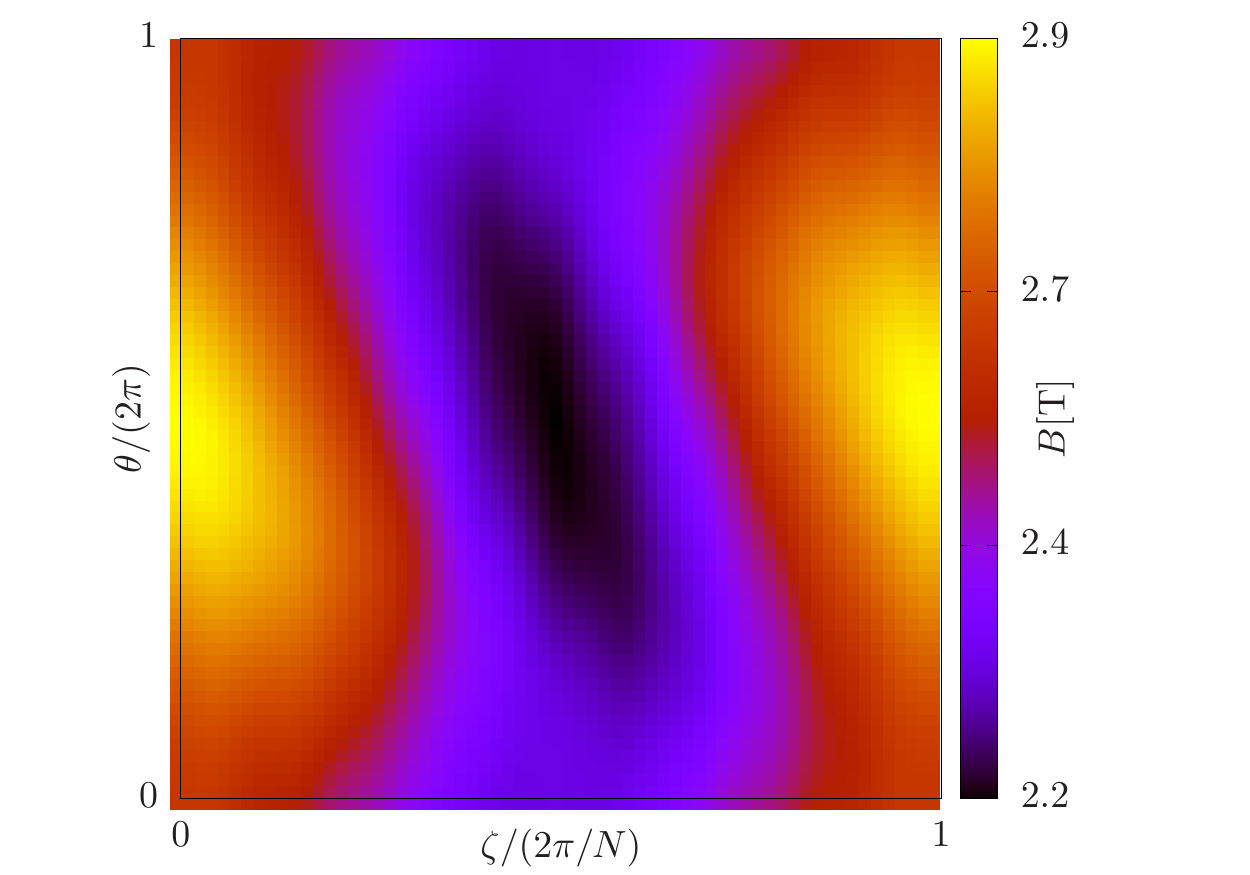}
\includegraphics[angle=0,width=0.45\columnwidth]{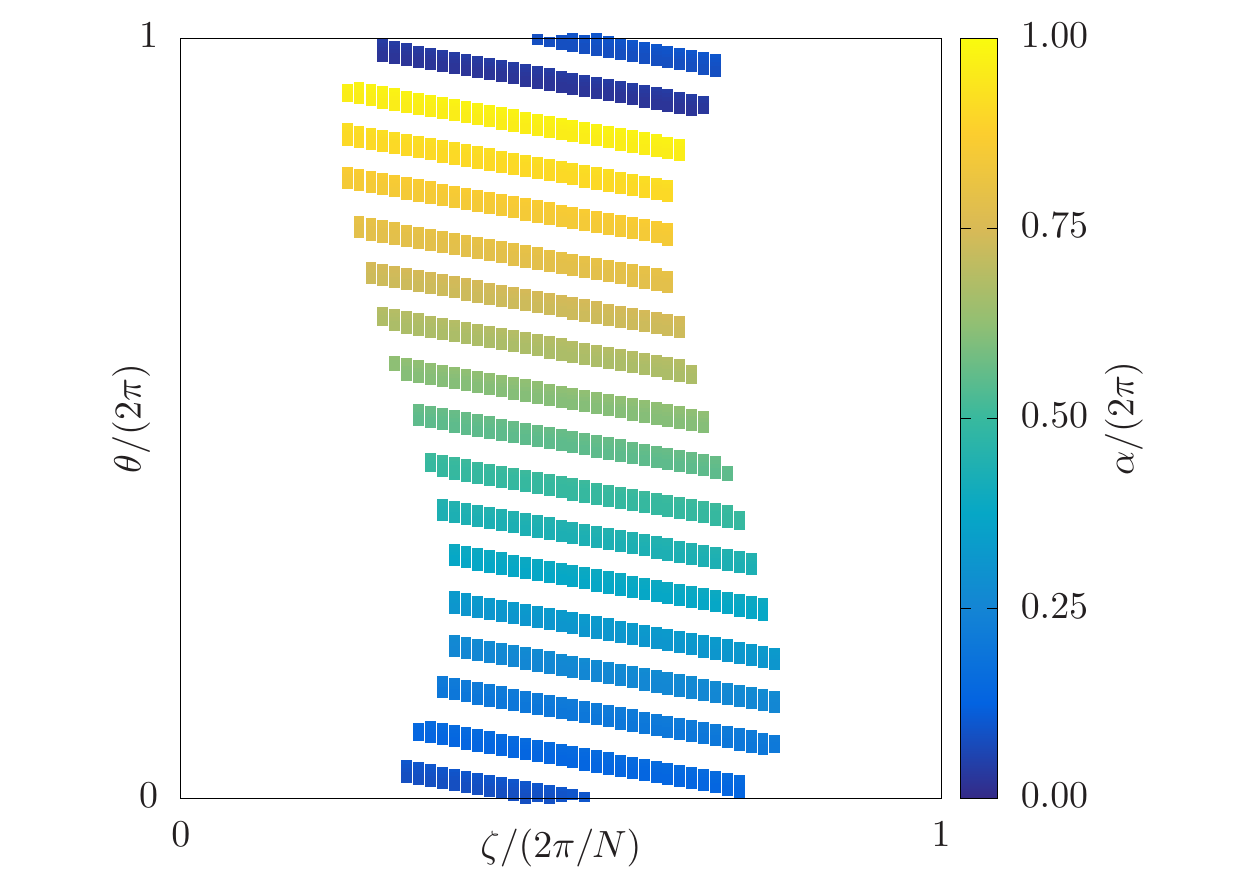}
\caption{Magnetic field strength on flux-surface $s=0.25$ of the  KJM configuration at $\beta=$0\% (left) and, for the same surface, several orbits with $\lambda=0.41\,$T$^{-1}$ and different $\alpha$.}
\label{FIG_B}
\end{figure}

Since passing particles are well confined even in stellarators, we are interested in the behaviour of the energetic ions that are trapped, i.e. those for which $v_\parallel=0$ at some point along their trajectories. Their back and forth motion along the magnetic field is much faster than collisions, and their distribution function, that we denote by $F(s,\alpha,l,v,\lambda,\sigma)$, does not depend either on the coordinate along the field line $l$ or on the sign of the parallel velocity $\sigma$. Figure~\ref{FIG_B} (left) represents the magnetic field strength $B$ on the flux-surface $s=0.25$ of the  KJM configuration of W7-X at $\beta=$0\%. The straight lines of figure~\ref{FIG_B} (right) correspond to orbits with $\lambda=0.41\,$T$^{-1}$ and different values of $\alpha$. By comparing the two figures, it can be seen that the most deeply trapped particles (i.e. with higher $\lambda$) live in $\alpha$ approximately (and slightly higher than) $\pi$: particles moving along these field lines explore the minimum values of $B$ on the flux surface. Conversely, particles moving along the field line labelled by $\alpha=0$ have a smaller range of accesible $\lambda$. These general features are common to all the configurations in this work.

The type of equation that we need to solve for the evaluation of the distribution function of trapped energetic ions is~\cite{calvo2017sqrtnu}
\begin{equation}
\partial_\alpha J \partial_s F - \partial_s J \partial_\alpha F = S\,, \label{EQ_DKE}
\end{equation}
where $S$ is a source term and collisions have been neglected. In our variables, the second adiabatic invariant $J$ reads
\begin{equation}
J(s,\alpha,v,\lambda) = 2v \int_{l_{b_1}}^{l_{b_2}}\sqrt{1-\lambda B}\,\mathrm{d}l\,,
\end{equation}
where the integral over the arc-length is taken between the bounce points $l_{b_1}$ and $l_{b_2}$, i.e., between the points where the parallel velocity of the particle is zero. 

It follows from equation~(\ref{EQ_DKE}) that energetic ions move in phase-space at constant $J$. It is clear then that $\partial_s J=0$ means that the bounce-averaged motion of the trapped ions is directed in the radial direction following a superbanana orbit, something to be avoided. This dynamics can be made more explicit by using the identities (in the absence of radial electric field)~\cite{calvo2017sqrtnu}
\begin{eqnarray}
\partial_\alpha J &=&  \frac{Ze\Psi_{LCMS}}{m}\tau_b \overline{\mathbf{v}_M\cdot\nabla s}\,, \nonumber\\
\partial_s J &=& -\frac{Ze\Psi_{LCMS}}{m}\tau_b \overline{\mathbf{v}_M\cdot\nabla\alpha}\,.\label{EQ_VDJ}
\end{eqnarray}
Here, $Ze$ is the ion charge, $m$ is the ion mass,
\begin{equation}
\mathbf{v}_{M} = \frac{mv^2}{Ze}\left(1-\frac{\lambda B}{2}\right)\frac{\mathbf{B}\times\nabla B}{B^3}\,
\end{equation}
is the magnetic drift, $\overline{f}$ denotes the bounce-average of a function $f(s,\alpha,l,v,\lambda)$ that does not depend on $\sigma$
\begin{equation}
\overline{f}=\frac{2}{v\tau_b}\int_{l_{b_1}}^{l_{b_2}} \frac{f}{\sqrt{1-\lambda B}}\mathrm{d}l\,,\label{EQ_BA}
\end{equation}
and
\begin{equation}
\tau_b=\frac{2}{v}\int_{l_{b_1}}^{l_{b_2}} \frac{\mathrm{d}l}{\sqrt{1-\lambda B}} \,\label{EQ_TAU}
\end{equation}
is the bounce time. According to equation~(\ref{EQ_VDJ}), $\partial_s J=0$ means that the magnetic drift has zero component on the flux surface. More generally, $|\partial_s J| \ll |\partial_\alpha J|$ is equivalent to $|\overline{\mathbf{v}_M\cdot\nabla\alpha}|\ll |\overline{\mathbf{v}_M\cdot\nabla s}|$.

%%%%%%%%%%%%%%%%%%%%%%%%%%%%%%%%%%%%%%%%%%%%%%%%%%%%%%%%%%%%%%%%%%%%%%%%%%%%%%%%%%%%%

\section{Trapped energetic ions and superbanana orbits}\label{SEC_SUPERBANANAS}

\begin{figure}
\centering
\includegraphics[angle=0,width=0.45\columnwidth]{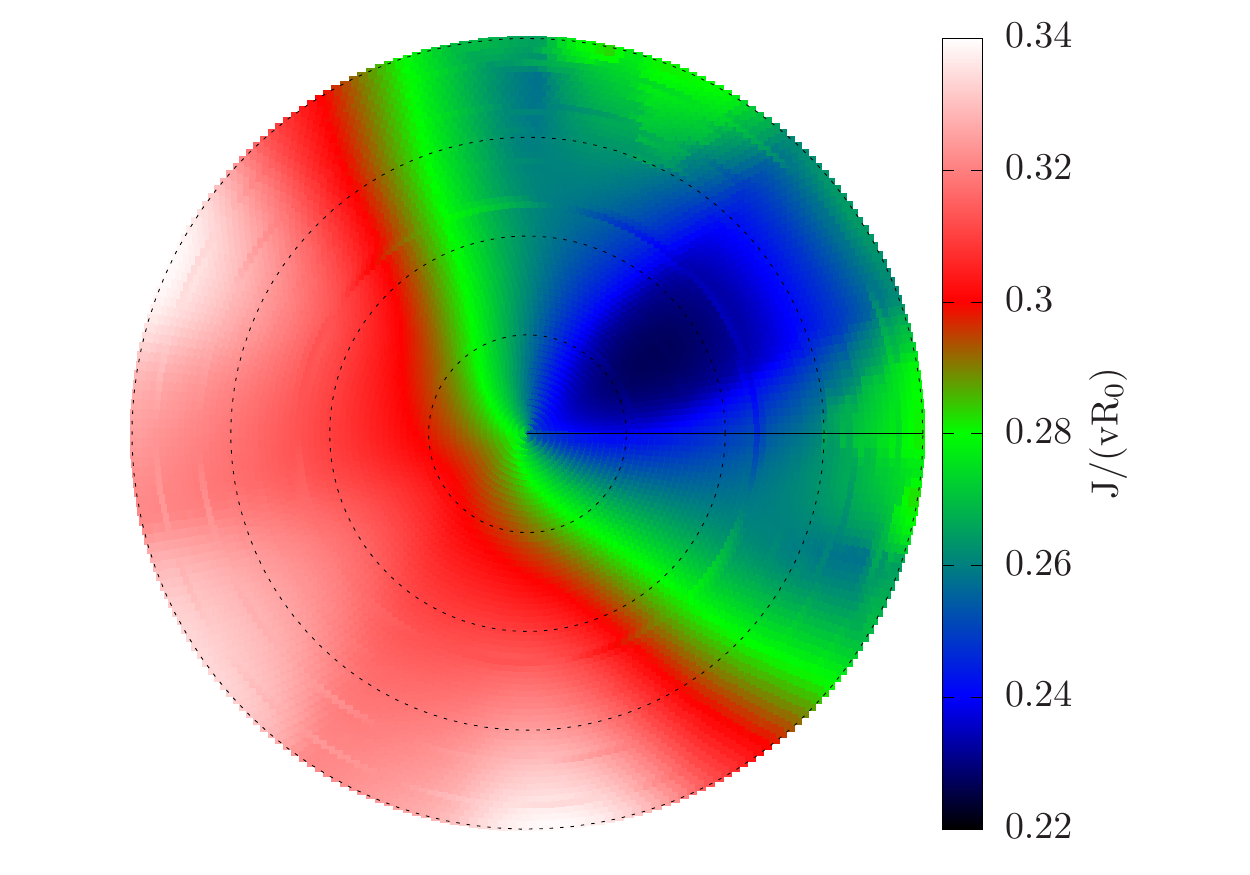}
\includegraphics[angle=0,width=0.45\columnwidth]{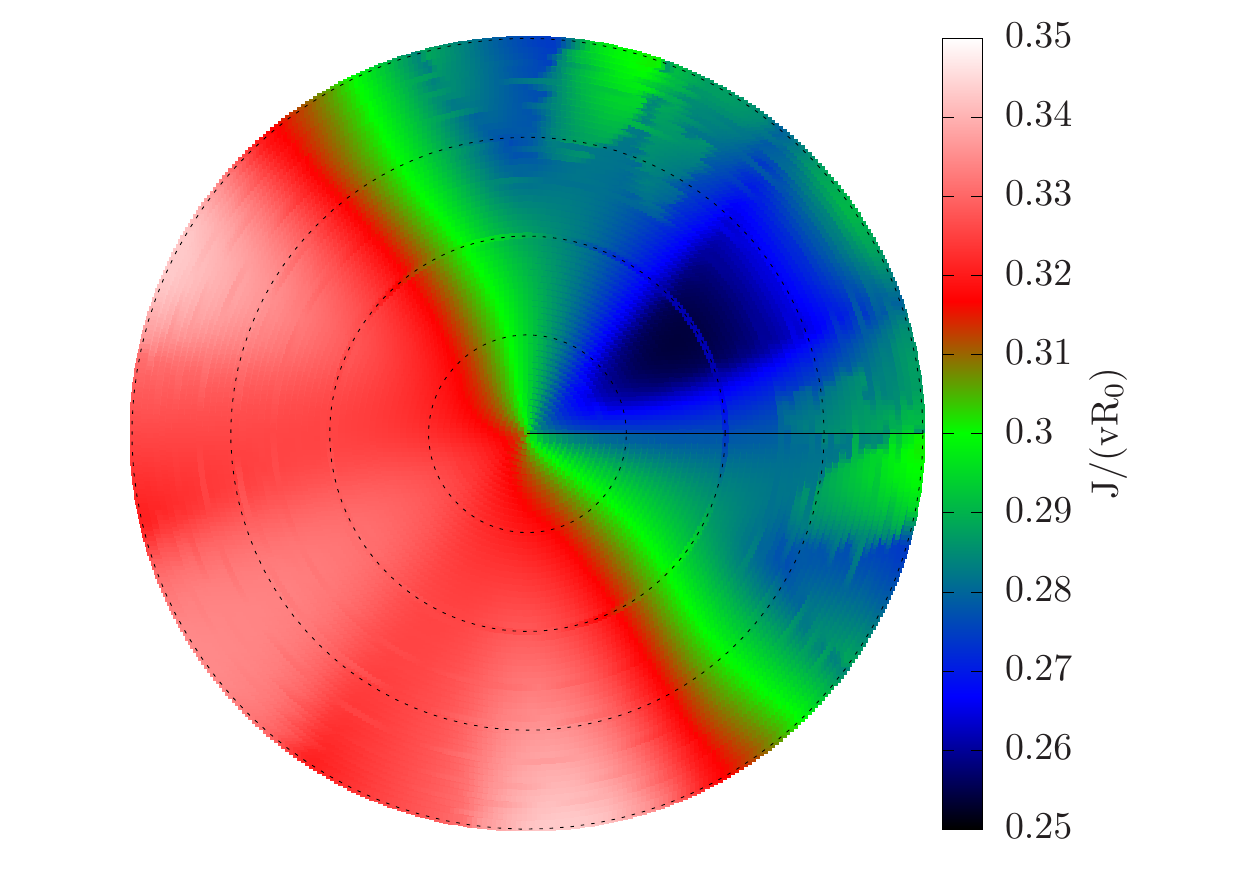}
\includegraphics[angle=0,width=0.45\columnwidth]{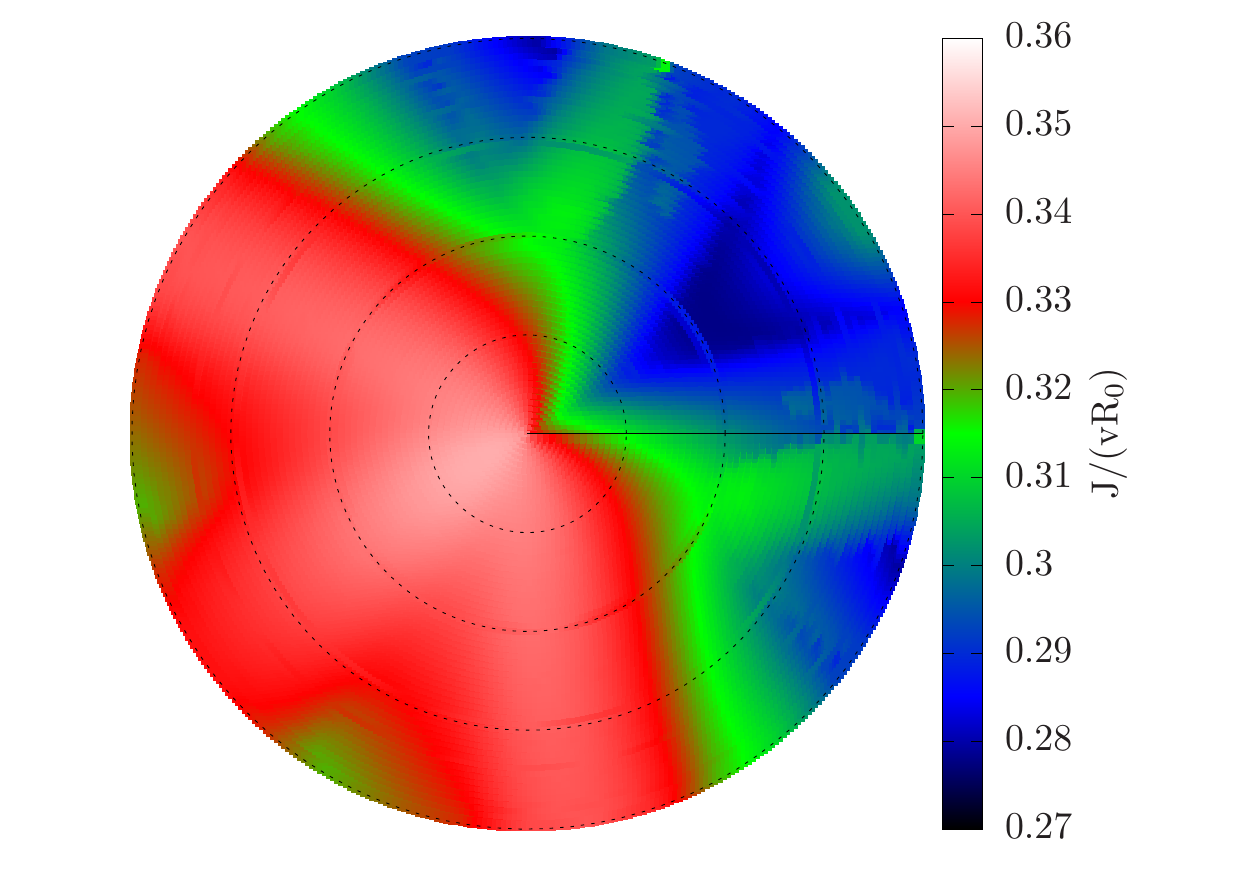}
\includegraphics[angle=0,width=0.45\columnwidth]{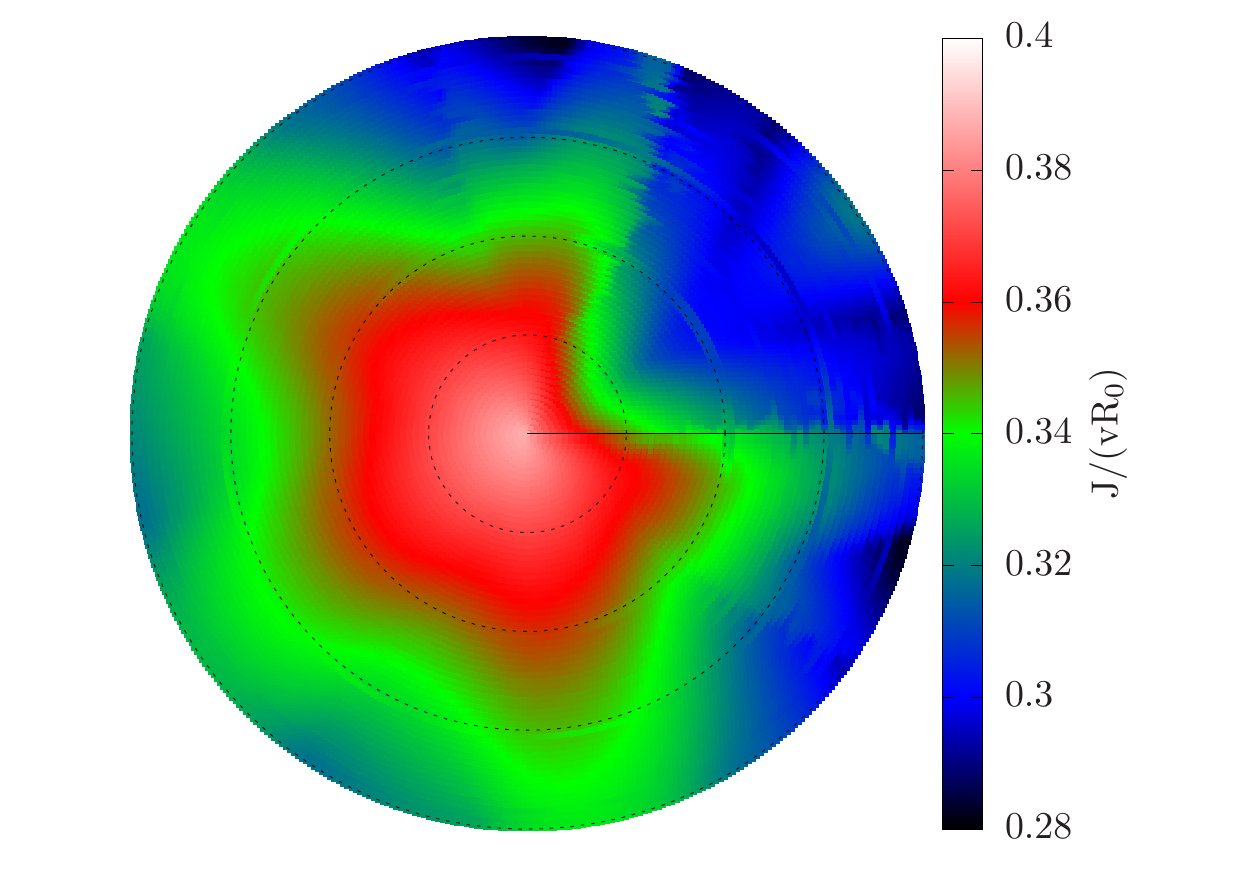}
\includegraphics[angle=0,width=0.45\columnwidth]{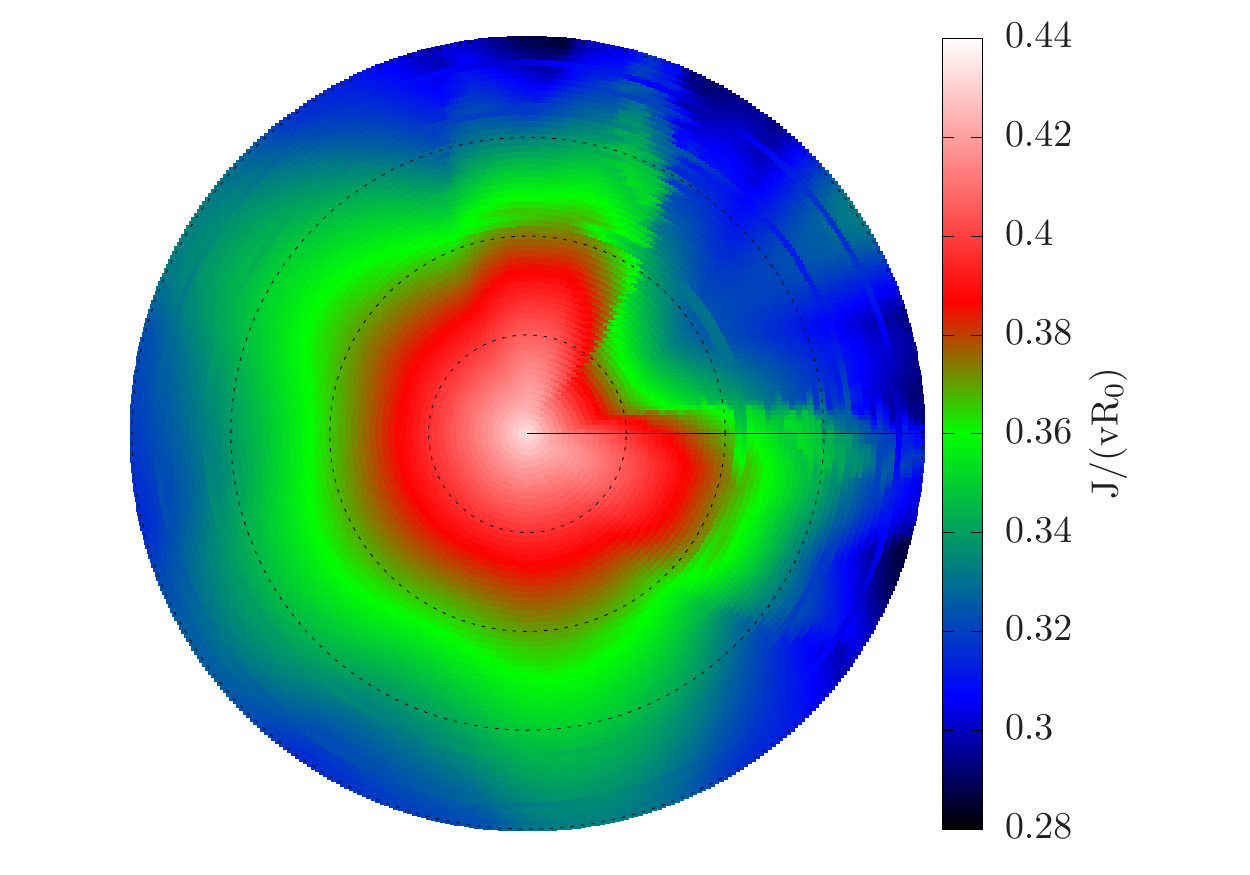}~~~~~~~~~~~~~~~~~~~~~~~~~~~~~~~~~~~~~~~~~~~~~~~~~~~~~~~~~
\caption{Color maps of $J/(vR_0)$ at $\lambda=0.41\,$T$^{-1}$ for the  KJM configuration of W7-X, using parabolic plasma profiles and $\beta=$0\% (top left), 1\% (top right), 2\% (center left),  3\% (center right), 4\% (bottom). Dashed circles represent $s=\,$0.25, 0.50, 0.75 and 1.00, and the horizontal black line corresponds to $\alpha=0$. Note that the colour range varies.}
\label{FIG_J_041}
\end{figure}

Once coordinates $l$ and $\sigma$ have been removed from the problem, we are left with two spatial coordinates, $s$ and $\alpha$, and two velocity coordinates, $v$ and $\lambda$. The two latter are constants of motion in the absence of collisions and electric fields. Additionally, $J$ is conserved if there are no collisions, and this can be employed to determine the trajectories $s(\alpha)$ for given $v$ and $\lambda$. Therefore, one of the clearest ways of depicting the collisionless trajectories of energetic ions are polar $s-\alpha$ maps of $J/v$ at fixed $\lambda$, see e.g.~\cite{mikhailov2002qipc}. This kind of representation is revisited in figure~\ref{FIG_J_041} for five different magnetic fields, corresponding to the KJM configuration of W7-X (also termed high-mirror), with parabolic plasma profiles and volume-averaged $\beta$ ranging from 0\% to 4\%. For this example, $\lambda$ is set to $0.41\,$T$^{-1}$, an intermediate value between very deeply trapped particles and the trapped/passing boundary. The general differences are clear: for small $\beta$, most contours of constant $J$ are open and intersect the last-closed flux-surface. As $\beta$ increases, this ceases to be true, and the region with open trajectories becomes smaller. One can focus, for $\beta=0\%$, on the ions born very close to the magnetic axis. These ions are promptly lost, since their $J$-contour ($J=0.28vR_0$, light green, where $R_0$ is the major radius) crosses almost orthogonally the flux-surfaces. Most ions belong to orbits that behave in a similar way. The only exception consists of a small fraction of orbits ($J=0.24vR_0$, blue) that are closed: their motion is highly localized at small values $\alpha$ and they explore the whole range $0<s<0.8$. At this value of $\lambda$, things do not change qualitatively for $\beta=1\%$ and $\beta=2\%$. The shape of orbits that contain the magnetic axis (light green, although corresponding different value of $J$ than for smaller values of $\beta$) changes slightly, and the region of closed orbits localized in the angular coordinate $\alpha$ (blue, although corresponding different value of $J$ than for smaller values of $\beta$) starts to shrink. Once $\beta=3\%$ is achieved, the map changes qualitatively, and the maximum-$J$ property starts to become apparent. All the orbits with $J>0.34vR_0$ (red and reddish green) are closed. The outermost closed-contour of constant $J$ is $J\approx 0.34vR_0$ (light green): particles born with this value of $J$ precess in $\alpha$ while moving radially from $s=0.2$ to $s\approx 1$. This means that all orbits starting at $s<0.2$ are confined, and that for outer flux-surfaces ($J<0.34vR_0$, bluish green and blue) open orbits exist. For $\beta=4\%$, $J$ becomes nearly independent of $\alpha$ and, as a consequence of that, the radial excursions of the confined orbits become smaller. Note however that there still exist orbits starting at $s\ge 0.6$, with small value of $J$, that are open.

\begin{figure}
\centering
\includegraphics[angle=0,width=0.45\columnwidth]{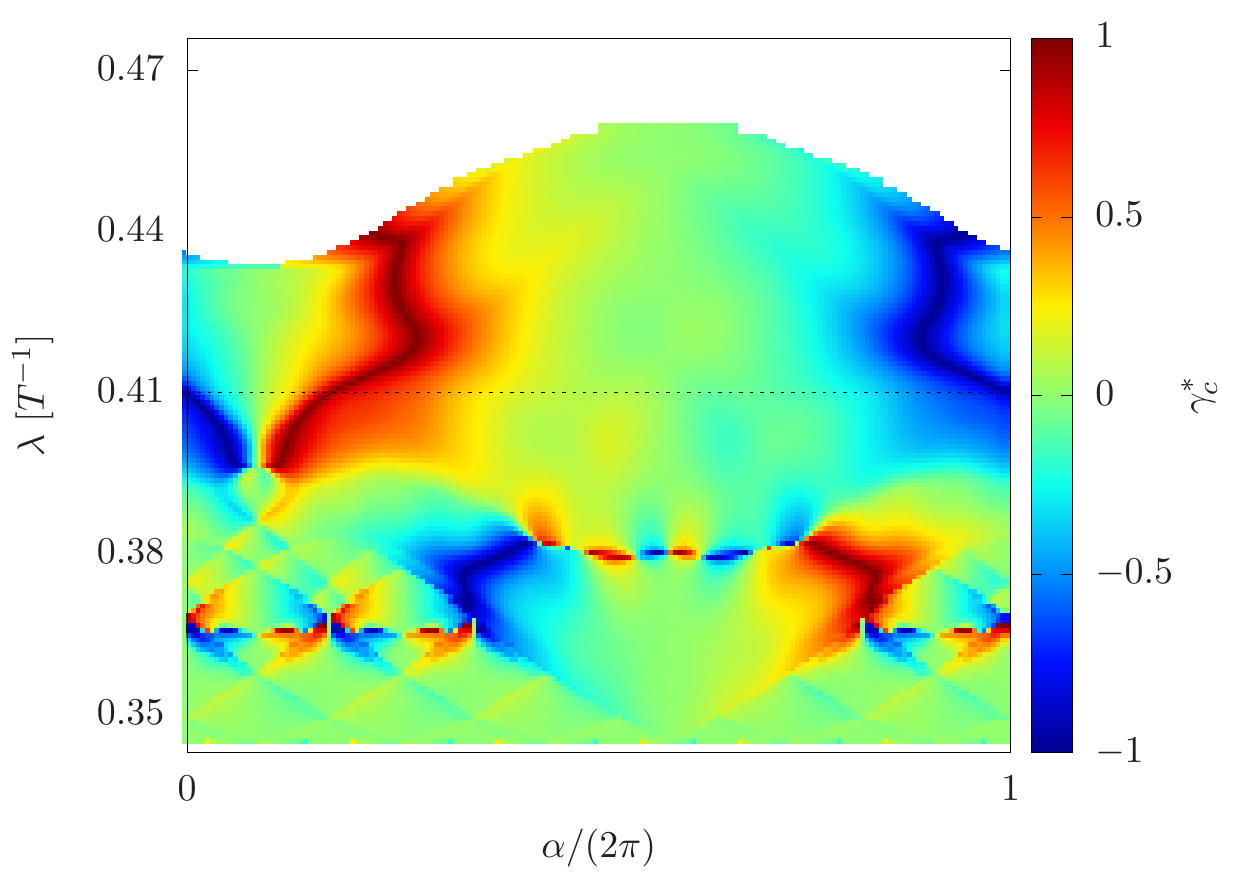}
\includegraphics[angle=0,width=0.45\columnwidth]{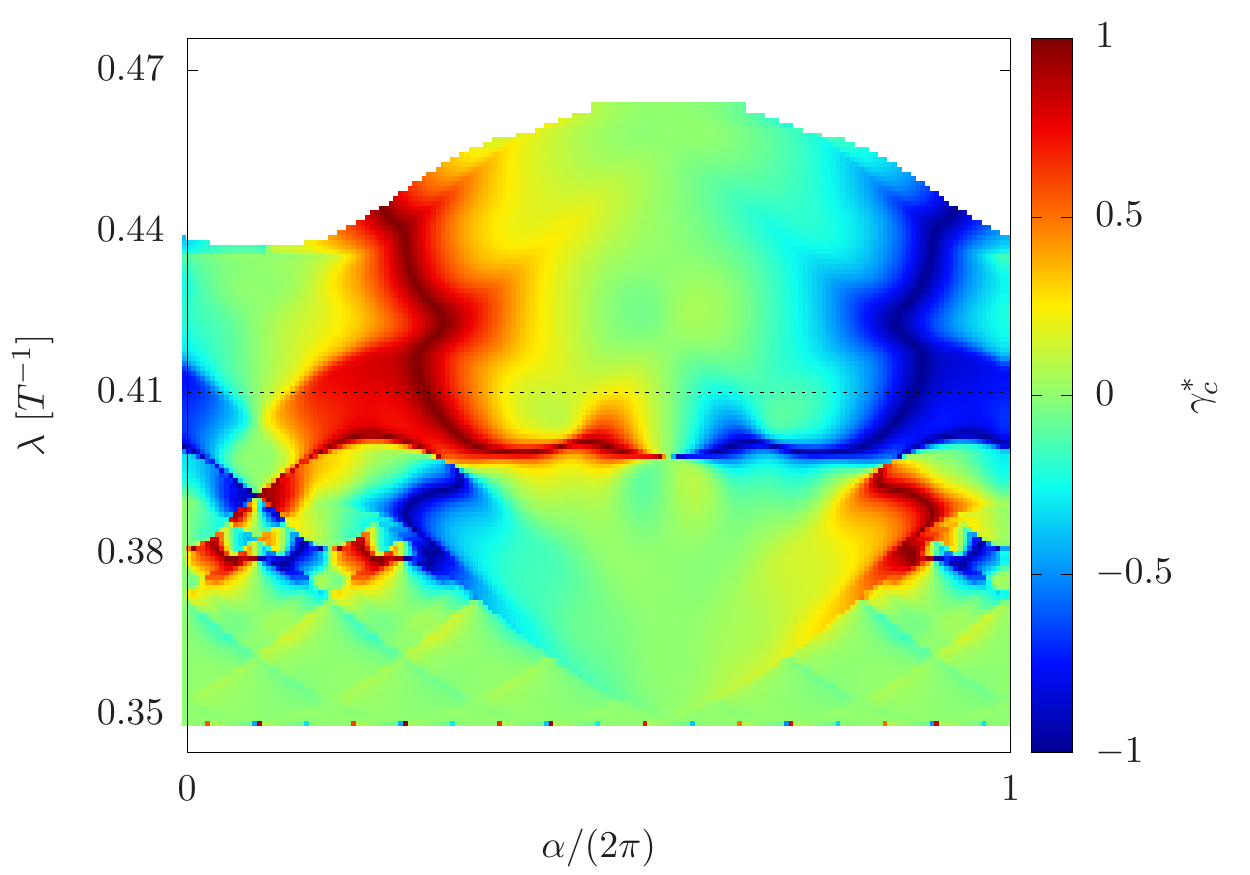}
\includegraphics[angle=0,width=0.45\columnwidth]{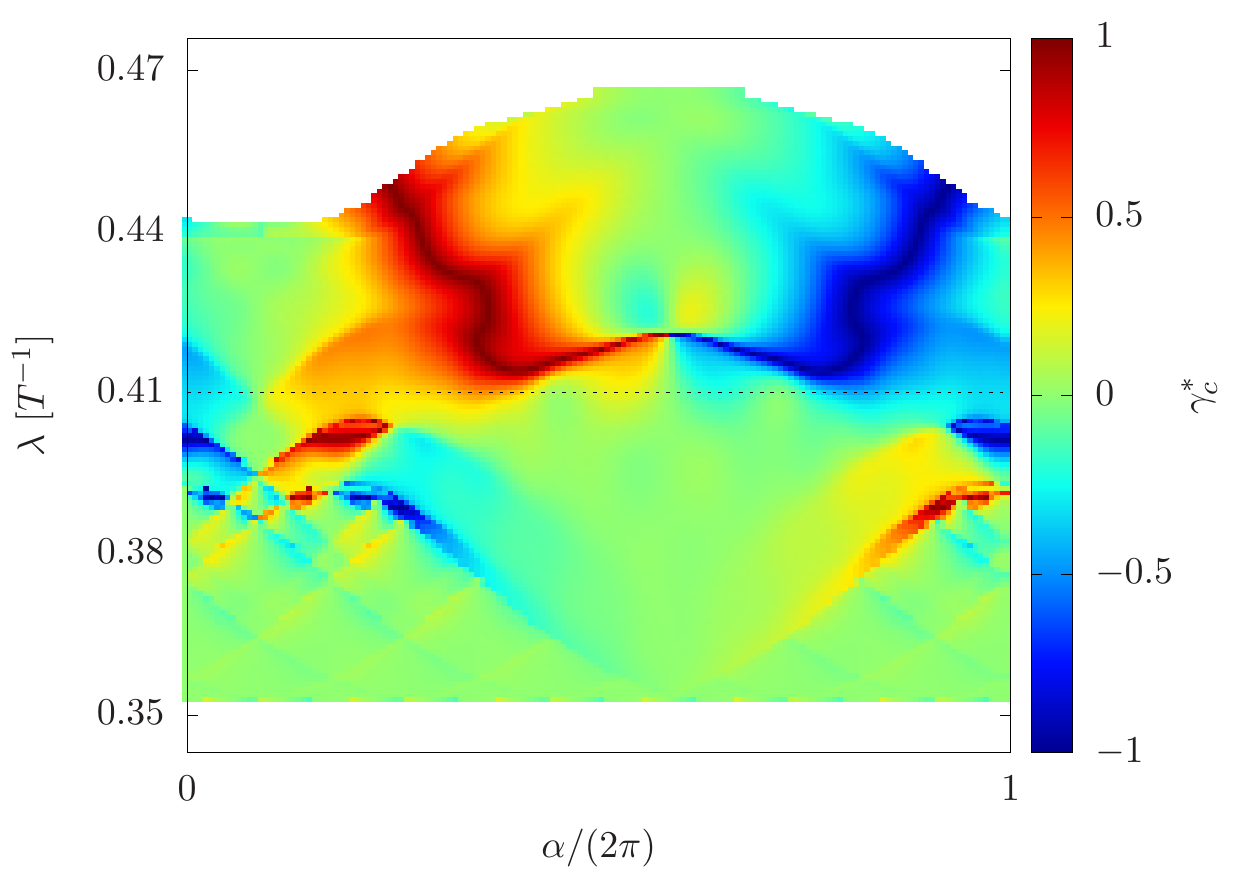}
\includegraphics[angle=0,width=0.45\columnwidth]{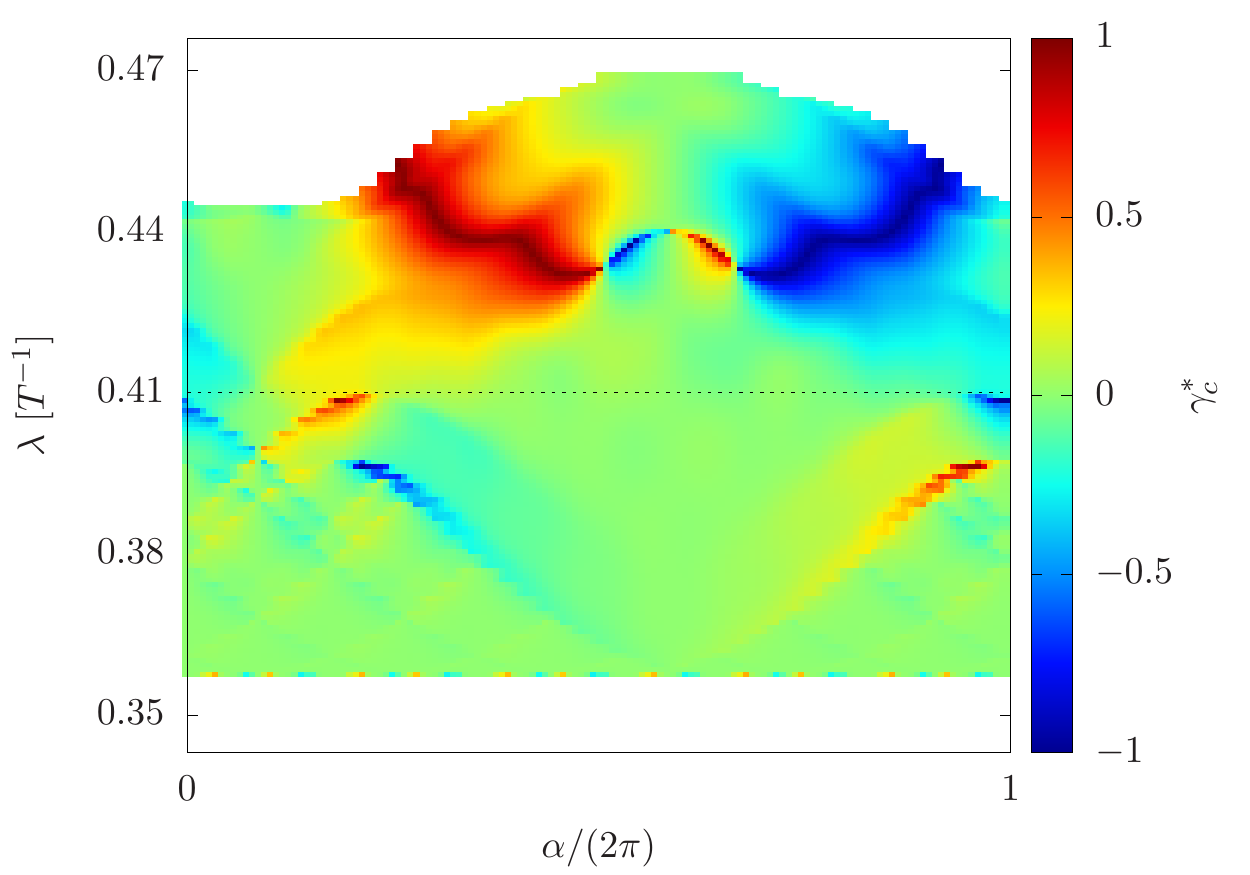}
\includegraphics[angle=0,width=0.45\columnwidth]{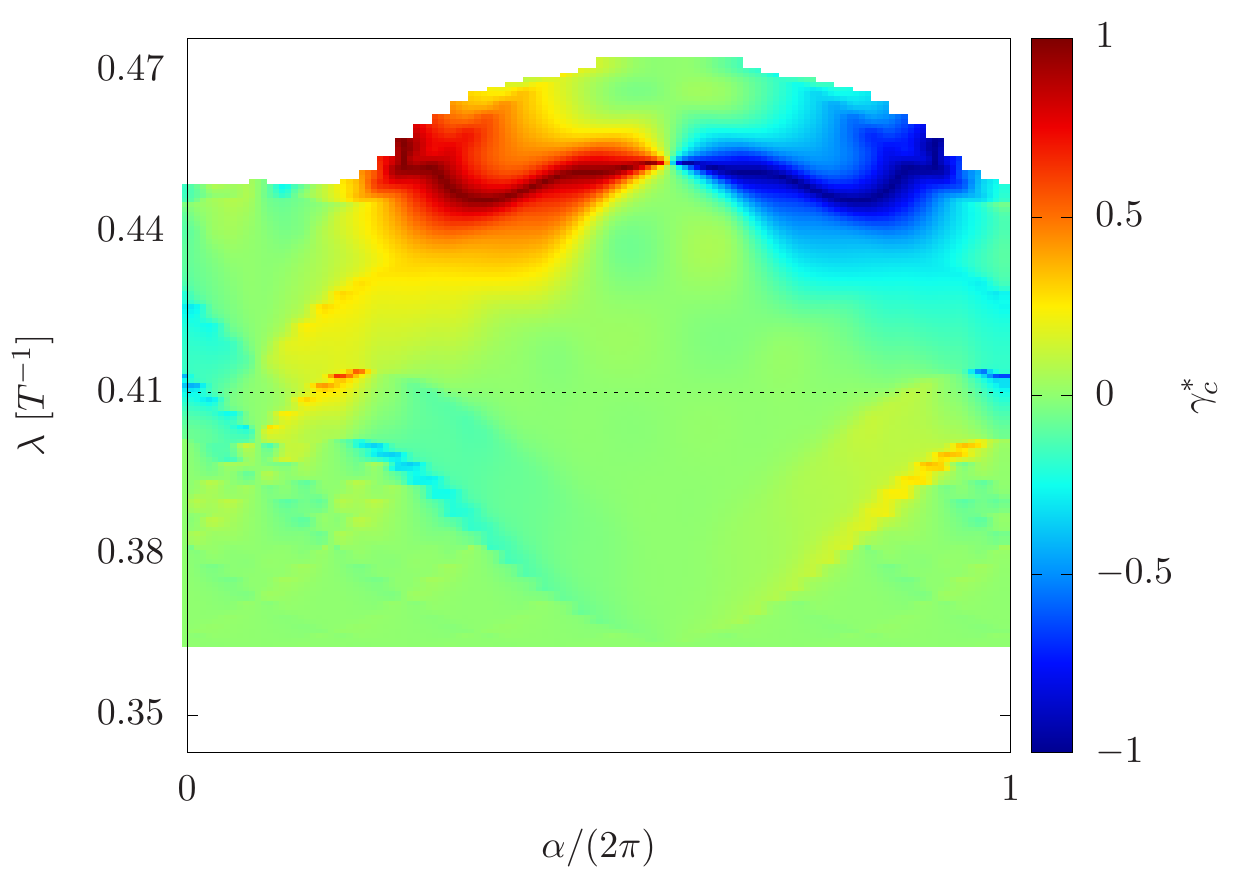}~~~~~~~~~~~~~~~~~~~~~~~~~~~~~~~~~~~~~~~~~~~~~~~~~~~~~~~~~
\caption{Angular and velocity dependence of $\gamma_{\mathrm{c}}^*$ at $s=0.25$ for the  KJM configuration of W7-X, using parabolic plasma profiles and $\beta=$0\% (top left), 1\% (top right), 2\% (center left),  3\% (center right), 4\% (bottom). The horizontal dashed line highlights the value of $\lambda$ employed at figure~\ref{FIG_J_041}, 0.41$\,$T$^{-1}$.}
\label{FIG_GAMMAC025}
\end{figure}

\begin{figure}
\centering
\includegraphics[angle=0,width=0.45\columnwidth]{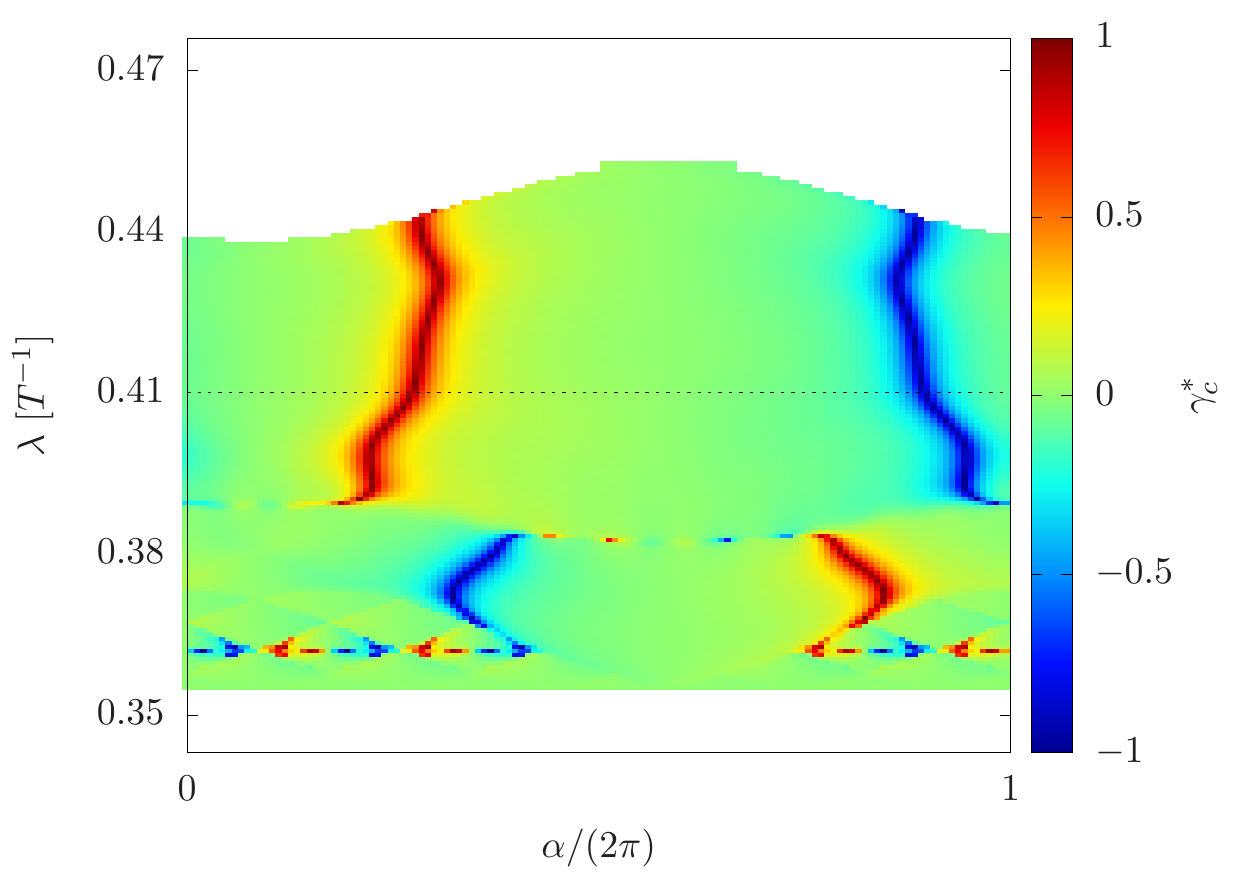}
\includegraphics[angle=0,width=0.45\columnwidth]{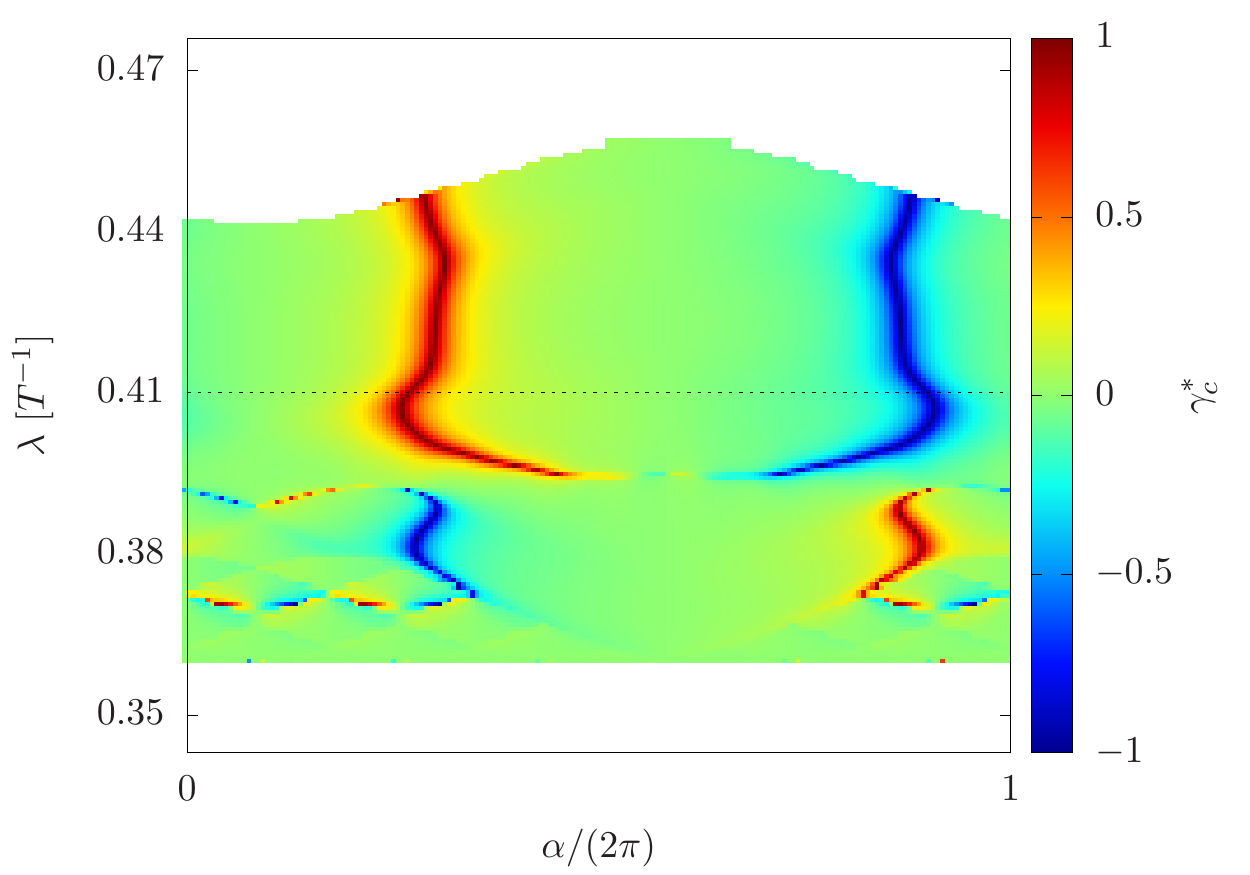}
\includegraphics[angle=0,width=0.45\columnwidth]{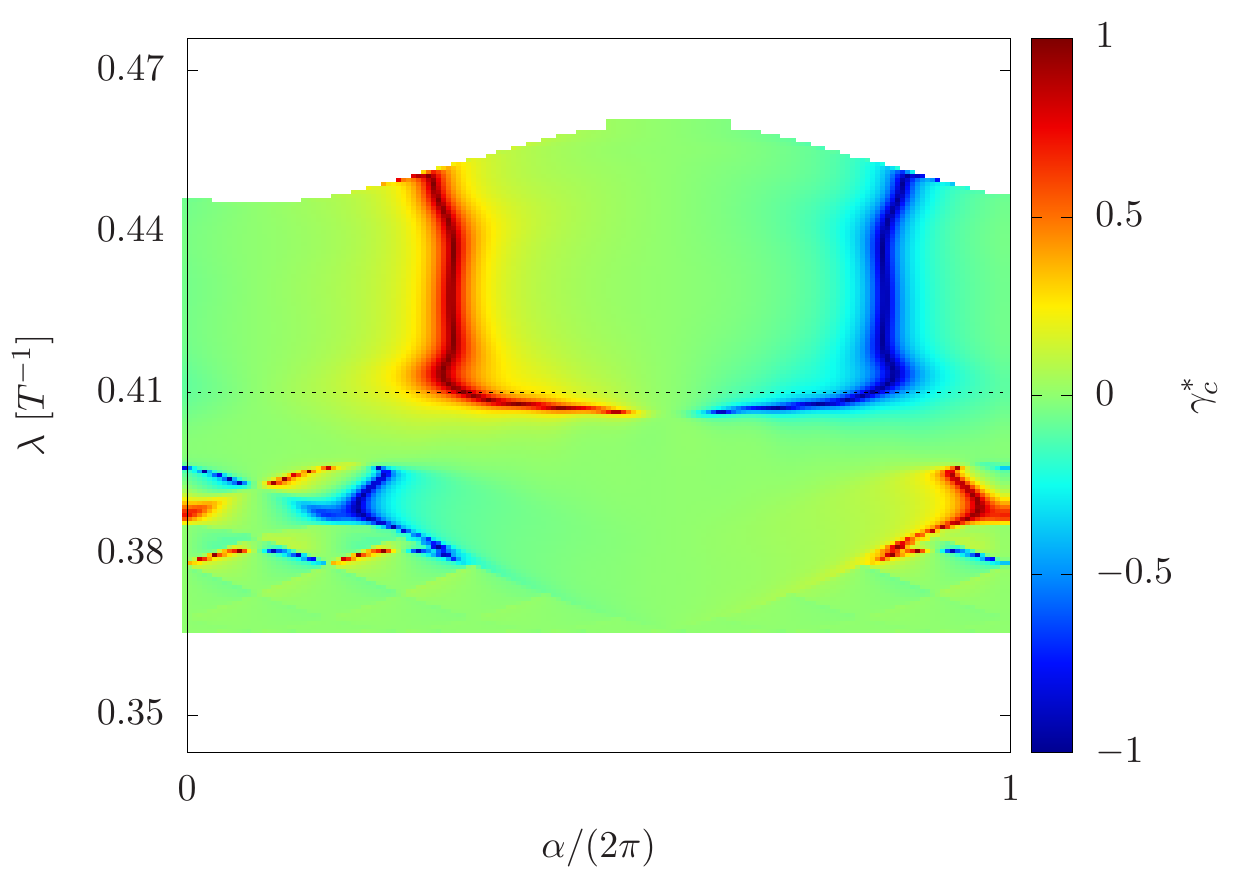}
\includegraphics[angle=0,width=0.45\columnwidth]{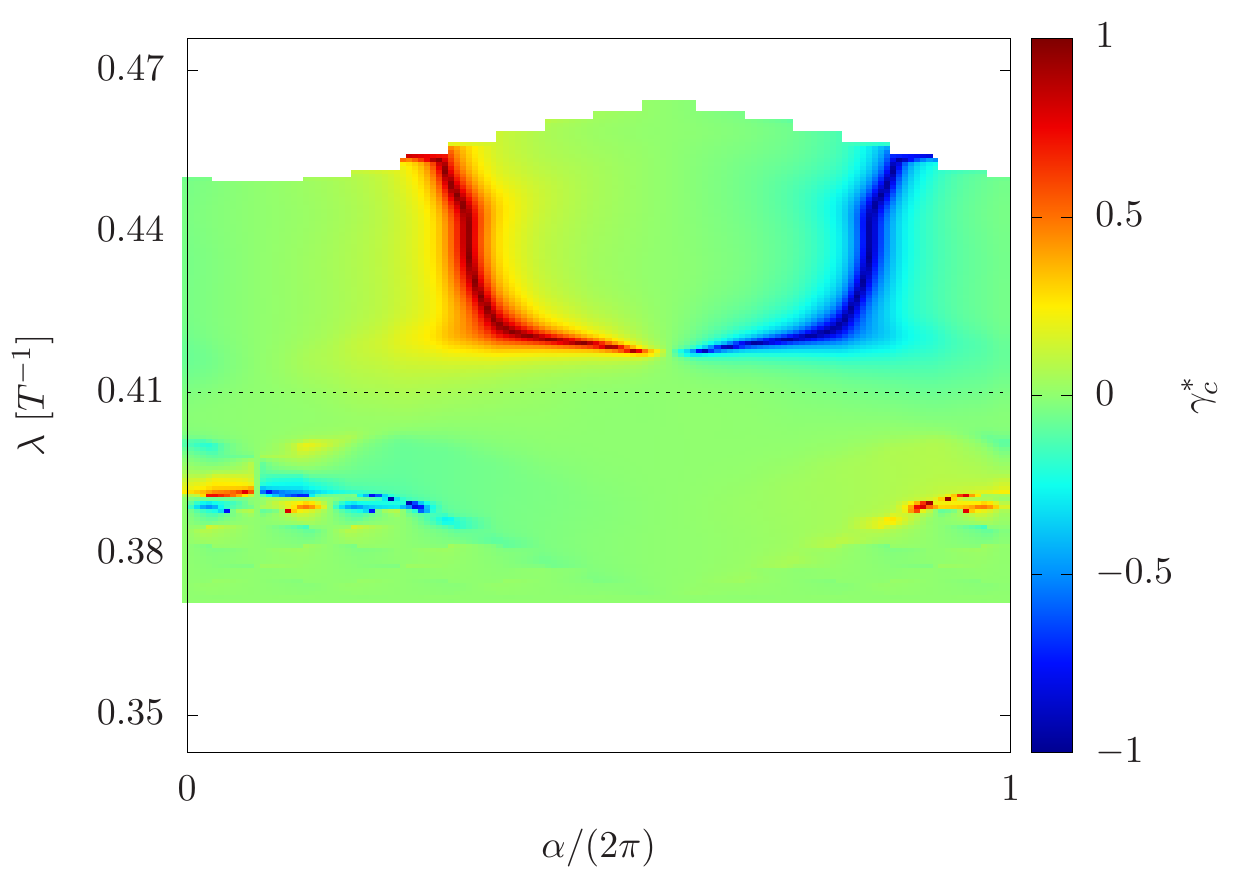}
\includegraphics[angle=0,width=0.45\columnwidth]{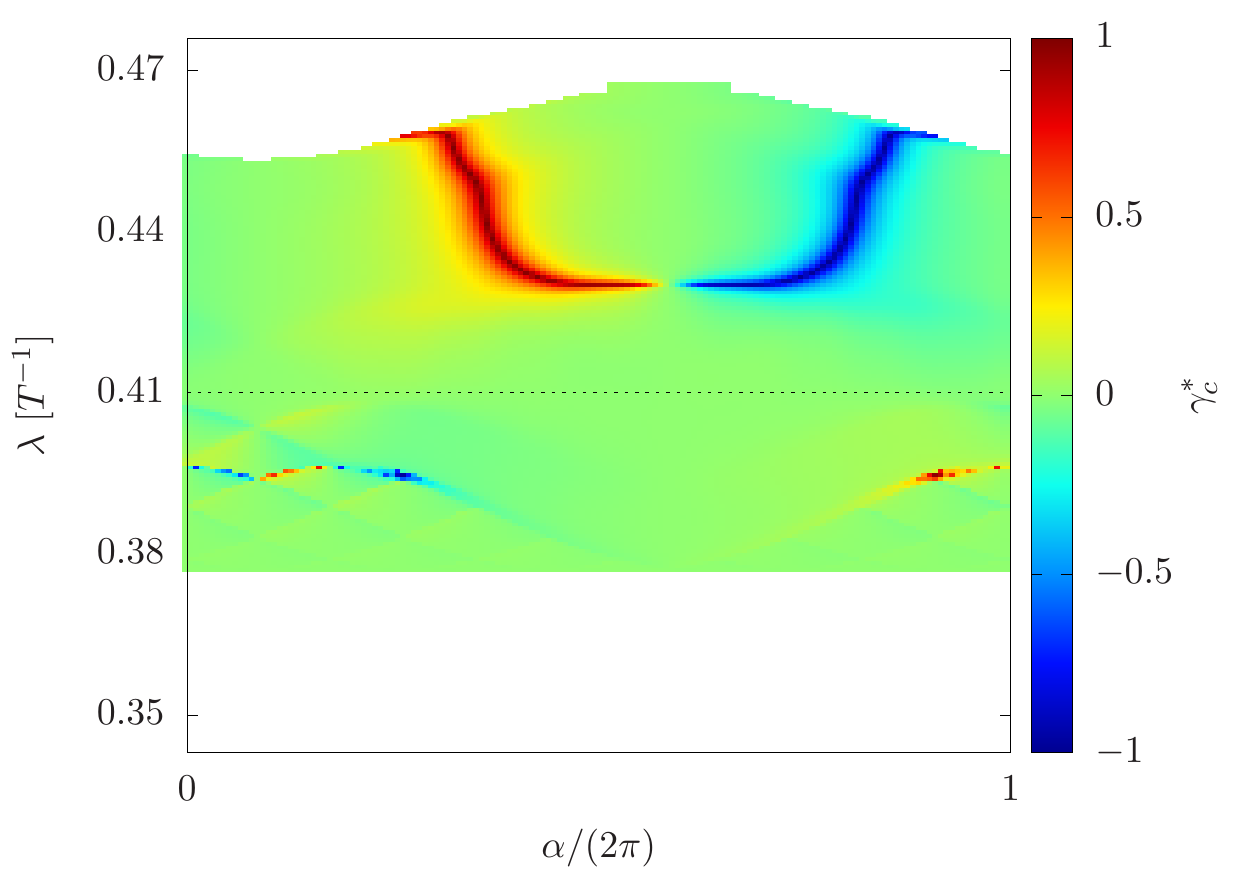}~~~~~~~~~~~~~~~~~~~~~~~~~~~~~~~~~~~~~~~~~~~~~~~~~~~~~~~~~
\caption{Angular and velocity dependence of $\gamma_{\mathrm{c}}^*$ at $s=0.06$ for the  KJM configuration of W7-X, using parabolic plasma profiles and $\beta=$0\% (top left), 1\% (top right), 2\% (center left),  3\% (center right), 4\% (bottom). The horizontal dashed line highlights the value of $\lambda$ employed at figure~\ref{FIG_J_041}, 0.41$\,$T$^{-1}$.}
\label{FIG_GAMMAC006}
\end{figure}

\subsection{Radially local description of superbanana orbits}

In this subsection, we undertake a radially local description of the $J$-maps. Even though, as we have seen, the orbits of energetic ions are inherently global, we will argue that a local description is enough for a reasonable characterization of what properties of the energetic ions cause them to be promptly lost, something not always easy to do with a (radially global) Monte Carlo guiding-center code. As a consequence of this, a local approach can provide figures of merit for energetic ion confinement that may be employed by stellarator optimization codes. After all, these codes tend to use a radially local approach: the properties of the magnetic configurations, specially within the optimization loop, are naturally evaluated at a discrete set of flux-surfaces. In some cases, there are solid theoretical reasons to do so in just a few surfaces: for instance, it is expected that perfect optimization with respect to quasisymmetry cannot be achieved in the full volume of a stellarator~\cite{garrenboozer1991qs}. This approach can be easily justified in the case of optimization with respect to energetic ion confinement: if the flux surface labelled by $s=s_0$ is perfectly optimized, all energetic ions born at $s\le s_0$ will be confined. The optimization of quasisymmetry on a single flux-surface can indeed produce configurations with very good energetic ion confinement~\cite{henneberg2019fastions}.

In what follows we rearrange the data from figure~\ref{FIG_J_041} in terms of radially local quantities. To that end, figure~\ref{FIG_GAMMAC025} represents 
\begin{equation}
\gamma_{\mathrm{c}}^*=\frac{2}{\pi} \arctan{\frac{\partial_\alpha J}{|\partial_s J|}}=\frac{2}{\pi} \arctan{\frac{\overline{\mathbf{v}_M\cdot\nabla s}}{|\overline{\mathbf{v}_M\cdot\nabla\alpha}|}}\,,
\label{EQ_GAMMACS}
\end{equation}
at the flux surface $s=0.25$. This quantity differs (for orbits with $\overline{\mathbf{v}_M\cdot\nabla\alpha}<0$) in sign from
\begin{equation}
\gamma_{\mathrm{c}}=\frac{2}{\pi} \arctan{\frac{\overline{\mathbf{v}_M\cdot\nabla s}}{\overline{\mathbf{v}_M\cdot\nabla\alpha}}}\,,
\end{equation}
already defined in~\cite{nemov2008gammac}, which means that $(\gamma_{\mathrm{c}}^*)^2=\gamma_{\mathrm{c}}^2${\footnote{{Nemov's $\gamma_c$ in has an additional dimensionless $\alpha-$dependent quantity ($|\nabla\psi_i|$ in equation (51) of ~\cite{nemov2008gammac}), in the argument of the arctan, that should not affect the discussion of this section.}}}. The function $\gamma_{\mathrm{c}}^*$ identifies clearly the superbanana orbits: where $\gamma_{\mathrm{c}}^*\approx -1$ (blue), the ions drift radially inwards with no tangential precession; where $\gamma_{\mathrm{c}}^*\approx 1$ (red), they do so outwards. These points correspond to points in figure~\ref{FIG_J_041} where the contours of constant $J$ intersect those of $s=0.25$ almost orthogonally. 

By comparing the maps of $\gamma_{\mathrm{c}}^*$ for different configurations, the effect of $\beta$  in velocity space can be studied in more detail (we note that a similar, less detailed exercise was done in~\cite{nemov2014ripple} to study qualitatively the effect of the coil ripple in HSX). At $\beta\le 1\%$, most of the $\lambda$-range contains superbananas. Things start to improve at $\beta$ around 2\%, where two separate regions with superbananas have developed: one at intermediate values of $\lambda$ and another one for deeply trapped ions, corresponding to the largest accessible values of $\lambda$. If $\beta$ is further increased, the former superbananas start to disappear, while the latter become concentrated in a narrow region of larger values of $\lambda$, where very deeply trapped ions live. Figure~\ref{FIG_GAMMAC006} represents the same information than figure~\ref{FIG_GAMMAC025} at an inner position, $s=0.06$. In this case, the angular extension of the superbananas is smaller, but their qualitative behaviour with $\beta$ is the same.

%%%%%%%%%%%%%%%%%%%%%%%%%%%%%%%%%%%%%%%%%%%%%%%%%%%%%%%%%%%%%%%%%%%%%%%%%%%%%%%%%%%%%

\section{Models for prompt ion losses}\label{SEC_MODEL_PL}

In this section, we discuss how the information contained in figures~\ref{FIG_GAMMAC025} and~\ref{FIG_GAMMAC006} can be employed to model the prompt losses of energetic ions. Indeed, the well-known $\Gamma_{\mathrm{c}}$ proxy is a simple surface integral of $(\gamma_{\mathrm{c}}^*)^2$ on the phase space~\cite{nemov2008gammac}:
\begin{equation}
\Gamma_{\mathrm{c}} =\frac{\pi}{4\sqrt{2}} \fsa{\int_{B^{-1}_{\mathrm{max}}}^{B^{-1}}\mathrm{d}\lambda\frac{B}{\sqrt{1-\lambda B}} (\gamma_{\mathrm{c}}^*)^2}\,,
\label{EQ_GAMMAC}
\end{equation}
where $\fsa{...}$ denotes flux-surface average, $B_\mathrm{max}$ is the maximum value of $B$ on the flux-surface and we have made use of 
\begin{equation}
\int\mathrm{d}^3v (...) = \pi \sum_\sigma\int_0^\infty\mathrm{d}v\,v^2\int_0^{B^{-1}}\mathrm{d}\lambda\frac{B}{\sqrt{1-\lambda B}} (...) \,.
\label{EQ_VELINT}
\end{equation}
Equation (\ref{EQ_GAMMAC}) can be more explicitly written as
\begin{equation}
\Gamma_{\mathrm{c}} =\frac{\pi}{16\sqrt{2}} \frac{\mathrm{d}\Psi_t}{\mathrm{d}V} \int_0^{2\pi}\mathrm{d}\alpha\sum_w\int_{B^{-1}_{\mathrm{M},w}}^{B^{-1}_{\mathrm{m},w}}\mathrm{d}\lambda v\tau_b (\gamma_{\mathrm{c}}^*)^2\,,
\end{equation}
where $V$ is the volume enclosed by the flux-surface, the summation is taken over all the wells $w$ at a given $\alpha$, $B^{-1}_{\mathrm{M},w}$ and $B^{-1}_{\mathrm{m},w}$ are the maximum and minimum values of $B$ on well $w$. In a generic stellarator, $\gamma_{\mathrm{c}}^*$ (as well as $\tau_b$) depends on $\lambda$, $\alpha$ and $w$. Only if it is close enough to omnigeneity (i.e., to perfect neoclassical optimization), just one minimum of $B$ along the field line exists for each $\alpha$, and no ripple wells (with the exception of configurations like those of~\cite{parra2015omni}). If this is the case (or if ripple wells can be ignored) equation (\ref{EQ_GAMMAC}) can be further simplified to
\begin{equation}
\Gamma_{\mathrm{c}} =\frac{\pi}{16\sqrt{2}} \frac{\mathrm{d}\Psi_t}{\mathrm{d}V} \int_0^{2\pi}\mathrm{d}\alpha\int_{B^{-1}_{\mathrm{max}}}^{B^{-1}_{\mathrm{min}}}\mathrm{d}\lambda v\tau_b (\gamma_{\mathrm{c}}^*)^2\,,
\end{equation}
where $B_\mathrm{max}$ and $B_\mathrm{min}$ are the ($\alpha$-dependent) maximum and minimum values of $B$ on the field line. 

Ripple wells typically have a limited angular extension. This means that, even if $\gamma_c^*$ is far from 0 at these locations, the particle may move in $\alpha$ and abandon the well before actually making a large radial excursion. Transitions between locally trapped and locally passing particles, that cause many small radial excursions, are usually associated to stochastic losses~\cite{beidler2001stochastic}. In the rest of this work about prompt losses, we will ignore ripple wells and $\gamma_{\mathrm{c}}^*$ will be a function of $\alpha$ and $\lambda$ for each flux-surface.

The $\Gamma_{\mathrm{c}}$ proxy has been employed in the optimization of quasisymmetric stellarators~\cite{bader2019fastions}. Its use in helias devices has however been less successful~\cite{drevlak2014fastions}. In this work, we will employ two variations over $\Gamma_{\mathrm{c}}$:
\begin{equation}
\check \Gamma_{\mathrm{c}} =\frac{1}{2} \fsa{\int_{B^{-1}_{\mathrm{max}}}^{B^{-1}}\mathrm{d}\lambda\frac{B}{\sqrt{1-\lambda B}} (\gamma_{\mathrm{c}}^*)^2} =\frac{\pi}{2\sqrt{2}} \Gamma_{\mathrm{c}}
\end{equation}
and
\begin{equation}
\hat \Gamma_{\mathrm{c}} =\frac{1}{2} \fsa{\int_{B^{-1}_{\mathrm{max}}}^{B^{-1}}\mathrm{d}\lambda\frac{B}{\sqrt{1-\lambda B}} |\gamma_{\mathrm{c}}^*|}\,.
\end{equation}
The first one is proportional to $\Gamma_{\mathrm{c}}$, normalized so that its range of variation is the same as that of other quantities defined in subsections~\ref{SEC_MODEL1}, \ref{SEC_MODEL2}; the second one will additionally guarantee, as we will see in section~\ref{SEC_COMP}, a more \textit{linear} relation with the fraction of prompt losses.

In the following subsections, we propose several models of increasing complexity that, building on $\gamma_{\mathrm{c}}^*$, try to model with greater detail the prompt losses of energetic ions in stellarators. The quantity $\Gamma_\delta$ will be presented in section~\ref{SEC_MODEL1} and the quantity $\Gamma_\alpha$ in section \ref{SEC_MODEL2}. The validation of $\Gamma_\alpha$ will be the main result of this work, but the different predictive capability of $\Gamma_\delta$ and $\Gamma_\alpha$ will provide a clear picture of the key aspects of prompt losses. 

It must be emphasized here that, even though they will use the same basic information, the philosophy of these models will be different from that of $\Gamma_{\mathrm{c}}$. The models presented in sections~\ref{SEC_MODEL1} and~\ref{SEC_MODEL2} will consist of a classification of orbits into confined or unconfined (each model using different criteria for this classification). This means that they can be employed to give a quantitative prediction of the loss fraction of energetic ions. No such interpretation is possible with the proxy $\Gamma_{\mathrm{c}}$, which is simply an estimate of how much the contours of constant $J$ deviate from the local flux-surface.

\subsection{Model I: existence of superbananas}\label{SEC_MODEL1}

\begin{figure}
\centering
\includegraphics[angle=0,width=0.45\columnwidth]{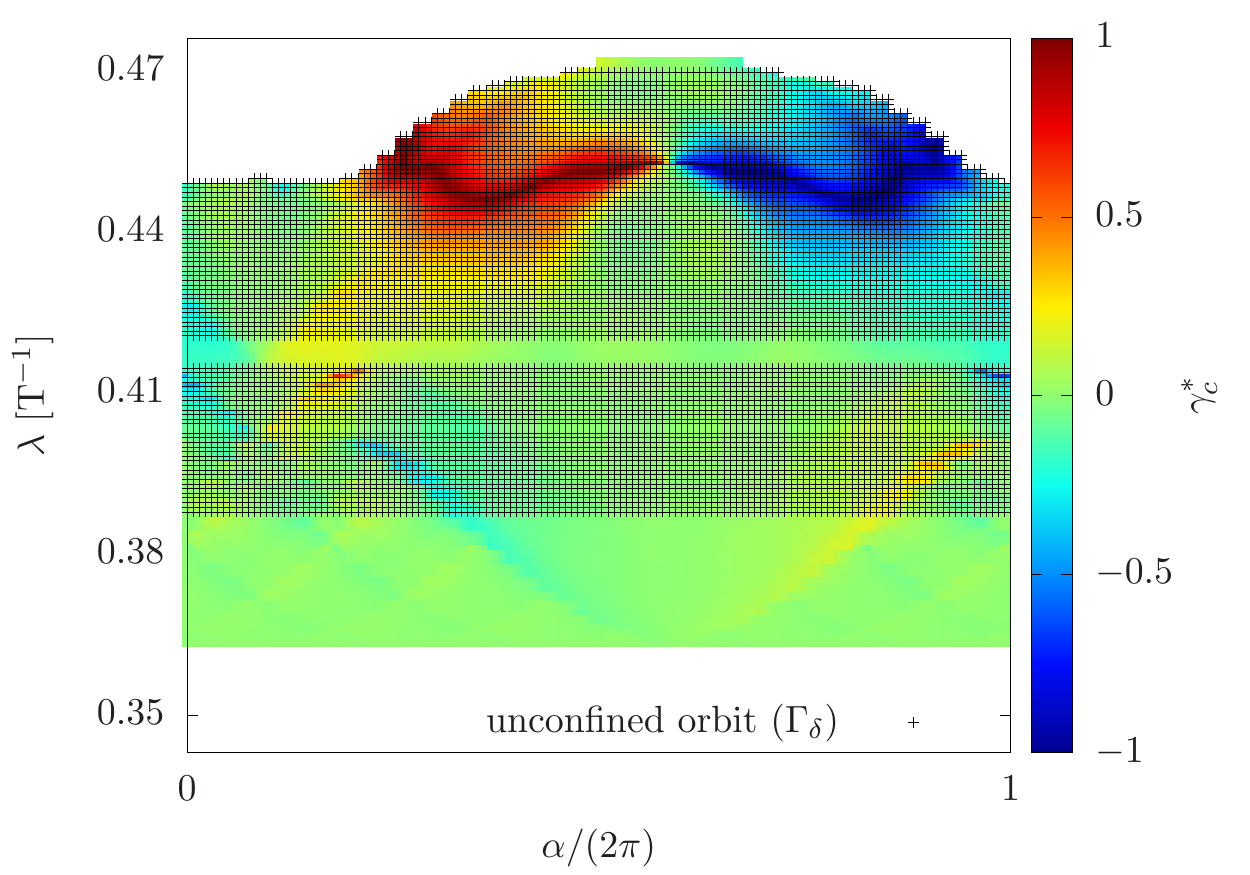}
\includegraphics[angle=0,width=0.45\columnwidth]{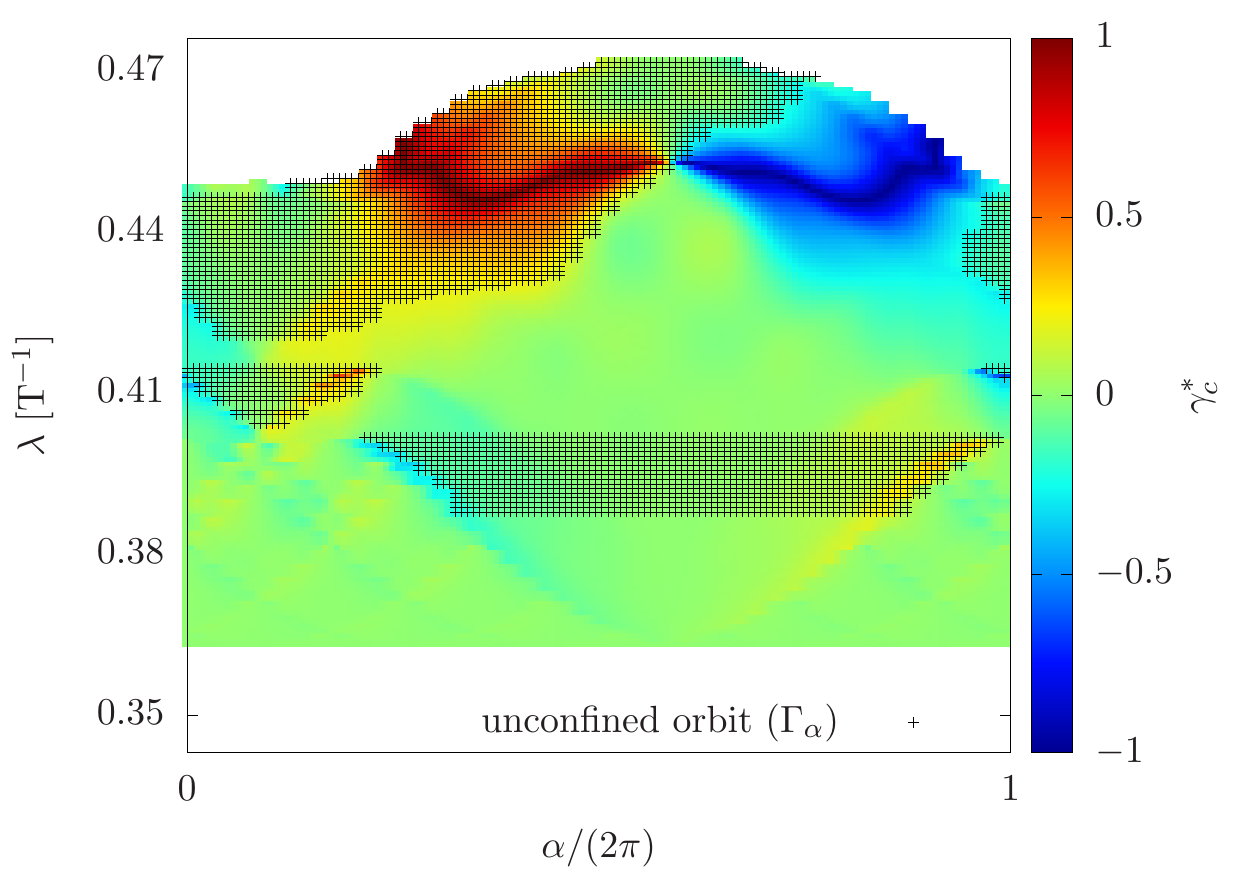}
\caption{Unconfined orbits (identified by grey crosses drawn over the color plot of $\gamma_{\mathrm{c}}^*$) at $s=0.25$ for the  KJM configuration of W7-X, using parabolic plasma profiles and $\beta=4$\% according to the classification of model I (left) and model II (right).}
\label{FIG_CLASS}
\end{figure}

\begin{figure}
\centering
\includegraphics[angle=0,width=0.9\columnwidth]{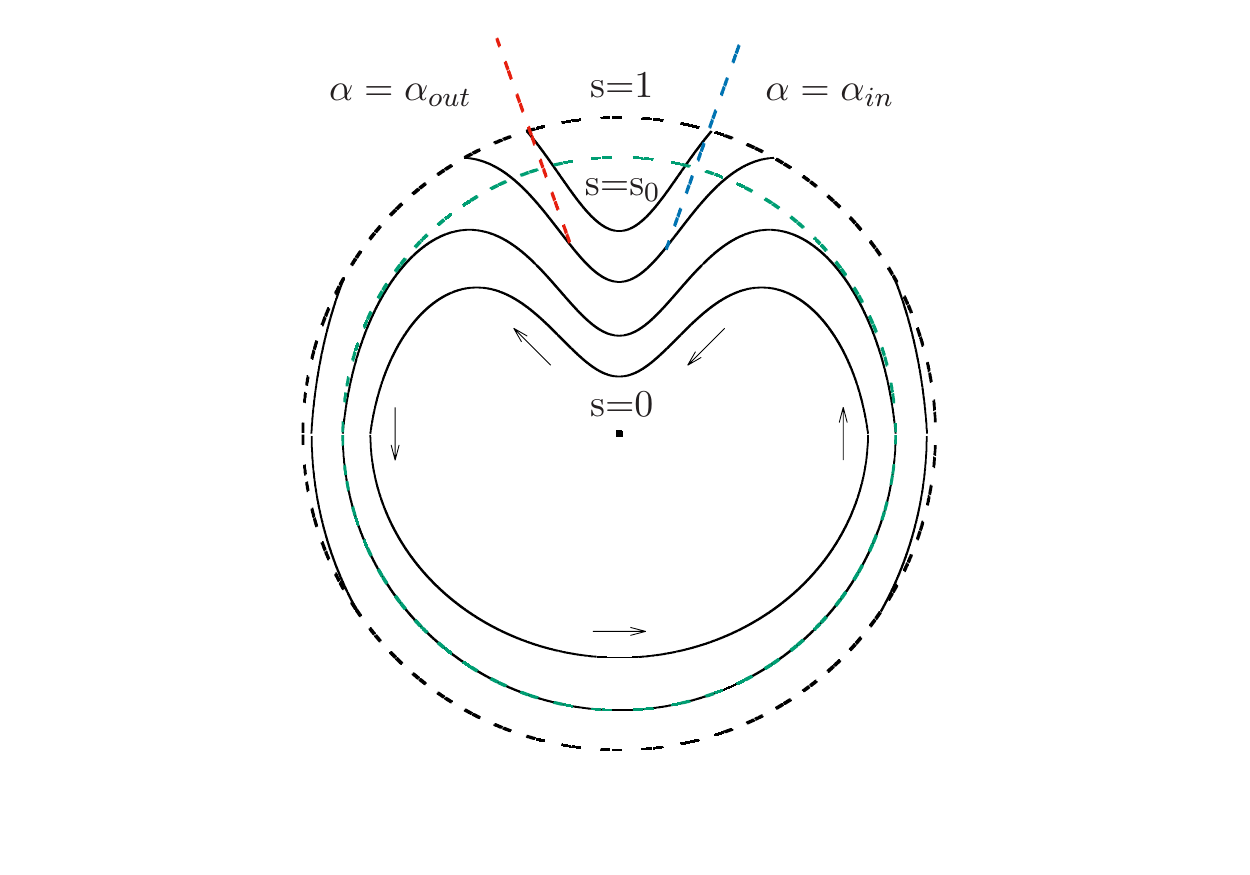}
\includegraphics[angle=0,width=0.9\columnwidth]{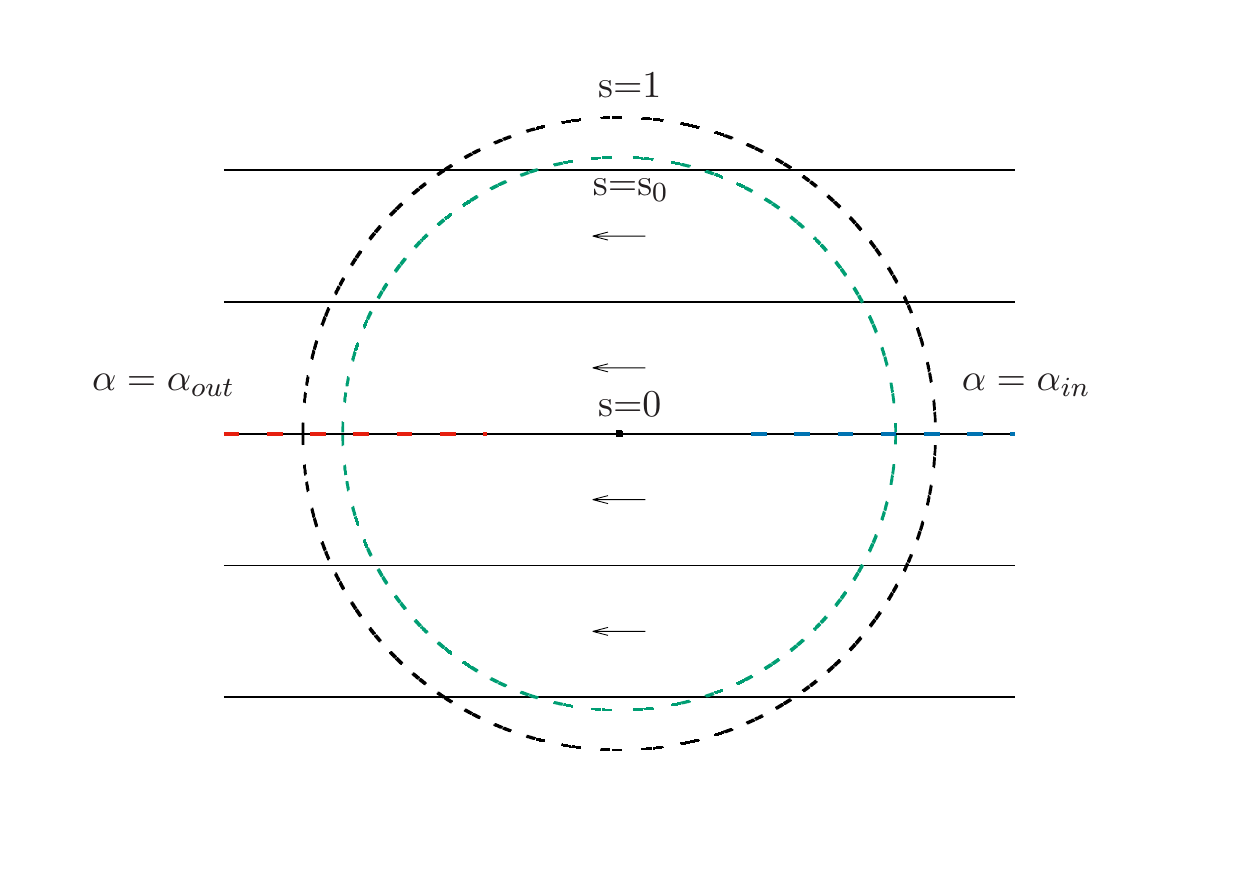}
\caption{Sketch of energetic ion orbits: example of partial optimization (top) and no optimization (bottom). Thin black lines and dashed circles represent contours of constant $J$ and $s$ respectively. Arrows indicate the direction of poloidal precession and the red and blue dashed lines mark the angular location of radial excursions from the original flux-surface, plotted in green.}
\label{FIG_SKETCH}
\end{figure}

Trapped ions do not move randomly in the ($\alpha$,$\lambda$) plane, but they precess periodically in $\alpha$ at constant $\lambda$. Because of this, a region of superbananas that is narrow in $\alpha$ and covers the whole $\lambda$-range should in principle be worse, for energetic ion confinement, than a region of superbananas that is narrow in $\lambda$ and fills completely $\alpha$. In the latter case, only the energetic ions born at that particular value of $\lambda$ would be lost, and the rest would stay confined (except for other transport mechanisms different than superbananas); in the former case, all the ions, after some precession in $\alpha$, would run into a superbanana and drift radially outwards. One could then propose as a figure of merit for energetic ion confinement the quantity
%\begin{equation}
%\Gamma_{\mathrm{lsb}} =\frac{\pi}{4\sqrt{2}} \fsa{\int_{B^{-1}_{\mathrm{max}}}^{B^{-1}}\mathrm{d}\lambda\frac{B}{\sqrt{1-\lambda B}} (\mathrm{max}( \gamma_{\mathrm{c}}^*(\alpha|\lambda)))^2}\,,
%\label{EQ_GAMMADELTA}
%\end{equation}
\begin{equation}
\Gamma_\delta = \frac{1}{2}\fsa{\int_{B^{-1}_{\mathrm{max}}}^{B^{-1}}\mathrm{d}\lambda\frac{B}{\sqrt{1-\lambda B}} H\bigg(\mathrm{max}( \gamma_{\mathrm{c}}^*(\alpha|\lambda))-\gamma_{\mathrm{th}}\bigg)}\,.
\label{EQ_GAMMADELTA}
\end{equation}
Here, the function $\mathrm{max}( \gamma_{\mathrm{c}}^*(\alpha|\lambda))$ denotes, for each $\lambda$, the maximum value of $\gamma_{\mathrm{c}}^*$ that can be found moving in $\alpha$ at fixed $\lambda$. The role of the Heaviside step function $H$ is to detect whether a superbanana exists or not for that value of $\lambda$. In all the calculations of this work, the threshold parameter $\gamma_\mathrm{th}$ is taken to be 0.2, corresponding approximately to
\begin{equation}
\frac{\overline{\mathbf{v}_M\cdot\nabla\alpha}}{\overline{\mathbf{v}_M\cdot\nabla s}} = \pi\,,
\end{equation}
i.e., to a trajectory that travels a distance 1 in the radial coordinate $s$ before moving $\pi$ in the $\alpha$ direction (varying this ratio between $\pi/2$ and $2\pi$ does not change significantly the results presented in the next section). Figure~\ref{FIG_CLASS} (left) contains one example (corresponding to the $\beta=4\%$ case of figure~\ref{FIG_GAMMAC025}) of classification of orbits by model I.

Equation~(\ref{EQ_GAMMADELTA}) is defined such that $\Gamma_\delta$ is bounded between 0 and the fraction of trapped particles,
\begin{equation}
f_\mathrm{trapped}=\frac{1}{2}\fsa{\int_{B_\mathrm{max}^{-1}}^{B^{-1}}\mathrm{d}\lambda'\frac{B}{\sqrt{1-\lambda' B}}}= \fsa{\sqrt{1-\frac{B}{B_\mathrm{max}}}}\,.
\end{equation}
%as a prediction of prompt losses should be.%\footnote{Note that the definition of $f_\mathrm{trapped}$ is different than the one of $f_\mathrm{t}=1-f_{\mathrm{c}}$, with $f_{\mathrm{c}}$ defined e.g. in~\cite{helander2017jpp}.}.

We will see in section~\ref{SEC_COMP} that $\Gamma_\delta$ is far from being a perfect predictor of prompt losses. The reason is that $\Gamma_\delta$ is still too rooted on the radially-local approach. In order to have a model that characterizes better the prompt losses in a stellarator, more global features need to be taken into account. We attempt to do so in the remainder of section~\ref{SEC_MODEL_PL}.

\subsection{Model II: local orbits and $\alpha-$loss cone}\label{SEC_MODEL2}

Figure~\ref{FIG_SKETCH} (top) is a sketch, based on figure~\ref{FIG_J_041}, of the contours of constant $J$ (thin black lines) and contours of constant $s$ (dashed circles) for a given value of $\lambda$. It roughly represents the orbits that start at $s=s_0=0.25$ with $\lambda=0.41\,$T$^{-1}$ at $\beta=4\%$. The flux-surface denoted by $s_0$ (green) contains a superbanana orbit: $\partial_s J$ is small around two angular positions $\alpha=\alpha_\mathrm{in}$ (blue) and $\alpha=\alpha_\mathrm{out}>\alpha_\mathrm{in}$ (red). At $\alpha_\mathrm{in}$, $\overline{\mathbf{v}_M\cdot\nabla s}<0$ and the radial excursion is directed inwards; at $\alpha_\mathrm{out}$, $\overline{\mathbf{v}_M\cdot\nabla s}>0$ and the radial excursion is directed outwards. As the arrows indicate, for any $\alpha$ different than $\alpha_\mathrm{in}$ or $\alpha_\mathrm{out}$, $\overline{\mathbf{v}_M\cdot\nabla\alpha}>0$.

It is now evident where the approach of subsection~\ref{SEC_MODEL1} stops being accurate: even though a superbanana exists at the same $\lambda$, energetic ions born at $\alpha<\alpha_\mathrm{in}$ and $\alpha>\alpha_\mathrm{out}$ describe orbits at constant $J$ that do not go beyond $s=s_0$: even if they do not move at constant $s$, because the $s$-contours and the $J-$contours are not perfectly aligned, the $J$-contours are closed and the motion in $s$ is bounded. Only those ions born at $\alpha_\mathrm{in}<\alpha<\alpha_\mathrm{out}$ do scape (the ones close to $\alpha_\mathrm{in}$, after an inwards excursion and later $\alpha$-precession). A better proxy for energetic ion confinement could then be
%\begin{eqnarray}
%\Gamma_{{\alpha}} =\frac{\pi}{4\sqrt{2}} \left\langle \int_{B^{-1}_{\mathrm{max}}}^{B^{-1}}\mathrm{d}\lambda\sum_c\frac{B}{\sqrt{1-\lambda B}} ~\times %\right.~~~~~~~~~~~~~~~~~~~~~~~~~~~~~~~~~~~ \nonumber\\ \left. \frac{1}{2}H\bigg((\alpha_{\mathrm{in},c}-\alpha)(\alpha-\alpha_{\mathrm{out},c})%\bigg)H\bigg((\alpha_{\mathrm{out},c}-{\alpha_{\mathrm{in},c}})~ \overline{\mathbf{v}_M\cdot\nabla\alpha}\bigg)\right\rangle \,.
%\label{EQ_GAMMAALPHA}
%\end{eqnarray}
\begin{eqnarray}
\Gamma_{{\alpha}} =\frac{1}{2}\left\langle \int_{B^{-1}_{\mathrm{max}}}^{B^{-1}}\mathrm{d}\lambda\frac{B}{\sqrt{1-\lambda B}} ~\times \right.~~~~~~~~~~~~~~~~~~~~~~~~~~~~~~~~~~~ \nonumber\\ \left. H\bigg((\alpha_\mathrm{out}-{\alpha})~ \overline{\mathbf{v}_M\cdot\nabla\alpha}\bigg)H\bigg((\alpha-{\alpha_\mathrm{in}})~ \overline{\mathbf{v}_M\cdot\nabla\alpha}\bigg)\right\rangle \,.
\label{EQ_GAMMAALPHA}
\end{eqnarray}
%\begin{eqnarray}
%\Gamma_{{\alpha}} =\frac{\pi}{4\sqrt{2}} \left\langle \int_{B^{-1}_{\mathrm{max}}}^{B^{-1}}\mathrm{d}\lambda\frac{B}{\sqrt{1-\lambda B}} ~\times %\right.~~~~~~~~~~~~~~~~~~~~~~~~~~~~~~~~~~~ \nonumber\\ \left. \frac{1}{2}H\bigg((\alpha_\mathrm{in}-\alpha)(\alpha-\alpha_\mathrm{out})\bigg)H\bigg((\alpha_\mathrm{out}-{\alpha_\mathrm{in}})~ \overline{\mathbf{v}_M\cdot\nabla\alpha}\bigg)\right\rangle \,.
%\label{EQ_GAMMAALPHA}
%\end{eqnarray}
Here, $\alpha_{\mathrm{out}}$ and $\alpha_{\mathrm{in}}$ are defined as two consecutive angular positions where $\gamma_{\mathrm{c}}^*(\alpha_{\mathrm{in}})<-\gamma_{\mathrm{th}}$ and $\gamma_{\mathrm{c}}^*(\alpha_{\mathrm{out}})>\gamma_{\mathrm{th}}$ (we employ $\gamma_\mathrm{th}=0.2$, as in section~\ref{SEC_MODEL1}). The Heaviside functions guarantee that the right portion of the $\alpha$-range is selected. We note that $\alpha$ is a periodic coordinate, and this has to be taken into account when computing $\alpha_\mathrm{out}-\alpha$ and $\alpha-\alpha_\mathrm{in}$. It is straightforward to generalize the formula for the case in which several superbananas, and thus several pairs of $\alpha_{\mathrm{out}}$ and $\alpha_{\mathrm{in}}$, exist at a given $s$ and $\lambda$. As in the case of $\Gamma_\delta$, $\Gamma_\alpha$ takes values between 0 and $f_\mathrm{trapped}$.

Figure~\ref{FIG_CLASS} (right) shows an example (also for $\beta=4\%$) of what orbits are identified by model II as unconfined. In this case, the tangential magnetic drift is negative for deeply trapped particles, which means that all those particles precess towards smaller values of $\alpha$ and end up in $\alpha_\mathrm{out}$, i.e., a region where $\gamma_{\mathrm{c}}^*>\gamma_{\mathrm{th}}$. For $\lambda$ smaller than 0.44$\,$T$^{-1}$,  the tangential magnetic drift is mainly positive, and only a fraction of the orbits born with a value of $\lambda$ such that $\mathrm{max}( \gamma_{\mathrm{c}}^*(\alpha|\lambda))>\gamma_\mathrm{th}$ are lost: those starting at $\alpha_\mathrm{in}\le\alpha\le\alpha_\mathrm{out}$ (and not those starting at $\alpha_\mathrm{out}\le\alpha\le\alpha_\mathrm{in}+2\pi$ or at  $\alpha_\mathrm{out}-2\pi\le\alpha\le\alpha_\mathrm{in}$).

%\begin{eqnarray}
%(\alpha_\mathrm{in}-\alpha)(\alpha_\mathrm{out}-\alpha)&<& 0\,,\nonumber\\
%(\alpha_\mathrm{out}-{\alpha_\mathrm{in}})~ \overline{\mathbf{v}_M\cdot\nabla\alpha}&>&0\,,\nonumber\\
%\gamma_{\mathrm{c}}^s(\alpha=\alpha_\mathrm{in})&=&-1\,,\nonumber\\
%\gamma_{\mathrm{c}}^s(\alpha=\alpha_\mathrm{out})&=&+1\,.\
%\label{EQ_CONE}
%\end{eqnarray}

\

The local description should break down entirely in situations in which the energetic ions are basically unconfined. Figure~\ref{FIG_SKETCH} (bottom) represents the situation of orbits that start at $s=s_0=0.25$ with $\lambda=0.41\,$T$^{-1}$ at $\beta\le2\%$. The flux-surface denoted by $s_0$ contains a superbanana orbit: $\partial_s J=0$ around two angular positions $\alpha_\mathrm{in}=0$ (blue) and $\alpha_\mathrm{out}=\pi$ (red). However, it is clear that all the energetic ions born at $s=0.25$ are promptly lost, even those at $\alpha=\pi$ or $\alpha=-\pi$, where $\partial_\alpha J/\partial_s J=0$. This could cause $\Gamma_{\mathrm{c}}$ to be inaccurate, although still able to distinguish between~\ref{FIG_SKETCH} (top)  and ~\ref{FIG_SKETCH} (bottom). However, it is possible that $\Gamma_{{\alpha}}$ does a good job, because, at $s_0$, for every value of $\alpha$ (except for the two discrete points $\alpha=\alpha_\mathrm{in}$ and $\alpha=\alpha_\mathrm{out}$ exactly), one has
%\begin{equation}
%H\bigg((\alpha_{\mathrm{in}}-\alpha)(\alpha-\alpha_{\mathrm{out}})\bigg)H\bigg((\alpha_{\mathrm{out}}-{\alpha_{\mathrm{in}}})~ \overline{\mathbf{v}_M\cdot\nabla\alpha}\bigg)=1\,
%\end{equation}
\begin{equation}
H\bigg((\alpha_\mathrm{out}-{\alpha})~ \overline{\mathbf{v}_M\cdot\nabla\alpha}\bigg)H\bigg((\alpha-{\alpha_\mathrm{in}})~ \overline{\mathbf{v}_M\cdot\nabla\alpha}\bigg)=1\,.
\end{equation}
In summary, $\Gamma_{{\alpha}}$ should be able to characterize at least stellarators with good energetic ion confinement whose map of $J$ is similar to figure~\ref{FIG_SKETCH} (top) and stellarators with poor confinement and a map of $J$ that looks like~\ref{FIG_SKETCH} (bottom). In section~\ref{SEC_VAL_PL} we demonstrate that this is the case, and that $\Gamma_\alpha$ also can be applied to intermediate situations.

\section{Model validation with full orbit simulations of prompt losses}\label{SEC_VAL_PL}

In this section, we characterize the transport of energetic ions, and validate our models, for the five configurations of the $\beta$-scan of section~\ref{SEC_SUPERBANANAS} and three additional vacuum configurations of W7-X: EIM (also known as standard), FTM (high-$\iota$) and DBM (low-$\iota$). Ions of $50\,$keV are launched from flux surfaces $s=0.06$, $s=0.25$ and $0.50$ (2$\times 10^4$ ions for each surface). The chosen energy guarantees that the Larmor radius is similar to that of a fusion-born alpha particle in a device of reactor size, and the radial electric field is set to zero, so that it does not obscure the role of the magnetic drift in making the ions precess. The distribution in the rest of variables of the phase-space is set to mimic that of alpha particles in the reactor. The orbits of these ions are then followed with the Monte Carlo code ASCOT for 0.1$\,$s. This is the time needed by a $50\,$keV ion to perform roughly as many toroidal turns in the W7-X configuration as an alpha particle does in a slowing-down time in a reactor-sized stellarator. Collisions are not included, since they are not relevant for prompt losses.

In the framework presented in section~\ref{SEC_MODEL_PL}, three qualitative predictions can be made for the prompt losses of energetic ions being born on an optimized flux surface $s$ with velocity $v$. First, their distribution in $\lambda$ should be concentrated in the regions where superbananas are identified, $\gamma_{\mathrm{c}}^*(\alpha,\lambda) > \gamma_\mathrm{th}$. This applies to both the distribution at birth (i.e., the starting points of their trajectories) and the final distribution (i.e., the points where they reach $s=1$), since $\lambda$ is a constant of motion. Second, the final distribution of prompt losses should be concentrated around $\alpha_\mathrm{out}(\lambda)$. In a bidimensional map of the flux-surface, this corresponds to a straight line of constant $\theta-\iota\zeta$. Third, the birth distribution of prompt losses should be concentrated around a generally wider region ($\alpha_\mathrm{in}(\lambda)<\alpha<\alpha_\mathrm{out}(\lambda)$, according to model II). For an unoptimized magnetic configuration, the velocity distribution will still be concentrated where superbananas exist, but this will likely be a very broad $\lambda$ distribution. According to model II, less localization can be expected in $\alpha$.

In this section, the first three subsections compare quantitatively the full-orbit simulations with several predictions of model II: in section~\ref{SEC_TIME}, the total fraction of prompt losses is described; section \ref{SEC_LAMBDA} shows the velocity distribution of prompt losses, and section \ref{SEC_ALPHA} their angular distribution. Finally, section~\ref{SEC_COMP} compares all the models in terms of usefulness for stellarator optimization.

%In this section, the first three subsections compare quantitatively the guiding-center simulations with the predictions of model II, given by
%\begin{equation}
%f_{\mathrm{pl}}=\frac{\fsa{\int_{B_\mathrm{max}^{-1}}^{B^{-1}}\mathrm{d}\lambda'\frac{B}{\sqrt{1-\lambda' B}}\hat f_n H\bigg((\alpha_\mathrm{out}-{\alpha})~ \overline{\mathbf{v}_M\cdot\nabla\alpha}\bigg)H\bigg((\alpha-{\alpha_\mathrm{in}})~ \overline{\mathbf{v}_M\cdot\nabla\alpha}\bigg)}}{\fsa{\int_{0}^{B^{-1}}\mathrm{d}\lambda'\frac{B}{\sqrt{1-\lambda' B}}\hat f_d}}\,.\label{EQ_FLOSS}
%\end{equation}
%Different choices of functions $\hat f_n$ and $\hat f_d$ will produce different quantities that can be compared to those calculated with ASCOT: in section~\ref{SEC_TIME}, the time evolution of prompt losses is described; section \ref{SEC_LAMBDA} shows the velocity distribution of prompt losses, and section \ref{SEC_ALPHA} their angular distribution. Finally, section~\ref{SEC_COMP} compares all the models in terms of usefulness for stellarator optimization.

%%%%%%%%%%%%%%%%%%%%%%%%%%%%%%%%%%%%%%%%%%%%%%%%%%%%%%%%%%%%%%%%%%%%%%%%%%%%%%%%%%%%%%

\subsection{Time scale and total fraction of prompt losses}\label{SEC_TIME}

The total fraction of prompt losses is given by $\Gamma_\alpha$ in equation (\ref{EQ_GAMMAALPHA}),
\begin{equation}
f_\mathrm{pl}= \Gamma_\alpha\,.
\end{equation}
We note that the upper limit for $f_\mathrm{pl}$ is the total fraction of trapped particles $f_\mathrm{trapped}$.

\begin{figure}
\centering
\includegraphics[angle=0,width=0.47\columnwidth]{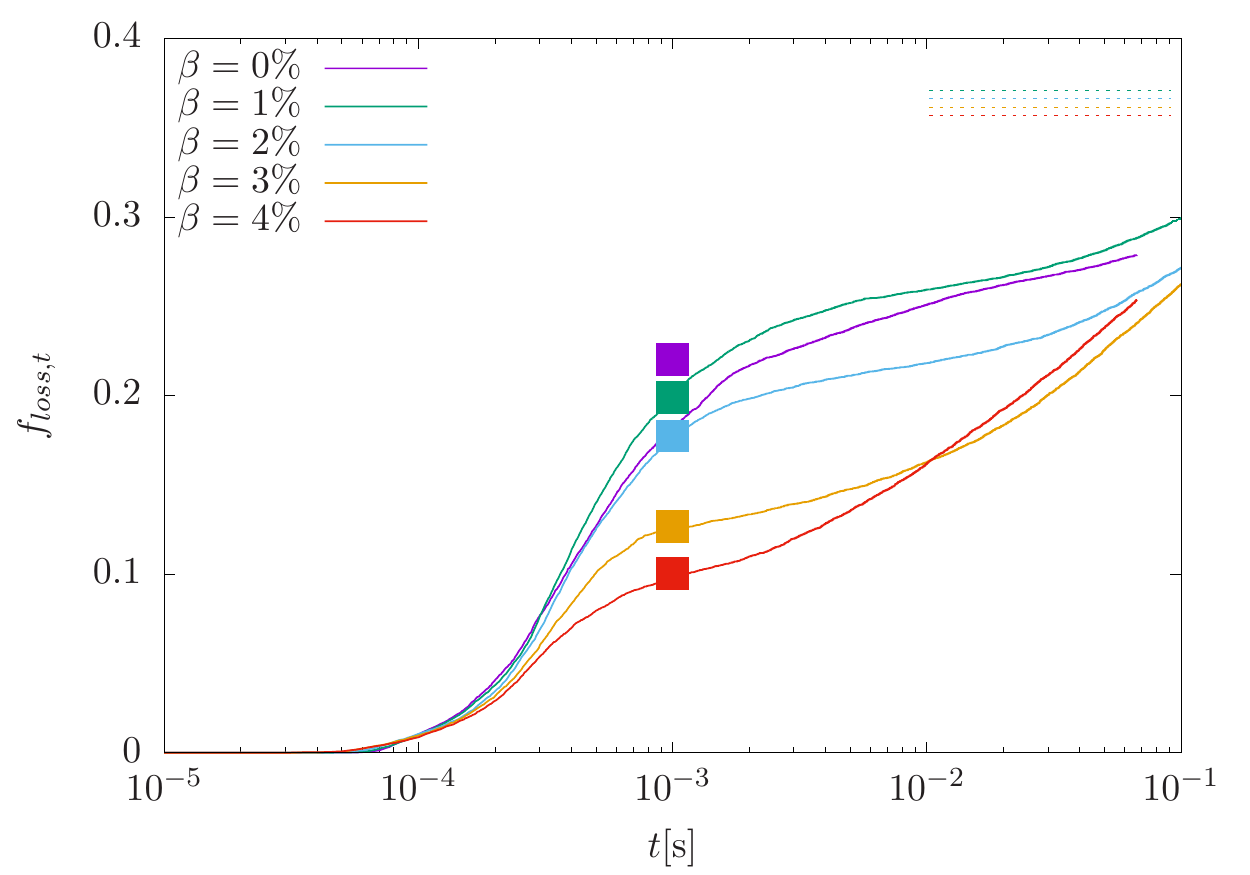}
\includegraphics[angle=0,width=0.47\columnwidth]{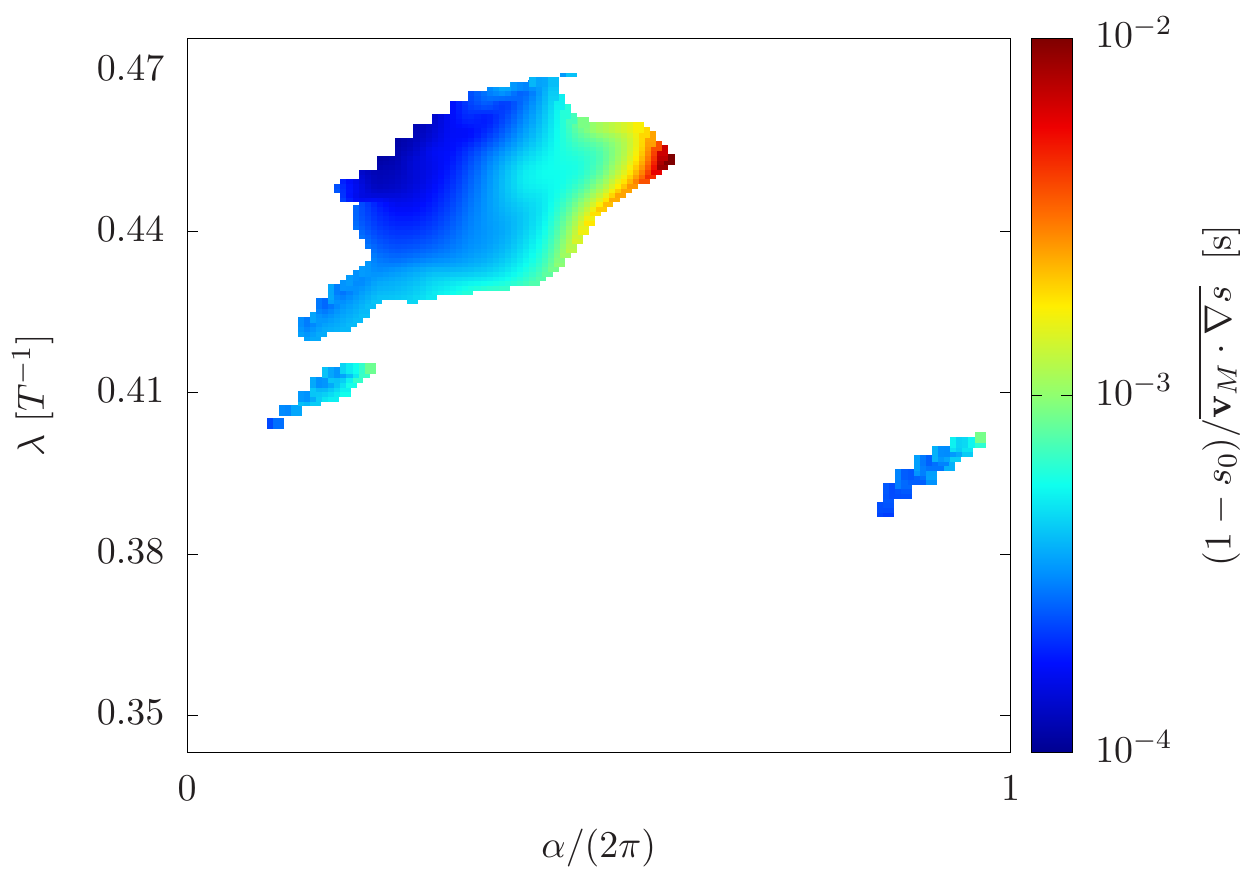}
\caption{Left: loss fraction of energetic ions born at $s=0.25$ as a function of time obtained with ASCOT (lines) and total fraction of prompt losses predicted by the model (squares). The dotted lines represent the fraction of trapped particles. Right: estimate of prompt loss time by the model}%, for the full time interval (top row) and very early times (bottom row).}
\label{FIG_LF025}
\end{figure}

Figure~\ref{FIG_LF025} (left) shows the time evolution of the loss fraction of energetic ions born at $s=0.25$ calculated with ASCOT, $f_{\mathrm{loss},t}(t)$, for all the configurations in the $\beta$-scan. The two time scales discussed in the introduction are present; prompt losses occur in the interval $10^{-4}\,$s$<t<10^{-3}\,$s. A slower loss of energetic ions takes place for $t>10^{-3}\,$s. As discussed throughout this paper, confinement is bad at  $\beta\le 1\%$, and larger $\beta$ leads to reduced prompt losses, with a clear improvement for $\beta\ge 3\%$. This feature is well captured by the predictions of the model, as shown in figure~\ref{FIG_LF025} (left). For $t>10^{-3}\,$s, no prediction can be attempted by the model, since it does not contain the relevant physical mechanism, stochastic diffusion. We note that the division between prompt and stochastic at $t=10^{-3}\,$s in the ASCOT data of figure~\ref{FIG_LF025} (left) is not rigorous, and therefore the comparison with the model can only be approximate.

Additionally, in our model, energetic ions escape drifting radially outwards once they reach $\alpha_\mathrm{out}$. The time it takes them to cover the radial distance $1-s_0$ (i.e., from the original surface to the last closed flux-surface), at a velocity given by $\overline{\mathbf{v}_M\cdot\nabla s}$ evaluated at $s_0$ and $\alpha_\mathrm{out}$, should provide a rough estimate of the time scale of the prompt losses. Figure~\ref{FIG_LF025} (right) shows $(1-s_0)/\overline{\mathbf{v}_M\cdot\nabla s}$ for the $\beta=4\%$ case at $s=0.25$ evaluated at the angular positions where $\gamma_{\mathrm{c}}^*$ is larger than $\gamma_\mathrm{th}$. The model indeed predicts the prompt losses at the right time scale.

%\begin{figure}
%\centering
%\includegraphics[angle=0,width=0.47\columnwidth]{fraction006ASCOT}
%\includegraphics[angle=0,width=0.47\columnwidth]{fraction006}
%%\includegraphics[angle=90,width=0.2\columnwidth]{fraction006zoom}
%%\includegraphics[angle=0,width=0.4\columnwidth]{fraction006zoom}
%\caption{Loss fraction of energetic ions born at $s=0.06$ predicted with ASCOT (left) and the model (right) as a function of time. The dotted lines represent the fraction of trapped particles.}%, for the full time interval (top row) and very early times (bottom row).}
%\label{FIG_LF006}
%\end{figure}

%%%%%%%%%%%%%%%%%%%%%%%%%%%%%%%%%%%%%%%%%%%%%%%%%%%%%%%%%%%%%%%%%%%%%%%%%%%%%%%%%%%%%%

\subsection{Velocity distribution of prompt losses}\label{SEC_LAMBDA}

We are also interested in the fraction of energetic ions born with a given pitch-angle $\lambda$ that are promptly lost, which reads
\begin{equation}
f_{\mathrm{pl},\lambda}(\lambda)=\frac{\fsa{\frac{B}{\sqrt{1-\lambda B}}H\bigg((\alpha_\mathrm{out}-{\alpha})~ \overline{\mathbf{v}_M\cdot\nabla\alpha}\bigg)H\bigg((\alpha-{\alpha_\mathrm{in}})~ \overline{\mathbf{v}_M\cdot\nabla\alpha}\bigg)}}{\fsa{\frac{B}{\sqrt{1-\lambda B}}}}\,.
\end{equation}
This quantity is 1 if no energetic ion of that pitch-angle velocity is confined. 

\begin{figure}
\centering
\includegraphics[angle=0,width=0.47\columnwidth]{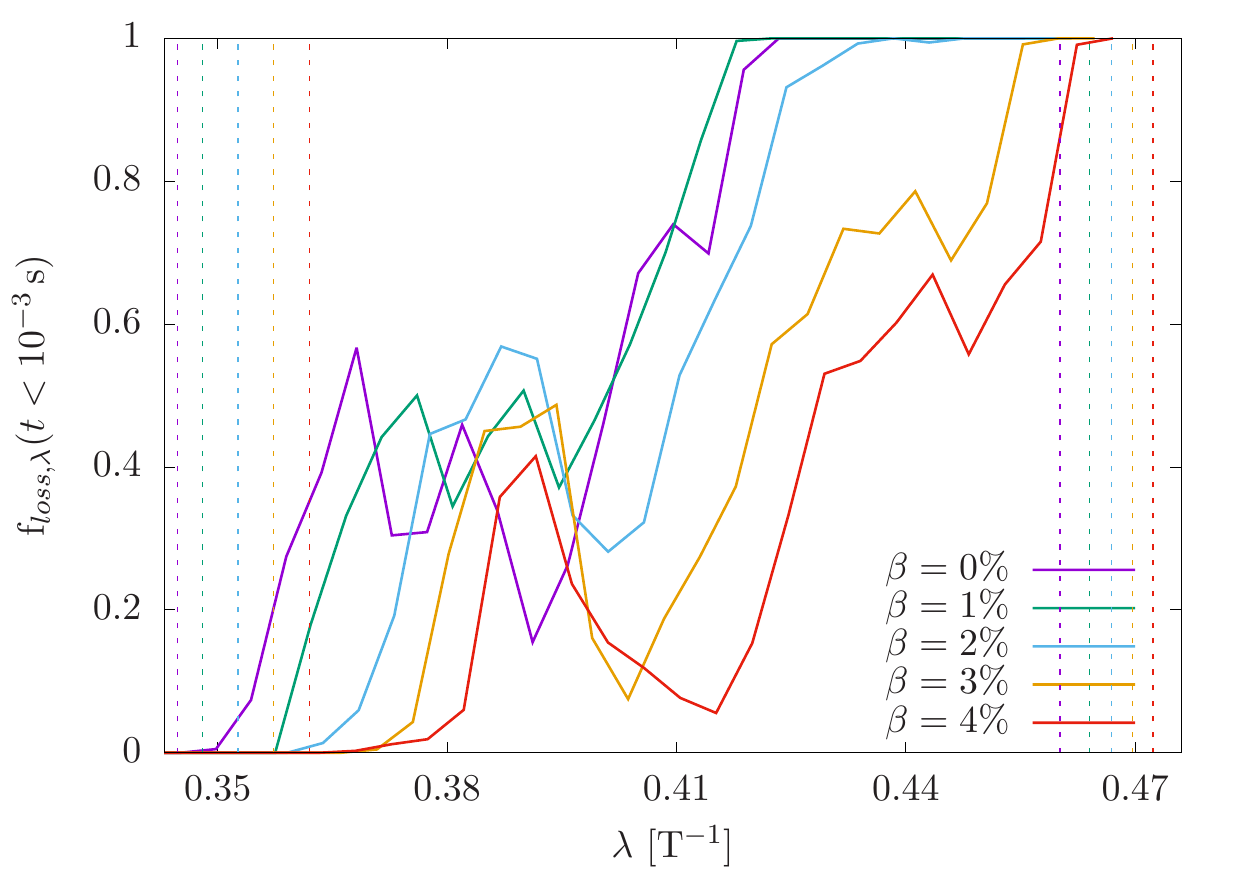}
\includegraphics[angle=0,width=0.47\columnwidth]{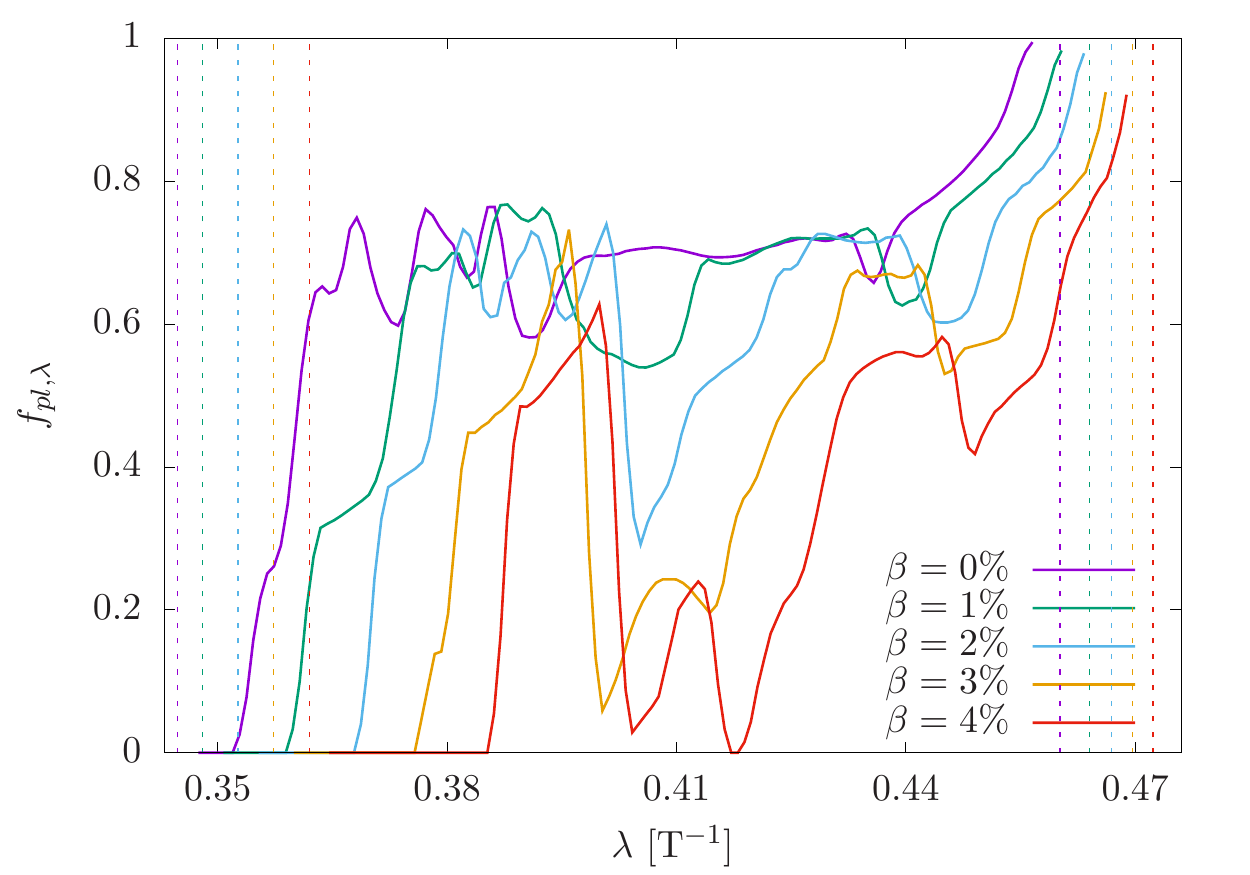}
\caption{Velocity-resolved prompt loss fraction of the energetic ions born at $s=0.25$ obtained with ASCOT (left) and the model (right). The dotted lines represent $1/B_\mathrm{min}$ and $1/B_\mathrm{max}$, being $B_\mathrm{min}$ the minimum of B on the flux surface.}
\label{FIG_PITCH025}
\end{figure}

Figure~\ref{FIG_PITCH025} (left) represents $f_{\mathrm{loss},\lambda}(t<10^{-3}\,$s), the fraction of energetic ions born at $s=0.25$, calculated with ASCOT, that are lost at $t<10^{-3}\,$ for each $\lambda$. It shows that, for small $\beta$, all values of $\lambda$ have bad confinement, but specially the region of deeply trapped particles. The effect of $\beta\ge 3\%$ is to push superbananas deeper into the trapped region (both around $0.38\,$T$^{-1}$, not far from the trapped/passing boundary, and $0.45\,$T$^{-1}$, in the deeply trapped region) and to remove the prompt losses of ions born around $0.42\,$T$^{-1}$. These features were partially discussed in section~\ref{SEC_SUPERBANANAS}, in terms of $\gamma_{\mathrm{c}}^*$, and the model captures them quite well, according to figure~\ref{FIG_PITCH025} (right), that shows $f_{\mathrm{pl},\lambda}$. The absolute values are globally slightly smaller for the deeply trapped region and slightly higher around $0.38\,$T$^{-1}$.

%%%%%%%%%%%%%%%%%%%%%%%%%%%%%%%%%%%%%%%%%%%%%%%%%%%%%%%%%%%%%%%%%%%%%%%%%%%%%%%%%%%%%%

\subsection{Angular distribution of prompt losses}\label{SEC_ALPHA}

\begin{figure}
\centering
\includegraphics[angle=0,width=0.4\columnwidth]{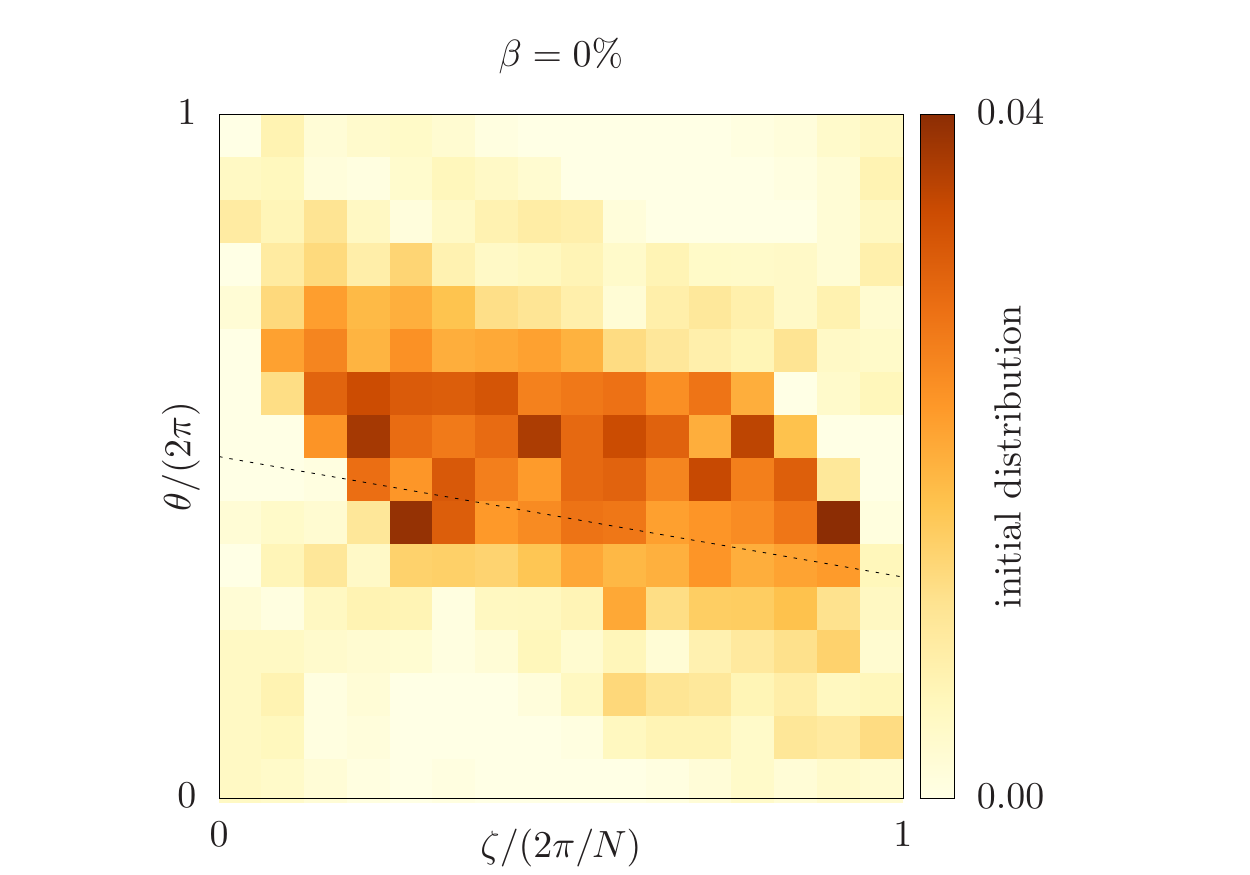}
\includegraphics[angle=0,width=0.4\columnwidth]{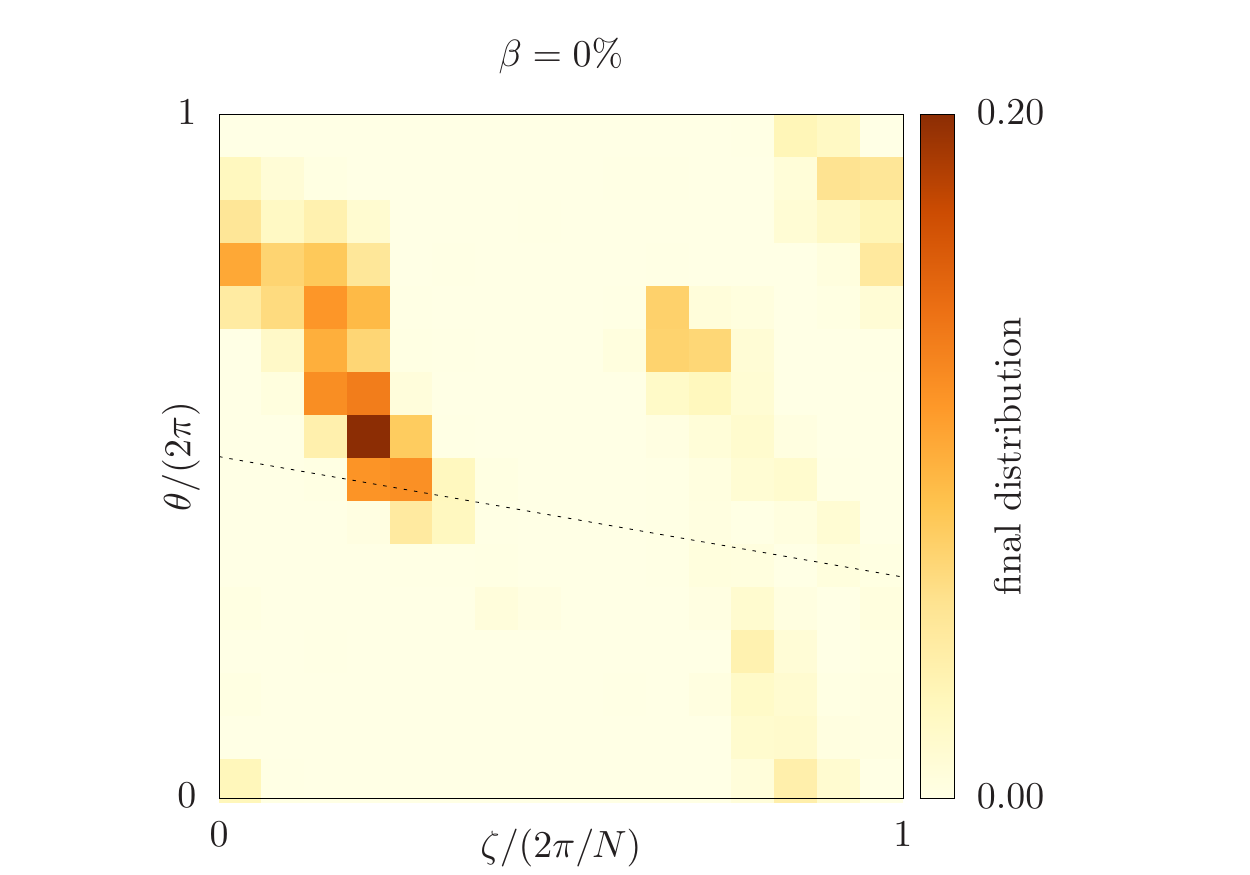}
\includegraphics[angle=90,width=0.4\columnwidth]{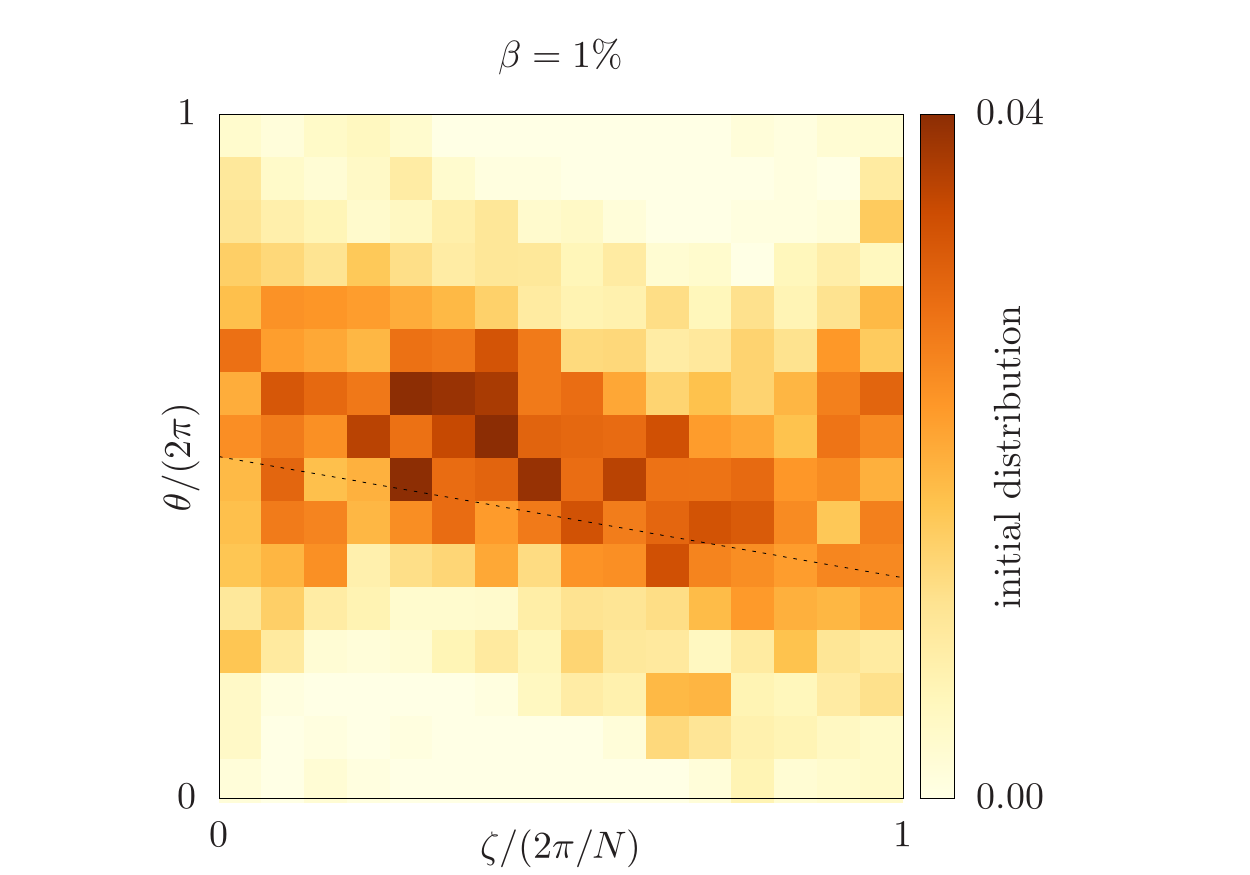}
\includegraphics[angle=0,width=0.4\columnwidth]{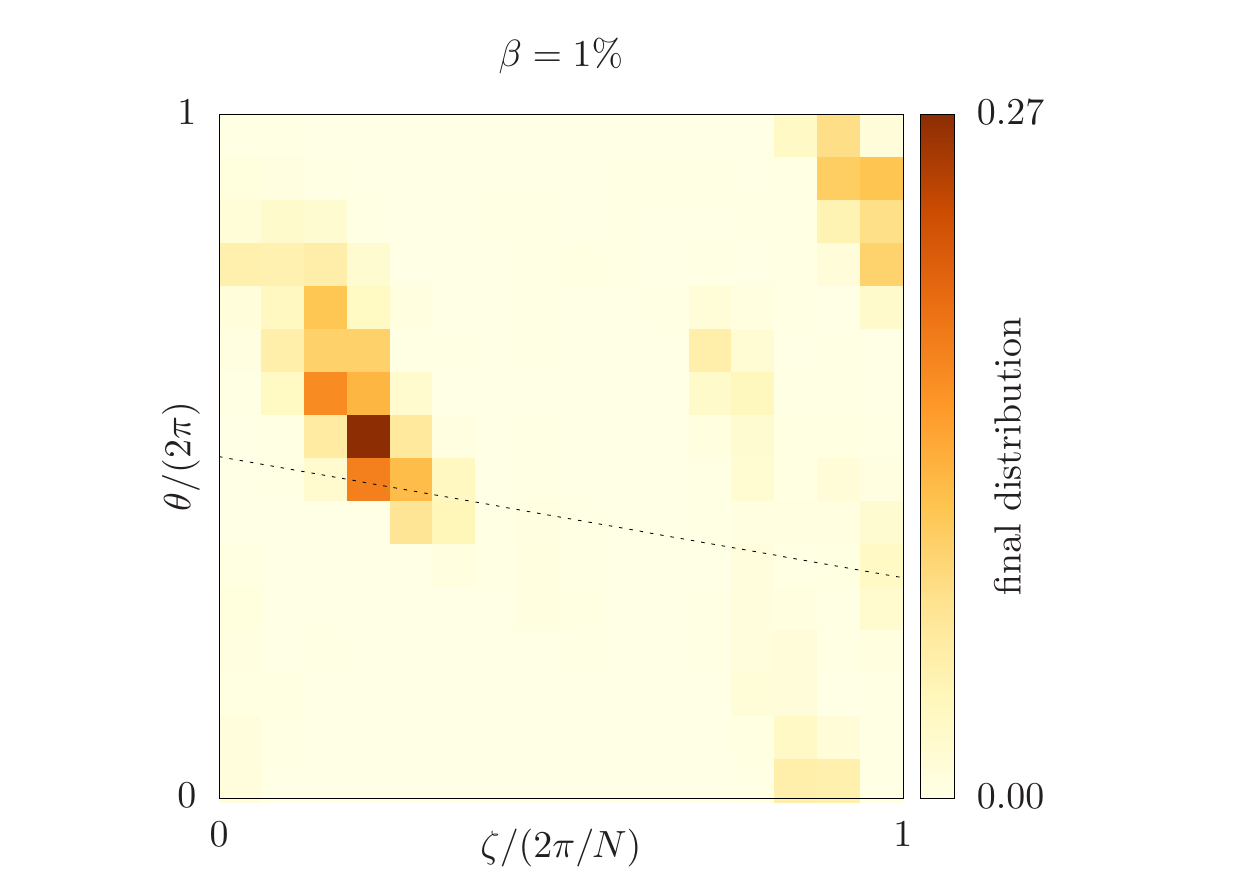}
\includegraphics[angle=90,width=0.4\columnwidth]{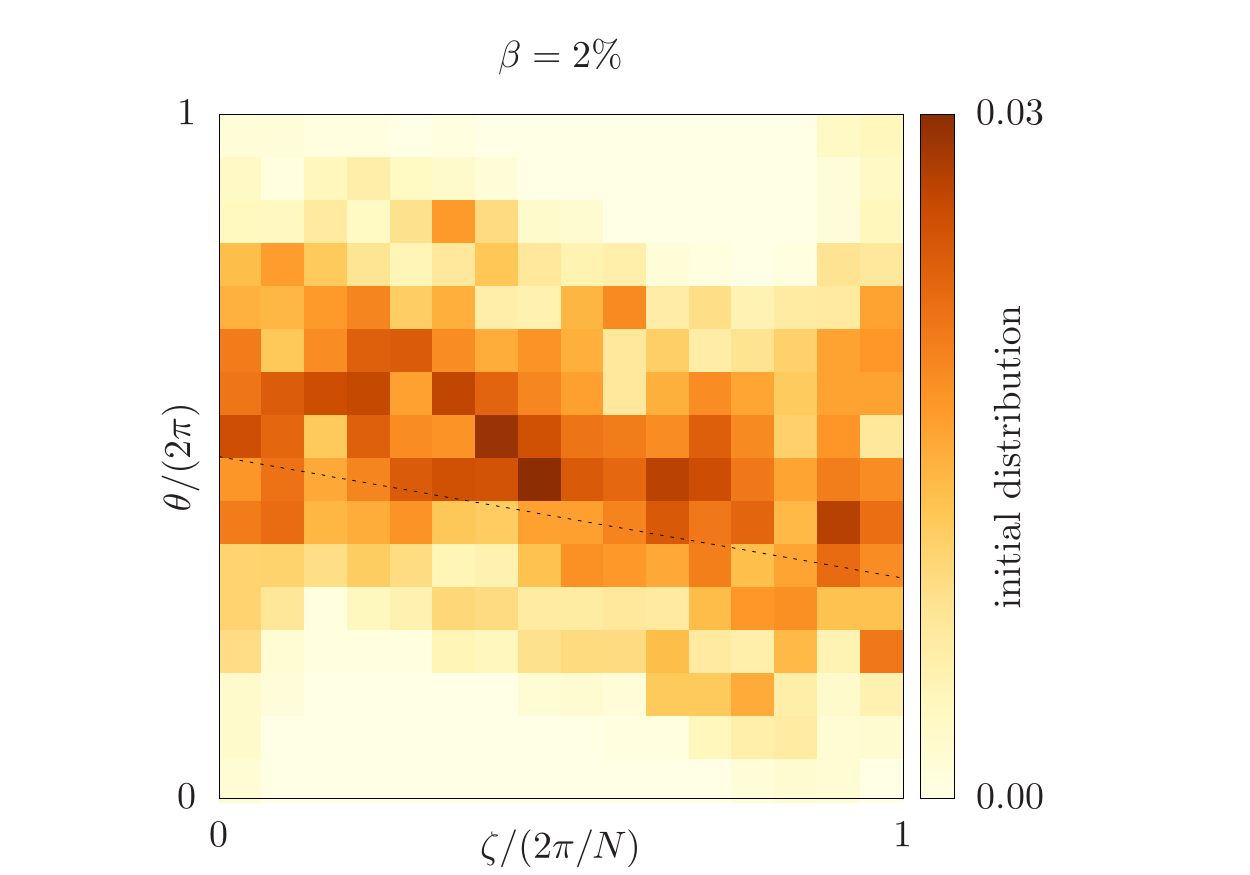}
\includegraphics[angle=0,width=0.4\columnwidth]{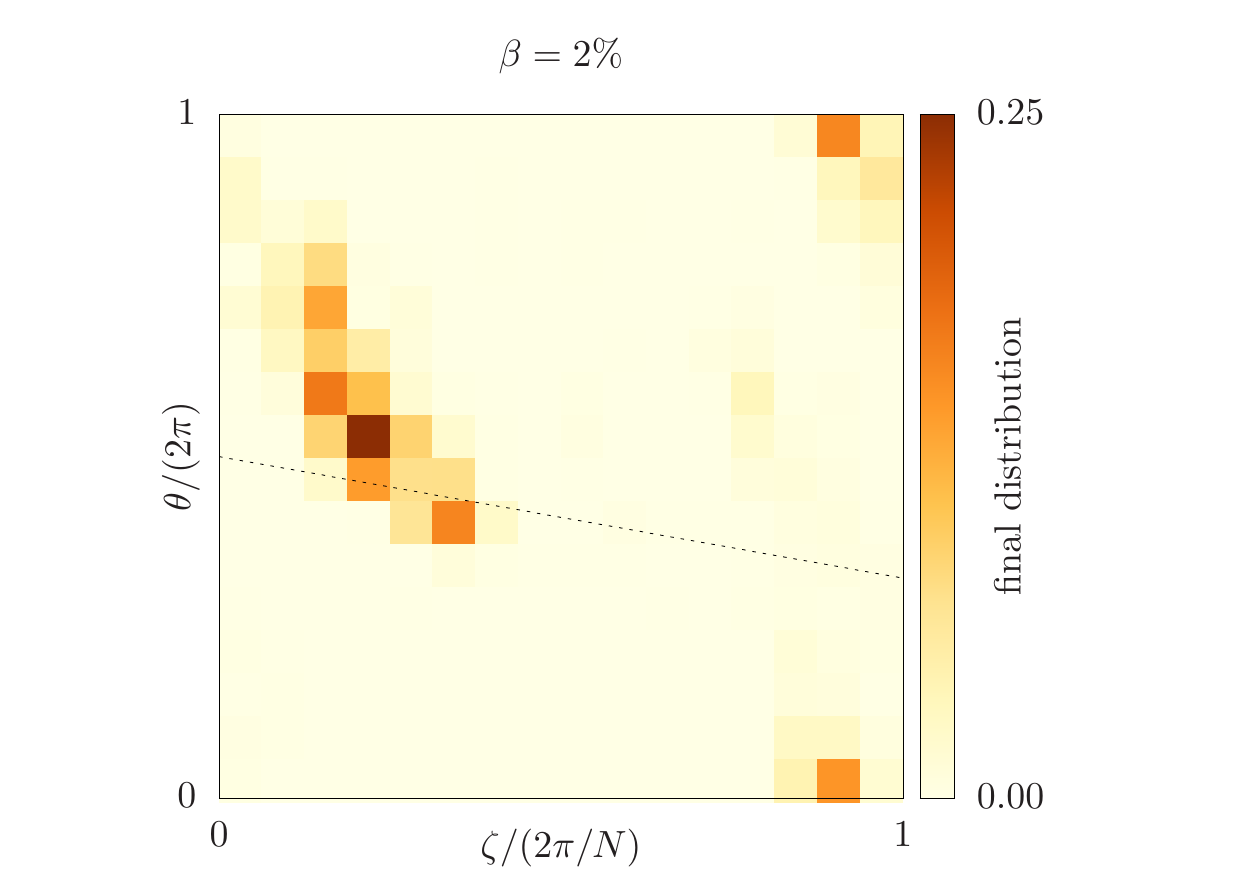}
\includegraphics[angle=90,width=0.4\columnwidth]{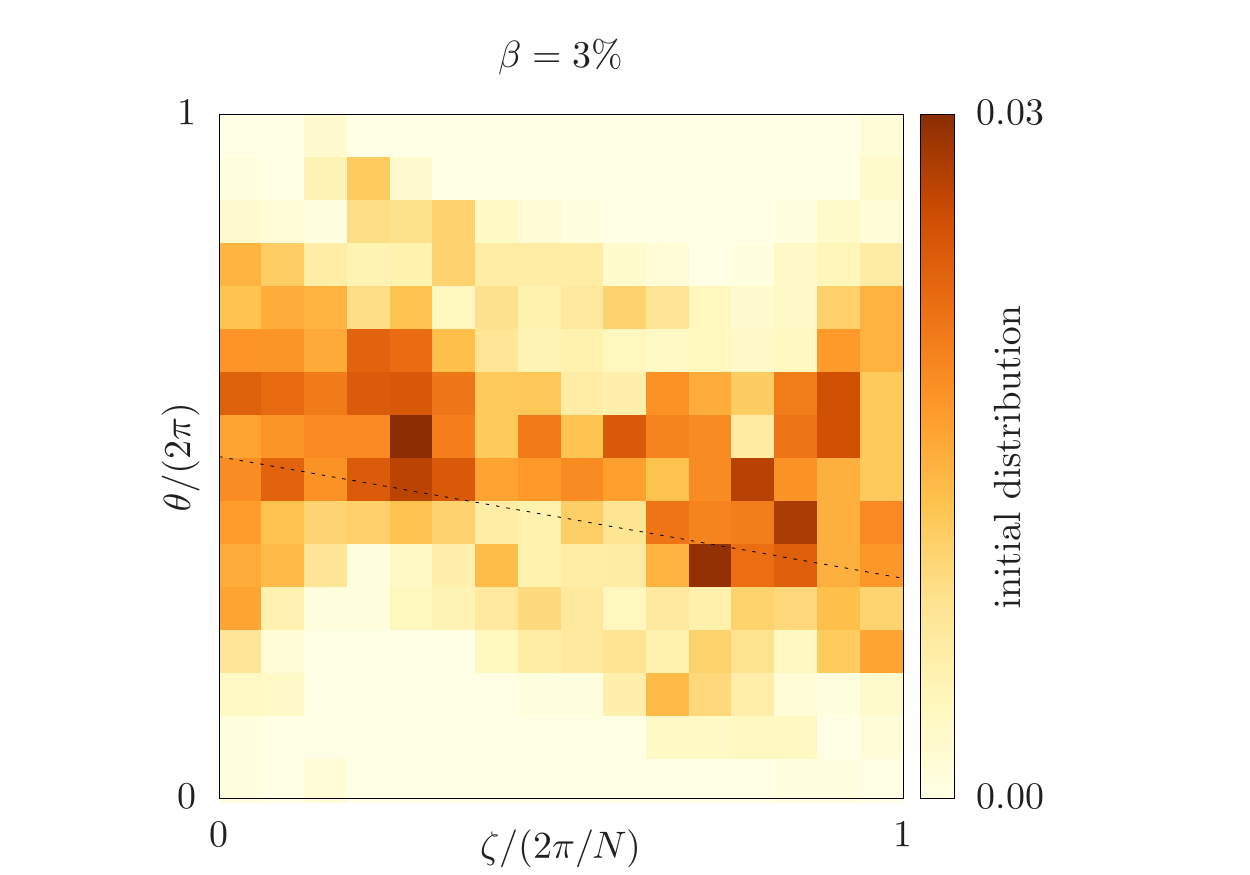}
\includegraphics[angle=0,width=0.4\columnwidth]{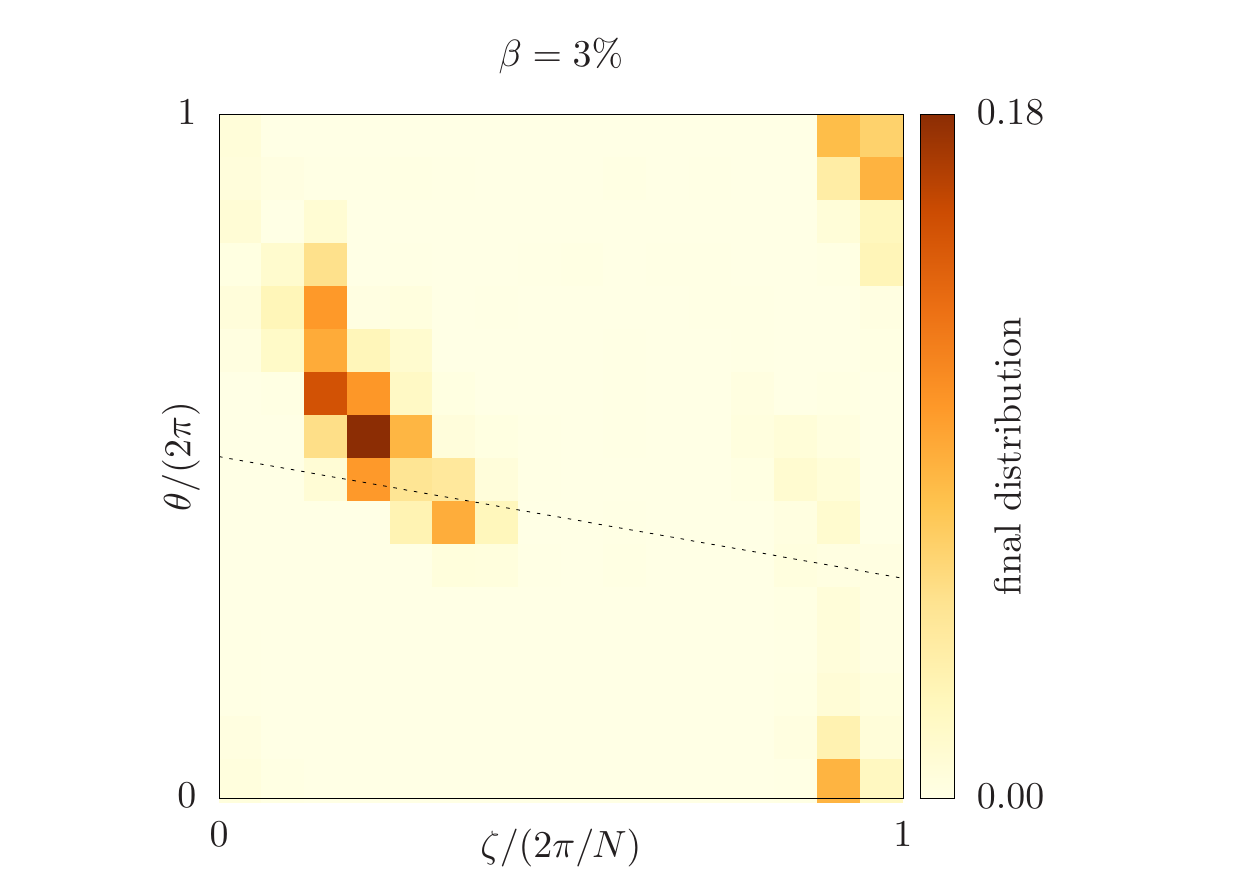}
\includegraphics[angle=90,width=0.4\columnwidth]{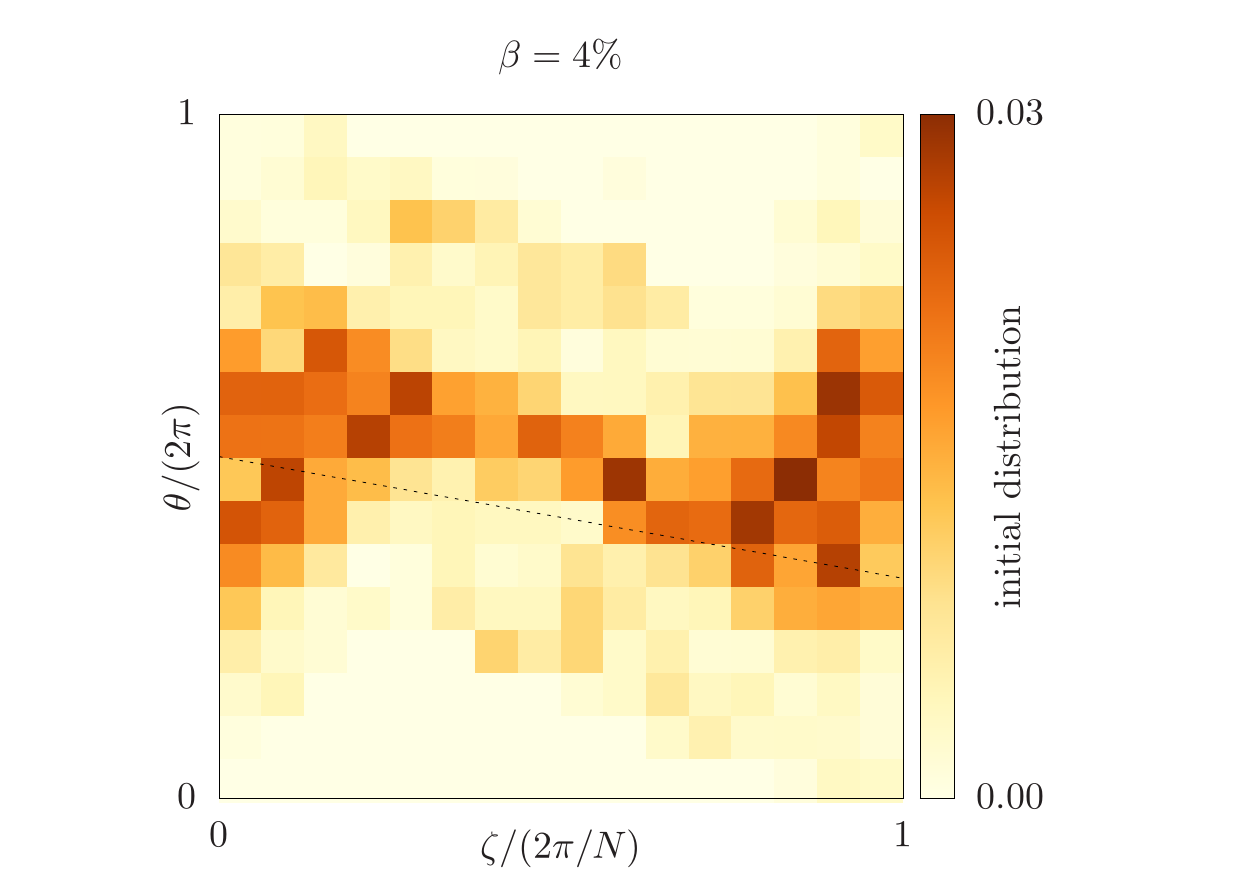}
\includegraphics[angle=0,width=0.4\columnwidth]{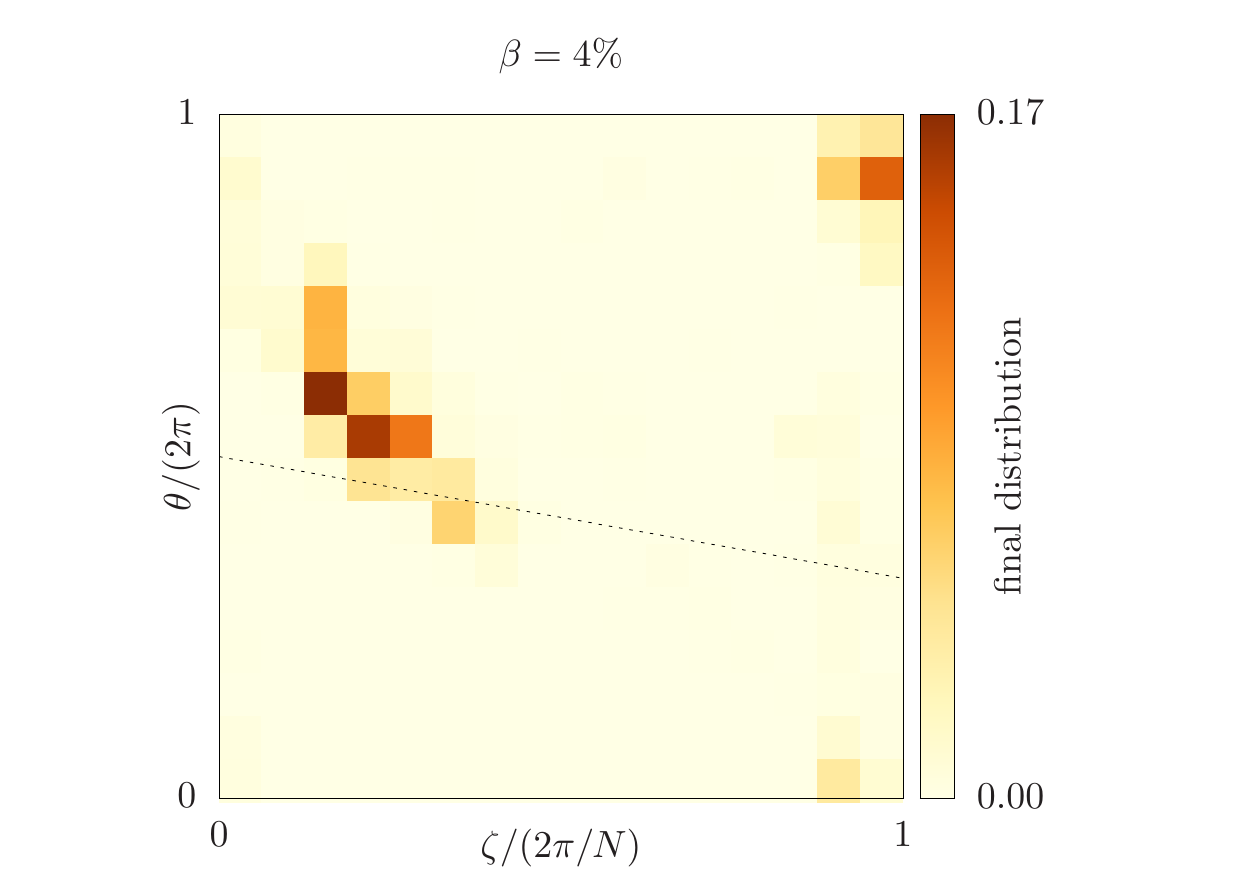}
\caption{Initial (left) and final (right) angular distribution of the orbits that are promptly lost after being born at $s=0.25$, obtained with ASCOT. The dotted line represents $\alpha=\pi$.}
\label{FIG_2D}
\end{figure}

%\begin{figure}
%\centering
%\includegraphics[angle=0,width=0.4\columnwidth]{KJMb4p_s025_mu_alpha_ini.png}
%\includegraphics[angle=0,width=0.4\columnwidth]{KJMb4p_s025_mu_alpha_final.png}
%\caption{Initial (left) and final (right) distribution of prompt losses calculated with ASCOT for $\beta=4\%$. \todo{A bit redundant? $\alpha$ and $\alpha_0$ are both $\theta_{\rm VMEC}-\iota\zeta_{\rm VMEC}$.}}
%\label{FIG_ANG}
%\end{figure}
%\begin{figure}
%\centering
%\includegraphics[angle=0,width=0.47\columnwidth]{end025ASCOT}
%\includegraphics[angle=0,width=0.47\columnwidth]{end025pl}
%\caption{Angular distribution of the initial and final positions of the orbits that are promptly lost after being born at $s=0.25$ obtained with ASCOT (left,~\todo{$\alpha_{VMEC}$ is not $\alpha_{Boozer}$, how different is it?}) and the model (right,~\todo{Numerical noise comes from non-uniform $\alpha$ grid, it can be improved}).}\label{FIG_END025}
%\end{figure}

\begin{figure}
\centering
\includegraphics[angle=0,width=0.47\columnwidth]{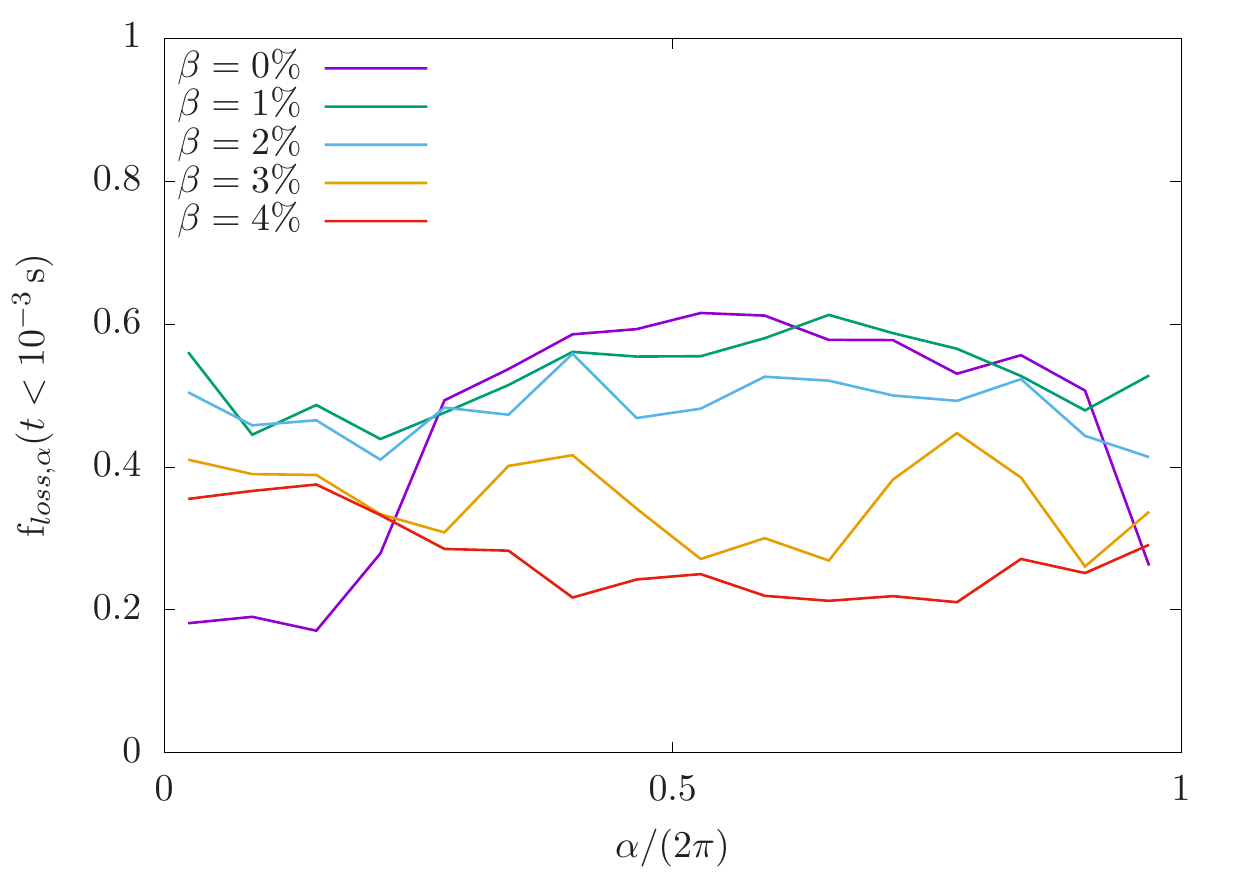}
\includegraphics[angle=0,width=0.47\columnwidth]{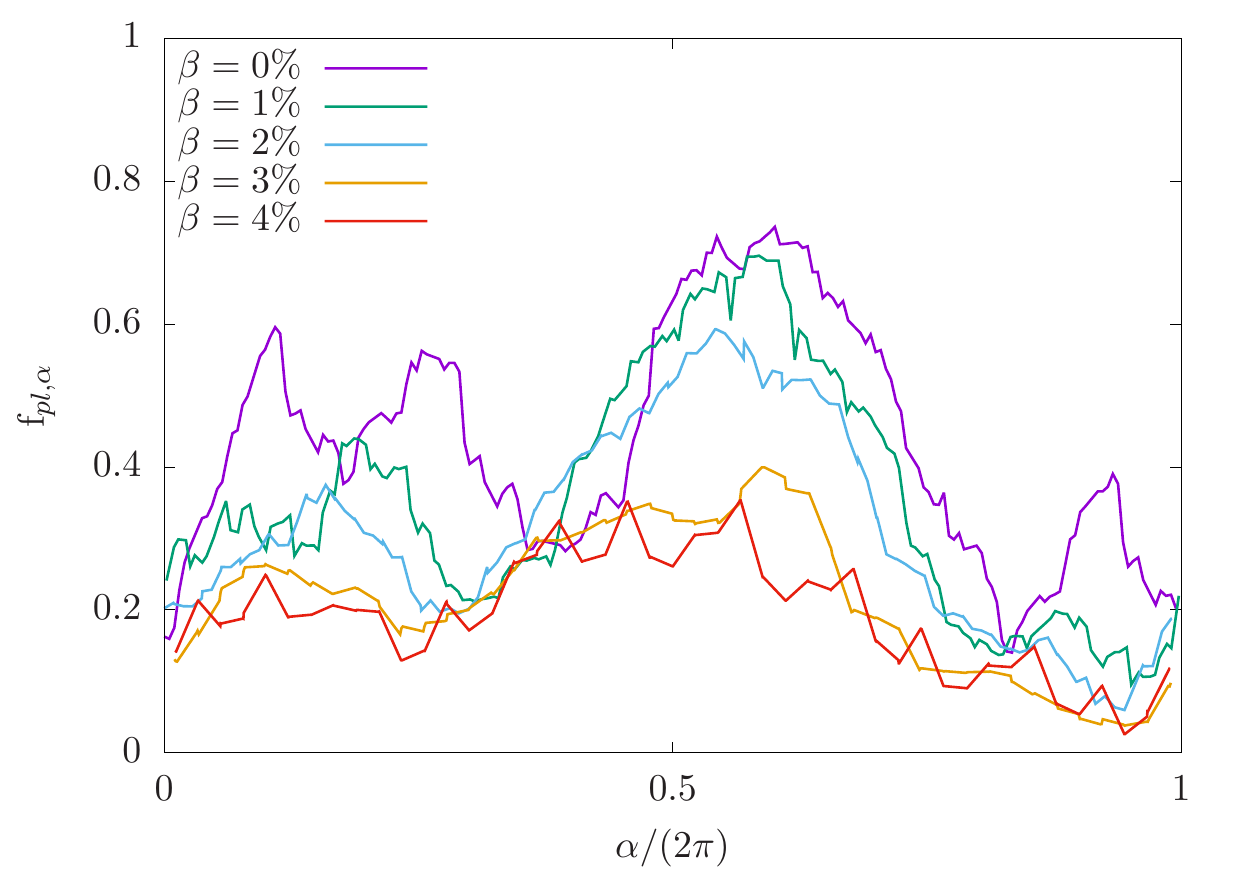}
\caption{Angular distribution of prompt loss fraction of energetic ions born trapped at $s=0.25$ obtained with ASCOT (left) and the model (right).}\label{FIG_ALPHA025}
\end{figure}

We start this subsection by discussing the angular distribution of the initial and final positions of the energetic ions that are promptly lost, as calculated with ASCOT, in order to validate the qualitative discussion around the sketches of figure~\ref{FIG_SKETCH}. Figure~\ref{FIG_2D} represents the number of energetic ions born at a given point of the flux-surface that are lost (left column) and the number of energetic ions that reach the last-closed flux-surface at a given angular position. In both cases, prompt losses (before $10^{-3}\,$s) are considered, the angular position is described by the toroidal and poloidal Boozer angles, and the data are normalized by the total quantity of prompt losses. As we expected, the distribution of initial points of the promptly lost ions, figure~\ref{FIG_2D} (left), is reasonably well-aligned with the field lines. Although the distribution of losses is rather broad, there appears to be a slight concentration of losses for values of $\alpha$ around $\pi$, which in the case of $\beta=4\%$ could correspond to deeply trapped particles, lost as predicted by figure~\ref{FIG_CLASS} (right). This will be studied in more detail in figure~\ref{FIG_ALPHA025}. Here, we are interested in comparing this angular distribution with the one of final positions of figure~\ref{FIG_2D} (right). As predicted, this distribution concentrates on a much smaller area, that for all the cases of the $\beta$ scan corresponds to values of $\alpha$ close to $\pi$. This is roughly the region where $\gamma_{\mathrm{c}}^*$ is maximum in figure~\ref{FIG_GAMMAC025}. In the same figure, $\gamma_{\mathrm{c}}^*\approx 1$ is also obtained in a small area around $\lambda=0.41\,$T$^{-1}$ for $\alpha$ slightly larger than 0, and around $\lambda=0.38\,$T$^{-1}$ and $\alpha$ slightly smaller than $2\pi$. This could correspond to the secondary peaks of the right corners of~\ref{FIG_2D} (right column). We conclude that the location of the superbananas at $s=0.25$ can provide predictions, at least of qualitative nature, at $s=1$.

The fraction of \textit{trapped} energetic ions born at a given angular position $\alpha$ that are promptly lost reads
\begin{eqnarray}\hskip-2.5cm
f_{\mathrm{pl},\alpha}(\alpha)= \frac{\fsa{\int_{B_\mathrm{max}^{-1}}^{B^{-1}}\mathrm{d}\lambda\frac{B}{\sqrt{1-\lambda B}}\delta(\alpha-\alpha')H\bigg((\alpha_\mathrm{out}-{\alpha})~ \overline{\mathbf{v}_M\cdot\nabla\alpha}\bigg)H\bigg((\alpha-{\alpha_\mathrm{in}})~ \overline{\mathbf{v}_M\cdot\nabla\alpha}\bigg)}}{\fsa{\int_{B_\mathrm{max}^{-1}}^{B^{-1}}\mathrm{d}\lambda\frac{B}{\sqrt{1-\lambda B}}\delta(\alpha-\alpha')}} \hfill \nonumber\\
\hskip-1cm = \frac{\int_{0}^{L}\mathrm{d}l\int_{B_\mathrm{max}^{-1}}^{B^{-1}}\frac{\mathrm{d}\lambda}{\sqrt{1-\lambda B}}H\bigg((\alpha_\mathrm{out}-{\alpha})~ \overline{\mathbf{v}_M\cdot\nabla\alpha}\bigg)H\bigg((\alpha-{\alpha_\mathrm{in}})~ \overline{\mathbf{v}_M\cdot\nabla\alpha}\bigg)}{\int_{0}^{L}\mathrm{d}l\int_{B_\mathrm{max}^{-1}}^{B^{-1}}\frac{\mathrm{d}\lambda}{\sqrt{1-\lambda B}}}\,.
\end{eqnarray}
It takes values between 0 and 1 and has been derived using that
\begin{equation}
\fsa{f}=\frac{\mathrm{d}s}{\mathrm{d}V}\int_0^{2\pi}\mathrm{d}\alpha\int_{0}^{L} \mathrm{d}lB^{-1}f\,,
\end{equation}
being $L(\alpha)$ the length of the field line. Figure~\ref{FIG_ALPHA025} (left) shows a broad angular profile of prompt losses: a similar proportion of the trapped energetic ions followed with ASCOT are lost for each $\alpha$, irrespective of $\beta$. This absence of clear features makes this aspect of the calculation not very useful for model validation. Figure~\ref{FIG_ALPHA025} (right) shows that, according to the model, the trapped energetic ions born at $\alpha$ slightly larger than $\pi$ should be slightly more likely to be promptly lost, corresponding to the deeply trapped region of figure~\ref{FIG_GAMMAC025}. A smaller peak exists around $\alpha=0.2$, corresponding to the superbananas at $\lambda\approx 0.41\,$T$^{-1}$ in figure~\ref{FIG_GAMMAC025}. Both are reduced by $\beta$. Although a similar structure of two peaks has been detected by guiding-center simulations of ions generated by neutral beam injection~\cite{faustin2016loss}, it cannot be seen in figure~\ref{FIG_ALPHA025} (left).

%%%%%%%%%%%%%%%%%%%%%%%%%%%%%%%%%%%%%%%%%%%%%%%%%%%%%%%%%%%%%%%%%%%%%%%%%%%%%%%%%%%%%%

\begin{figure}
\centering
\includegraphics[angle=0,width=0.32\columnwidth]{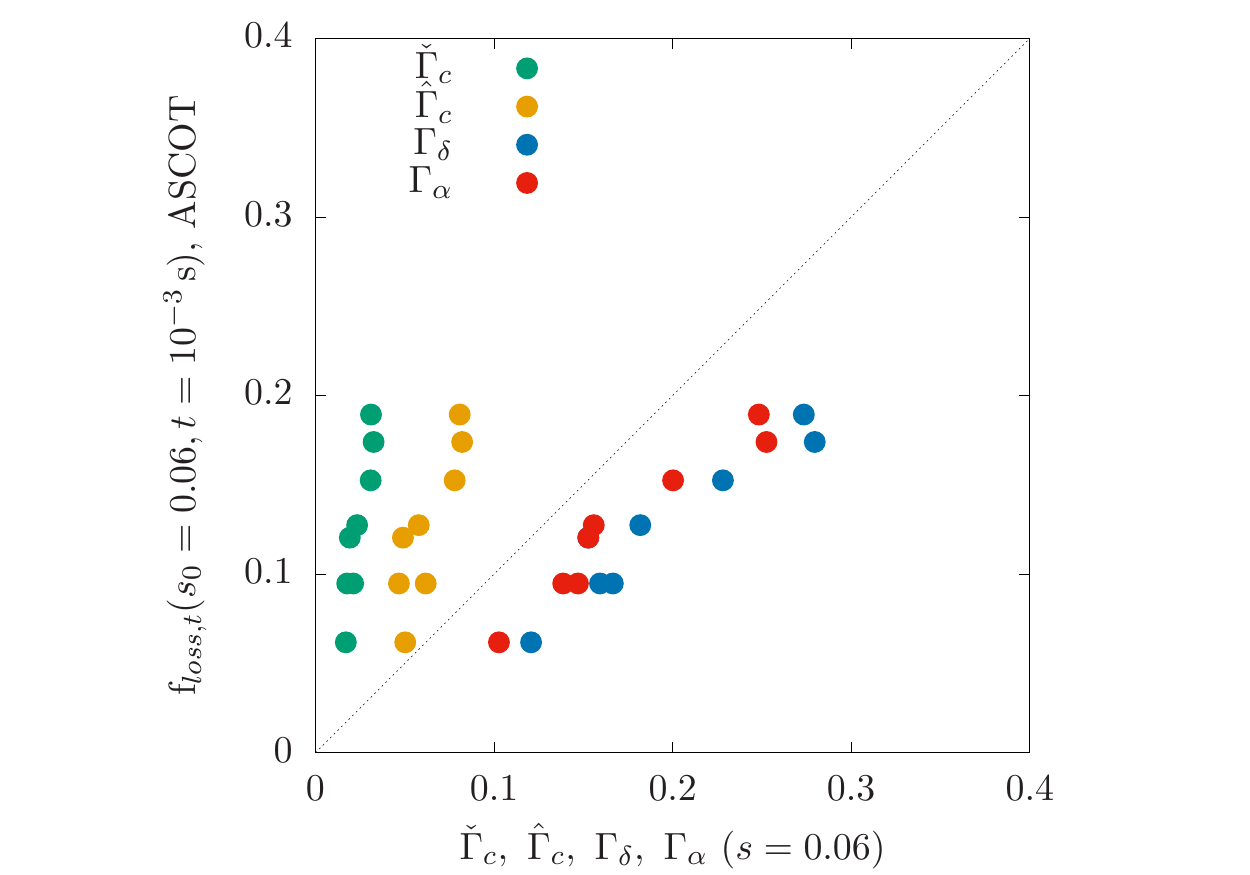}
\includegraphics[angle=0,width=0.32\columnwidth]{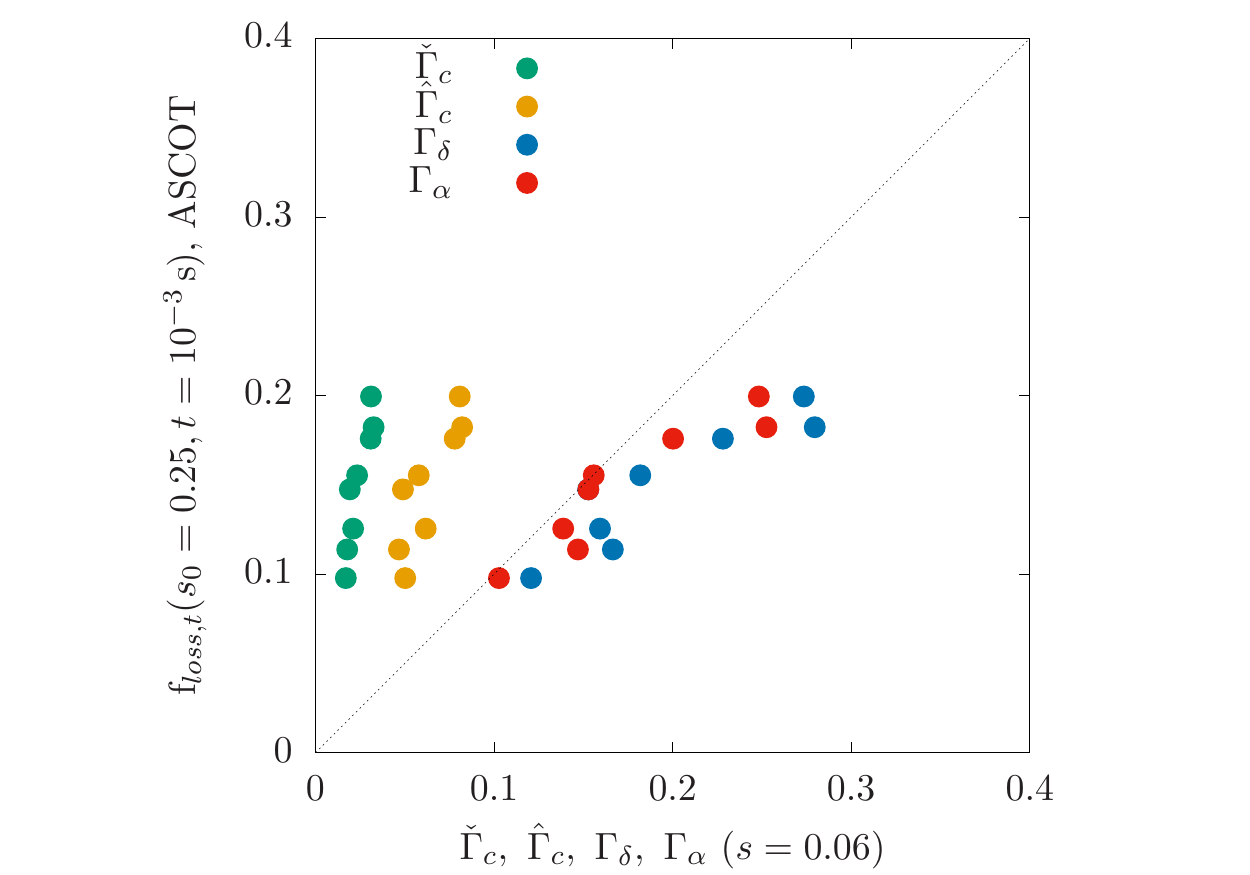}
\includegraphics[angle=0,width=0.32\columnwidth]{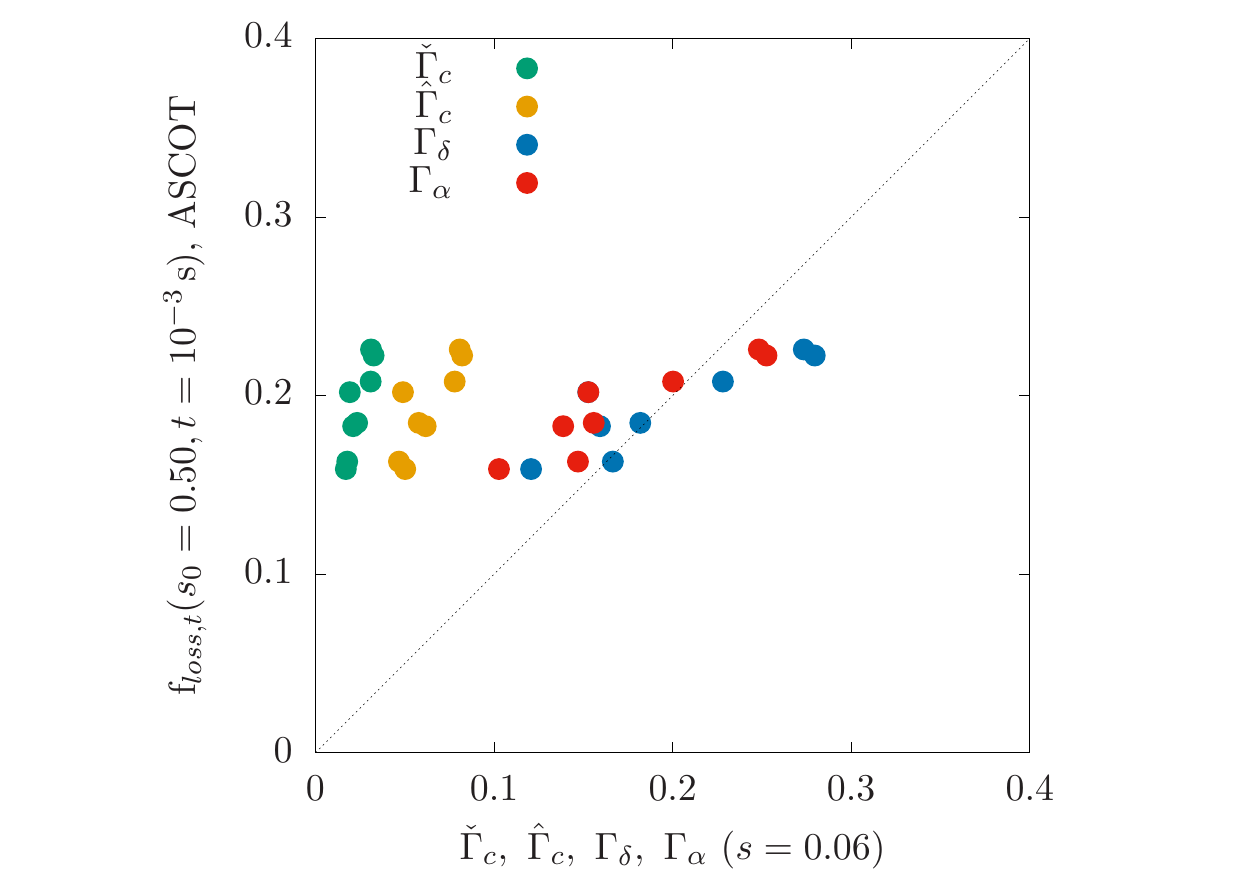}
\includegraphics[angle=0,width=0.32\columnwidth]{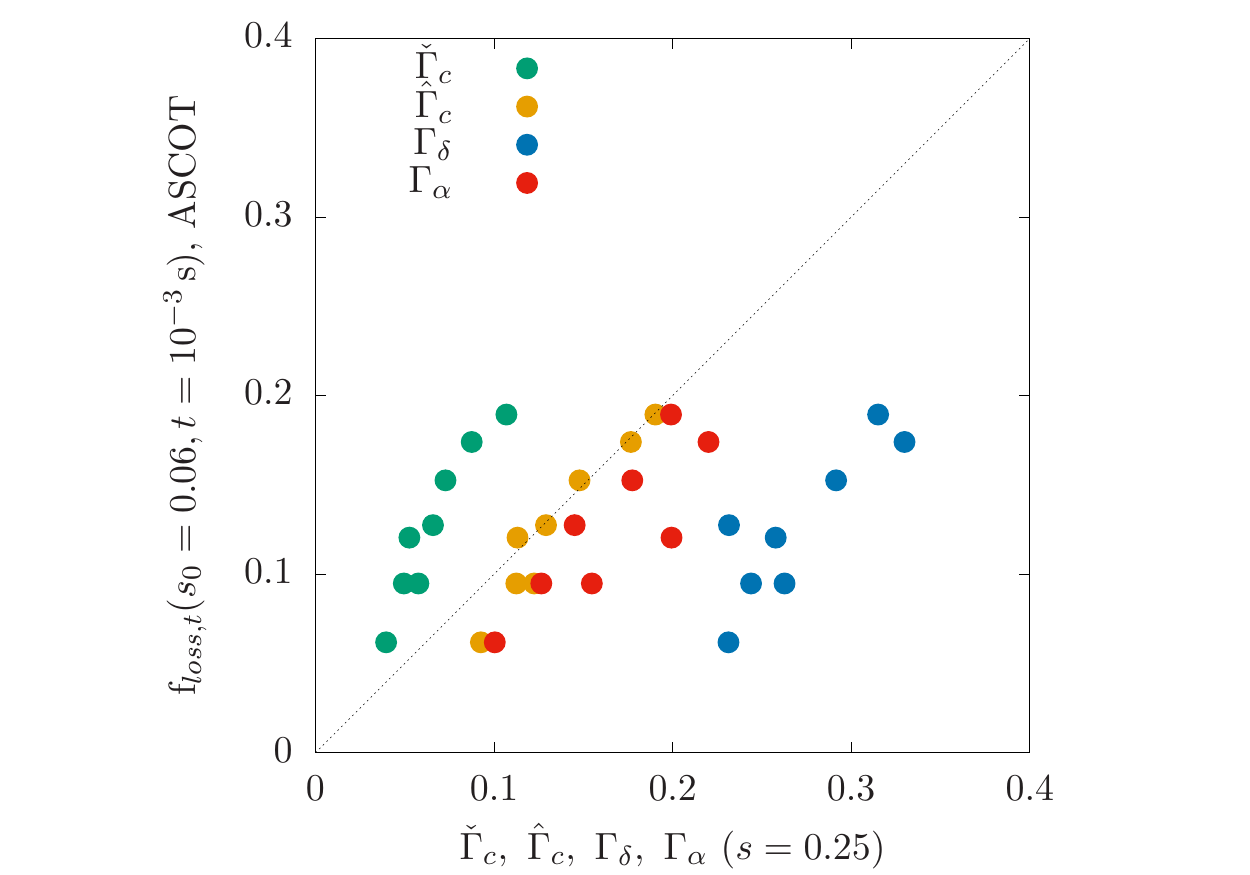}
\includegraphics[angle=0,width=0.32\columnwidth]{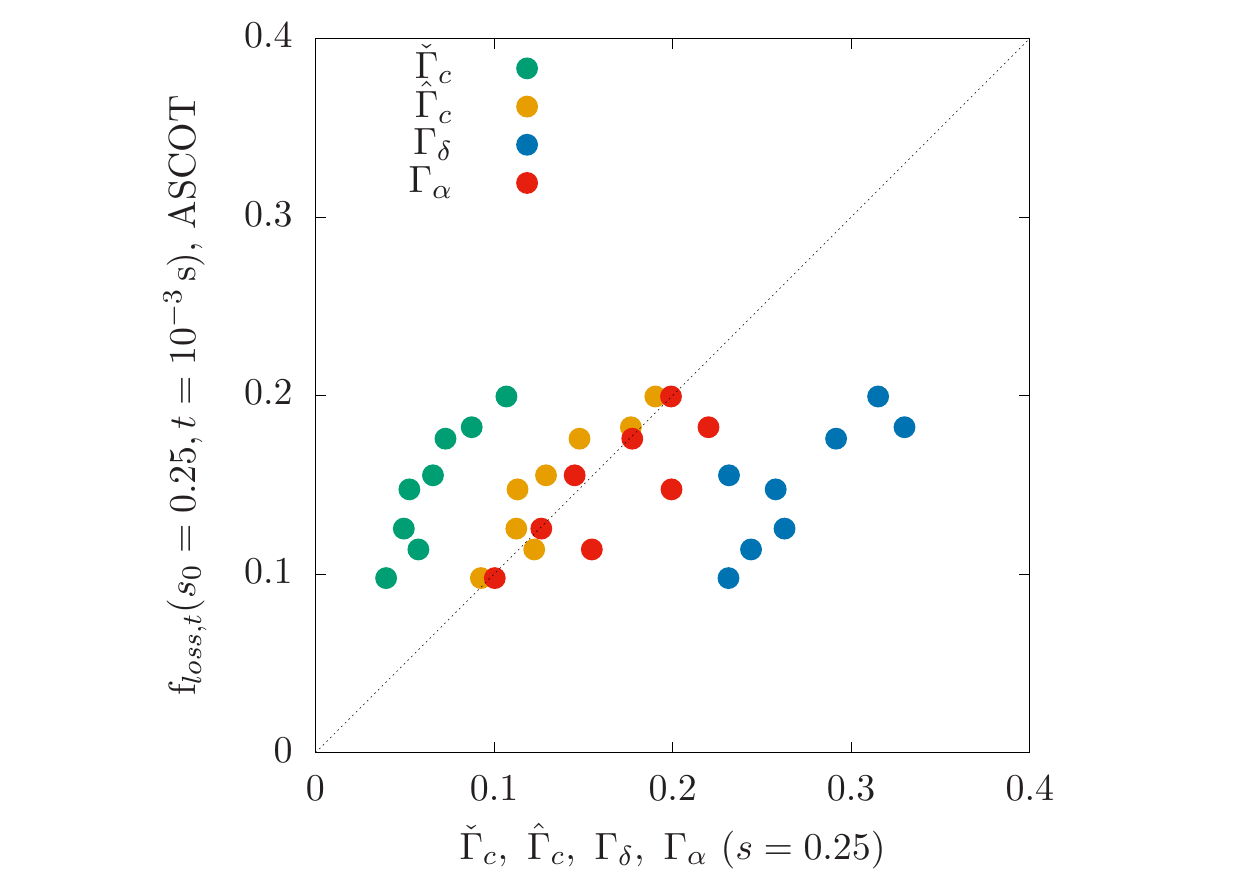}
\includegraphics[angle=0,width=0.32\columnwidth]{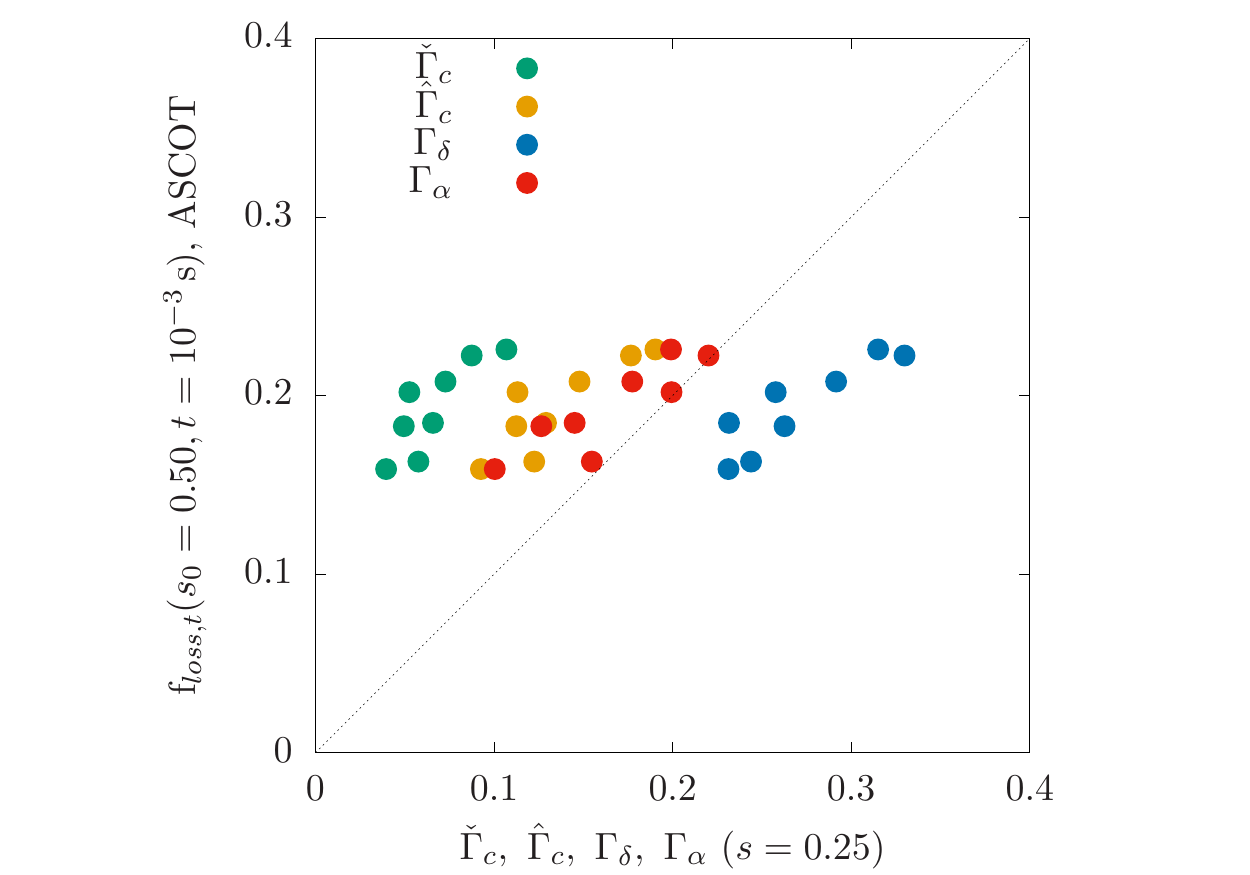}
\includegraphics[angle=0,width=0.32\columnwidth]{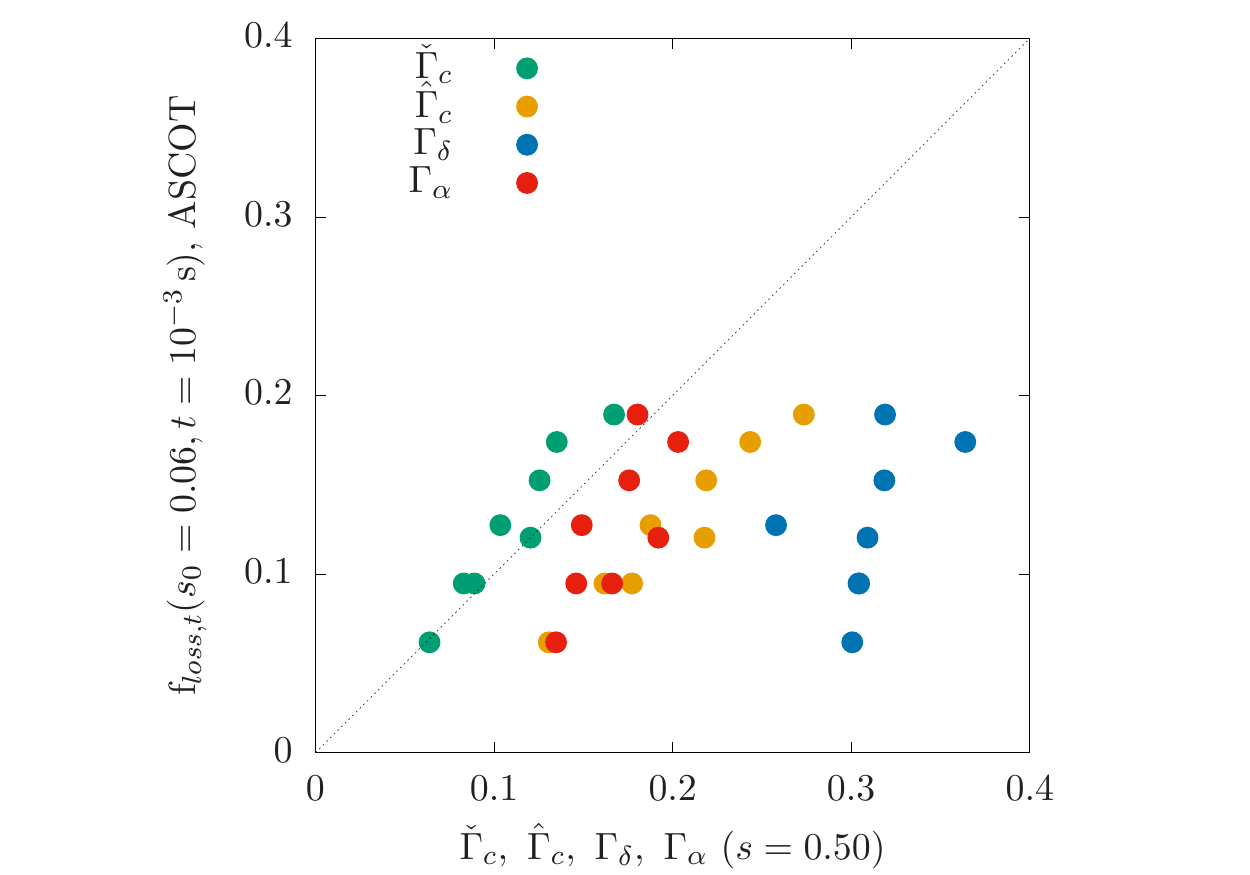}
\includegraphics[angle=0,width=0.32\columnwidth]{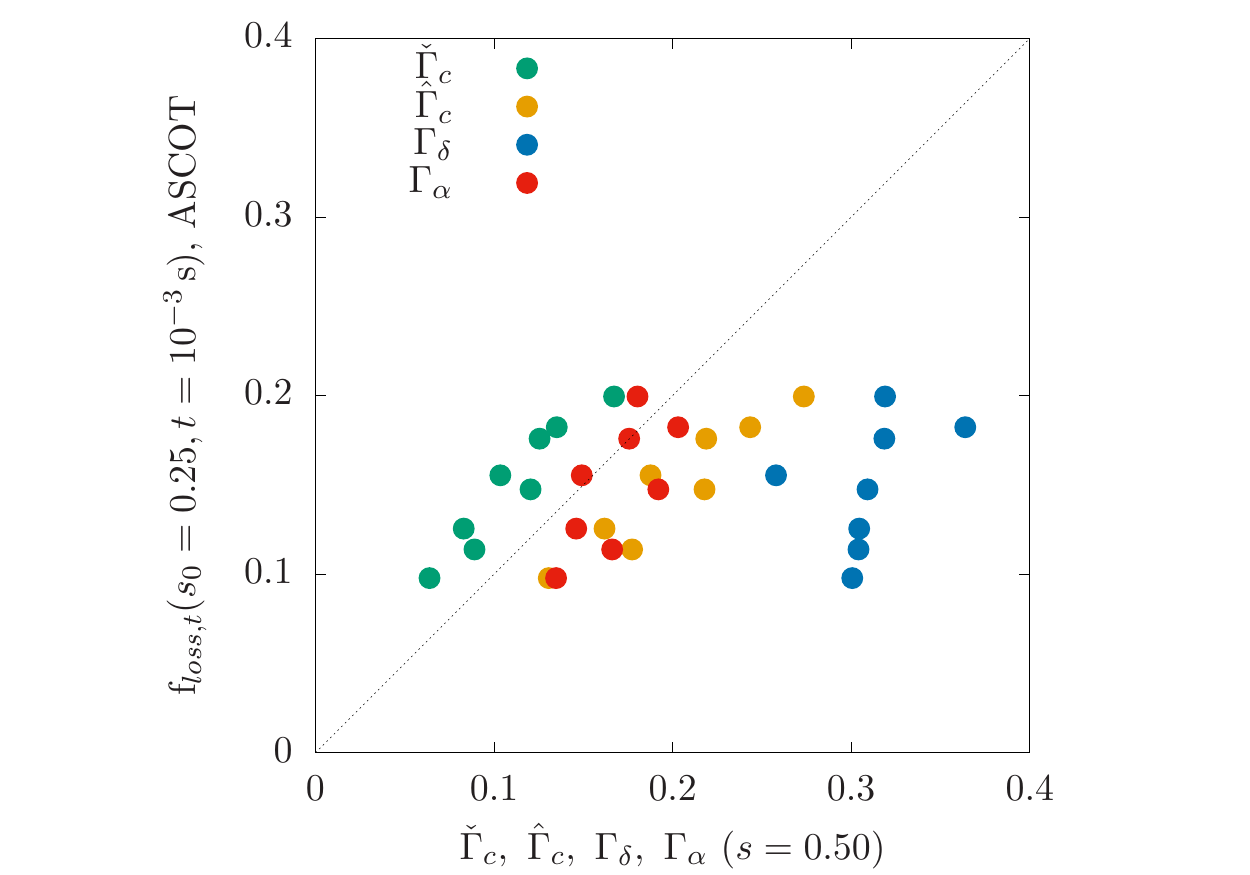}
\includegraphics[angle=0,width=0.32\columnwidth]{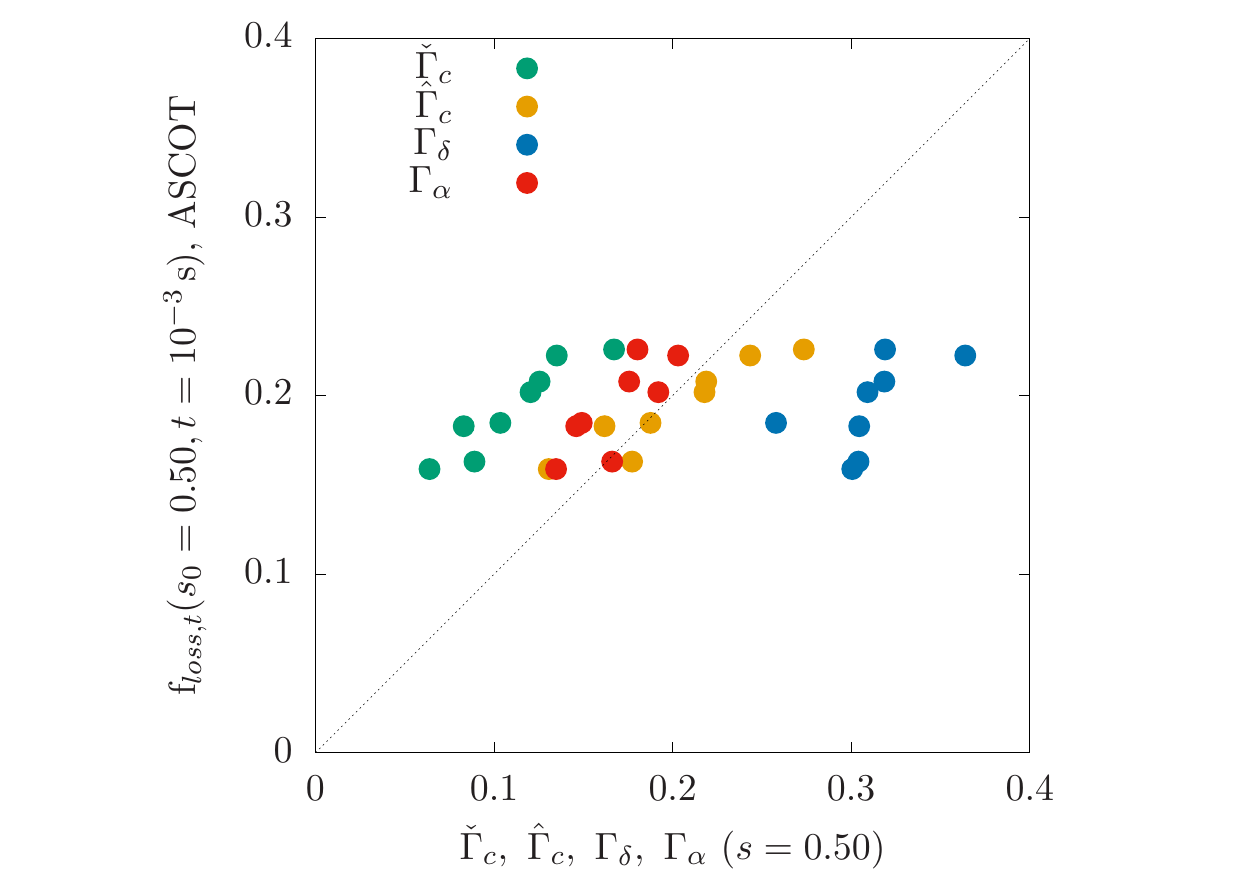}
\caption{Comparison between proxies and ASCOT calculations of prompt losses.}
\label{FIG_COMP}
\end{figure}

\subsection{Comparison of proxies}\label{SEC_COMP}

We have seen that the model of section~\ref{SEC_MODEL2} succeeds in predicting important features of the prompt losses of energetic ions, some of them configuration-dependent. This is a crucial property if it is to be employed for stellarator optimization, i.e., for the design of new magnetic configurations with reduced energetic ion losses. We now go one step further, try to use our model as proxy for energetic ion confinement, and check quantitatively how good it is predicting the configuration dependence of the prompt losses. This is done in figure~\ref{FIG_COMP}, that includes data from the eight W7-X configurations of the study and three different radial positions: $s=0.06$, $s=0.25$ and $s=0.50$. 

The diagonal subplots (top left, center and bottom right) in figure~\ref{FIG_COMP}  study the correlation between the fraction of prompt losses of ions born at a given flux-surface, calculated with ASCOT, and the predictions of the models evaluated at the same flux-surface. Overall, $\Gamma_\alpha$ performs better at predicting the prompt ion losses $f_{\mathrm{loss},t}$: the points lie closer to the diagonal of each subplot; $\Gamma_\delta$ generally overestimates them, something to be expected from figure~\ref{FIG_CLASS}. At some flux-surfaces, $\hat\Gamma_{\mathrm{c}}$ could be used to predict $f_{\mathrm{loss},t}$, even though it is not its purpose. This cannot be done with $\check\Gamma_{\mathrm{c}}$ (or $\Gamma_{\mathrm{c}}$, which is not shown). %{\todo{\footnote{\todo{Should we even test $\hat\Gamma_{\mathrm{c}}$?}}}}

The subplots below the diagonal (bottom left, center left and bottom center) in figure~\ref{FIG_COMP} correspond to the comparison between the fraction of prompt losses of ions born at a given flux-surface, calculated with ASCOT, and the predictions of the models evaluated at an outer flux-surface. It makes sense that the proxies still succeed in predicting qualitatively (not quantitatively) energetic ion confinement, since the ions followed by ASCOT have to go through the flux-surface characterized by the models if they are to escape the plasma. Indeed we see good correlation between $\Gamma_\alpha$ and ASCOT, which means that the former could be used for optimization with respect to prompt losses of energetic ions born at inner flux-surfaces, not only at the flux-surface where the formula is evaluated. Indeed, one could perform a linear fitting of the data, and the intercept would be negative. This suggests that this could be a more effective optimization strategy, since the losses approach zero faster than the proxy.

Finally, most neoclassical properties, and the optimization with respect to neoclassical prompt losses could be one of these properties, vary smoothly with the radial coordinate of a magnetic equilibrium. It is natural then that a model evaluated on a flux-surface is able to predict reasonably well the properties of a non distant flux-surface. In order to distinguish this effect from the one discussed in the previous paragraph, we turn our attention to the subplots above the diagonal (top right, top center and center right) in figure~\ref{FIG_COMP}. These ones compare the fraction of prompt losses of ions born at a given flux-surface, calculated with ASCOT, and the predictions of the models evaluated at an inner flux-surface. Even though the correlation is worse than in the plots above the diagonal, it is larger than what one would obtain if no relation existed between the level of optimization of different flux-surfaces (we are disregarding in this discussion possible inwards excursions of the ions, but these should be scarce). In this case, the intercept of a linear fit would be positive, indicating that optimizing for prompt losses only at the innermost flux-surfaces is probably not a good strategy.

Let us finally mention that the computation of each of the points of \ref{FIG_COMP} took a few seconds on a single desktop computer (and most of the computing time was devoted to the calculation of the bounce averages, that could be trivially parallelized). This computing time is clearly small enough for the model presented in this work to be included within the loop of a stellarator optimization suite.

\begin{figure}
\centering
\includegraphics[angle=0,width=0.32\columnwidth]{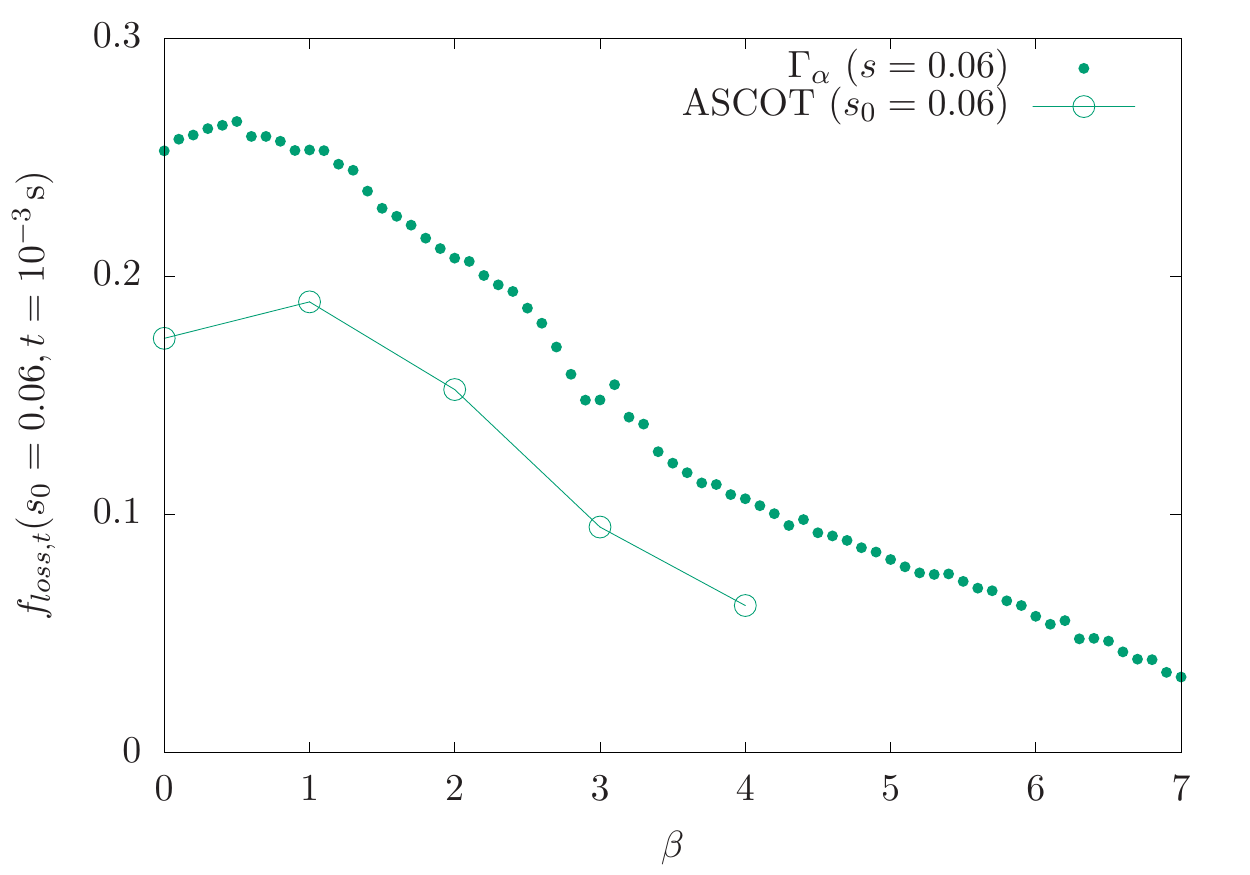}
\includegraphics[angle=0,width=0.32\columnwidth]{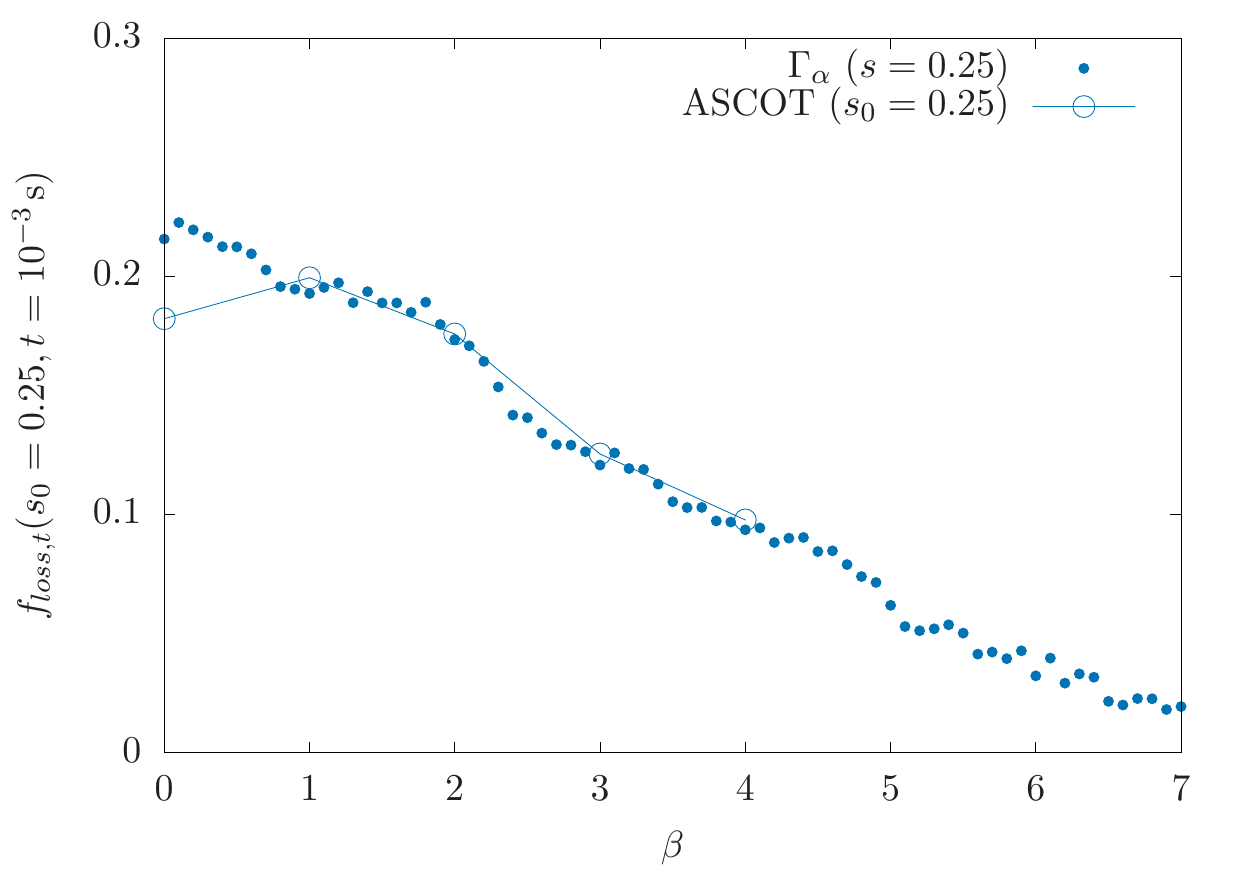}
\includegraphics[angle=0,width=0.32\columnwidth]{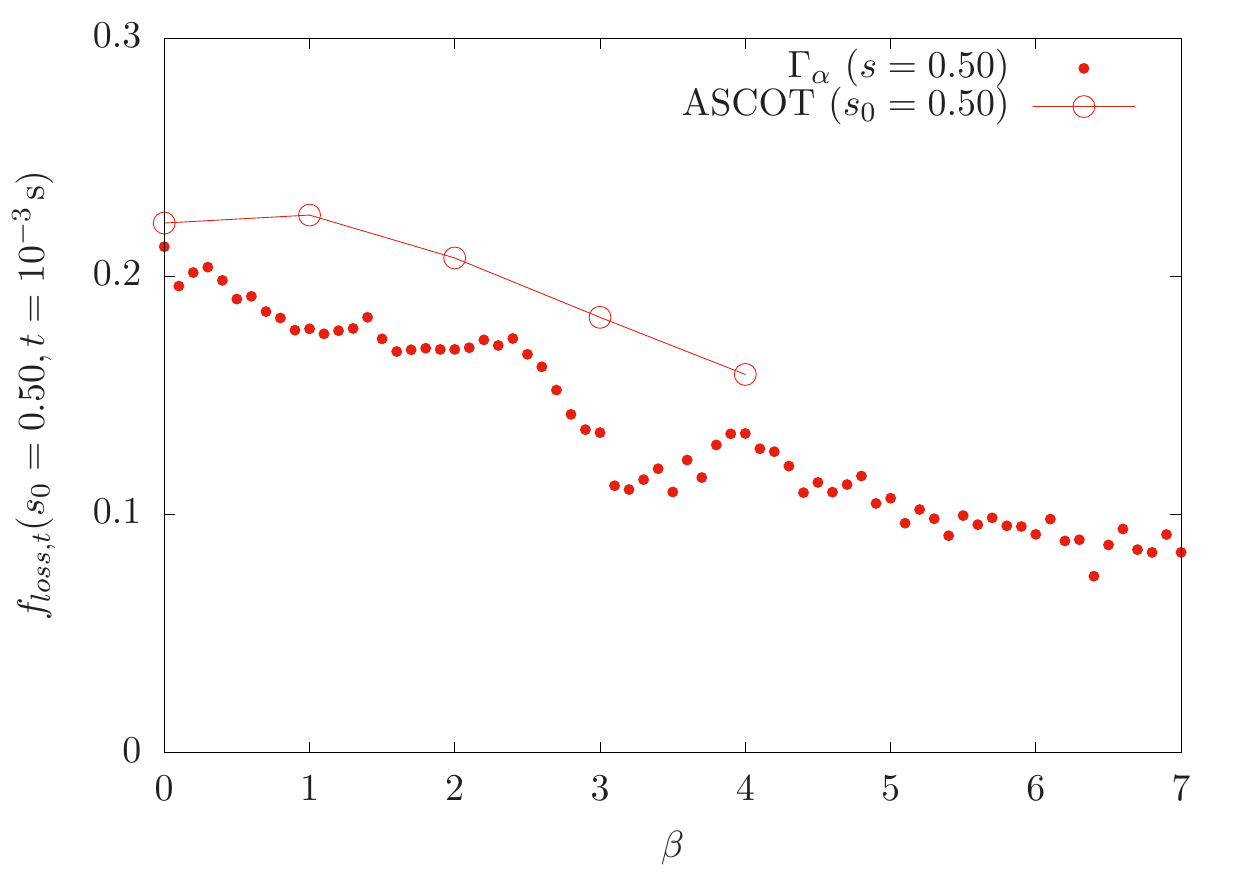}
\caption{Dependence of the prompt losses of the KJM configuration as a function of $\beta$ for energetic ions born at $s=0.06$ (left), $s=0.25$ (center) and $s=0.50$ (right).}
\label{FIG_BETA}
\end{figure}

\section{Discussion}\label{SEC_DISCUSSION}

We have derived a model, encapsulated in the quantity $\Gamma_\alpha$, that succeeds in predicting, even quantitatively, configuration-dependent features of the prompt losses of energetic ions in stellarator configurations. It is fast enough to be part of any stellarator optimization strategy. This model is based on the neoclassical code \texttt{KNOSOS}, which is already integrated in the optimization suite \texttt{STELLOPT}~\cite{lazerson2020stellopt}, and is planned to be included in other optimization codes in the near future.%the optimization code~\texttt{ROSE}~\cite{drevlak2018rose} in the near future.

This application should not be limited to the optimization of stellarators of the helias type. The results presented here could be applied for other kinds of stellarators, such as heliotrons and stellarators close to quasisymmetry. While the role of $\beta$ in increasing the toroidal precession and leading to the maximum-$J$ property is unique to quasi-isodynamic stellarators, the maps of $\gamma_{\mathrm{c}}^*$ and the conclusions drawn from them should be general. Actually, ~\cite{nemov2014ripple} contains figures with similar information that are employed to argue qualitatively on the different level of optimization of two quasisymmetric configurations.

The tools derived in this work may be useful not only for stellarator optimization but for an extensive search of the parameter space of an already designed stellarator. For instance, figure~\ref{FIG_BETA} shows a finer version of the scans in $\beta$  discussed so far in this work. The prompt losses are estimated with $\Gamma_\alpha$ for 71 magnetic equilibria corresponding to the KJM configuration with values of $\beta$ between 0\% and 7\% and parabolic pressure profiles. This fine scan allows us to assess whether the $\beta$-dependence is smooth or it shows some kind of threshold behaviour. The prompt losses are shown to decrease with approximately constant slope, with some small corrections: for the two innermost flux-surfaces, the slope reaches a (relatively small) minimum value around $2\%<\beta<3\%$; at small values of $\beta$, the diamagnetic effect may be even detrimental, a fact well captured by the model at $s=0.06$, although not at $s=0.25$. These results are consistent with figures~\ref{FIG_J_041},~\ref{FIG_GAMMAC025}  and~\ref{FIG_LF025} (right). More detailed studies are left for the future.

More physics could be included in the model. The radial electric field tends to improve the confinement of partially thermalized energetic ions. In our framework, electric fields (also those that are tangent to the flux-surface~\cite{regana2017phi1,calvo2018jpp}) can be included by replacing equation (\ref{EQ_GAMMACS}) with
\begin{equation}
\gamma_{\mathrm{c}}^*=\frac{2}{\pi} \arctan{\frac{\partial_\alpha J}{|\partial_s J|}}=\frac{2}{\pi} \arctan{\frac{\overline{(\mathbf{v}_M+\mathbf{v}_E)\cdot\nabla s}}{|\overline{(\mathbf{v}_M+\mathbf{v}_E)\cdot\nabla\alpha}|}}\,,
\label{EQ_GAMMACSE}
\end{equation}
where $\mathbf{v}_E$ represents the $E\times B$ drift. Collisions tend to increase the energetic ion losses~\cite{henneberg2019fastions,patten2018nbi}. In our approach, only trapped ions born with particular ranges of $\lambda$ escape, and the rest remain confined. Pitch-angle-scattering collisions will produce diffusion into those values of $\lambda$ from neighbouring regions in the velocity space, and, consequently, additional \textit{prompt} ion losses at longer time scales (this proccess has been characterized, for ions generated by neutral beam injection, in~\cite{faustin2016loss}). This could be added to our model as a diffusive term in $\lambda$. Energy diffusion does not change the character of confined or unconfined of orbits in the absence of radial electric field, since $v$ does not appear on equation  (\ref{EQ_GAMMACS}). However, it does when the radial electric field is relevant, since  equation (\ref{EQ_GAMMACSE}) is energy-dependent.

%The losses at longer time scales, associated to stochastic diffusion, have been modelled in~\cite{beidler2001stochastic} for a simplified magnetic field. It might be possible to model them by following radially-local orbits, at constant $J$, for many poloidal turns.

The study of this work (except for the possible upgrades mentioned in the previous paragraphs) basically exhausts what can be modelled with a local approach. The underlying assumption in any radially-local description of energetic ion losses, such as the one presented here, is that the structure of superbananas does not change qualitatively when moving in the radial coordinate. This is what happens in figure~\ref{FIG_SKETCH} (top), where the blue and red dashed lines are able to reproduce qualitatively well the open $J$-contours of the sketch. This property, which can be roughly expressed as
 \begin{equation}
\gamma_{\mathrm{c}}^*(s,\alpha,\lambda)\approx\gamma_{\mathrm{c}}^*(s_0,\alpha,\lambda)\,,
\label{EQ_LOCAL}
\end{equation}
is reasonably fulfilled for our configurations, as indicated by the comparison between figures~\ref{FIG_GAMMAC025} and \ref{FIG_GAMMAC006}. There are however a few exceptions: for instance, the above-mentioned closed orbits of $J=0.24vR_0$ at $\beta=0\%$ would probably be interpreted as prompt losses by any local model.

%One can introduce radial non-locality by not making use of equation~(\ref{EQ_LOCAL}). Precisely,  we one should replace equations~(\ref{EQ_MODEL}),~(\ref{EQ_MODEL2}) and~(\ref{EQ_MODEL2b}) with
%\begin{eqnarray}
%\dot s &=& u_s(s,\alpha)\,,\nonumber\\
%\dot \alpha &=& u_a(s,\alpha) \,,
%\label{EQ_MODELbis}
%\end{eqnarray}
%and
%\begin{eqnarray}
%u_s &\equiv & \overline{\mathbf{v}_M\cdot\nabla s}|_{\lambda=\lambda_0,v=v_0}\,,\nonumber\\
%u_a &\equiv & \overline{\mathbf{v}_M\cdot\nabla\alpha}|_{\lambda=\lambda_0,v=v_0}\,.
%\label{EQ_MODEL2bis}
%\end{eqnarray}

The good agreement between our local model and the global Monte Carlo simulations can be at the basis of more efficient radially global guiding-center (or full-orbit) simulations by means of existing Monte Carlo codes: it implies that many features of a particle trajectory are not determined by its initial point in phase space $(s,\alpha,l,\lambda,v)$ but specifically by the initial trapped-orbit in which it lies, $(s,\alpha,\lambda,v)$. Computationally speaking, an initial set of markers distributed accordingly may be significantly more efficient than a uniform distribution on the flux-surface. This is likely to be the case even in simulations with collisions. 

However, an even more relevant theoretical finding can be extracted from the results of this paper: a bounce-averaged drift-kinetic equation is likely able to describe, even quantitatively, the neoclassical transport of energetic ions. This equation needs to be radially global, but could in principle be solved much faster than the one solved by guiding-center codes, since the motion along the magnetic field lines would not need to be resolved. The code \texttt{KNOSOS} has been adapted in order to solve rigorously the radially global bounce-averaged drift kinetic equation. First results will be presented elsewhere.

%%%%%%%%%%%%%%%%%%%%%%%%%%%%%%%%%%%%%%%%%%%%%%%%%%%%%%%%%%%%%%%%%%%%%%%%%%%%%%%%%%%%%%

\section*{Acknowledgments}

%%%%%%%%%%%%%%%%%%%%%%%%%%%%%%%%%%%%%%%%%%%%%%%%%%%%%%%%%%%%%%%%%%%%%%%%%%%%%%%%%%%%%%

%Acknowledgements should be included at the end of the paper, before the References \Sor any appendicies, and should be a separate paragraph without a heading. Several anonymous individuals are thanked for contributions to these instructions.

The authors are grateful for instructive discussions with A. Bader, C.D. Beidler, M. Drevlak and M. Landreman. This work has been carried out within the framework of the EUROfusion Consortium and has received funding from the Euratom research and training programme 2014-2018 and 2019-2020 under grant agreement No. 633053. The views and opinions expressed herein do not necessarily reflect those of the European Commission. This research was supported in part by grant PGC2018-095307-B-I00, Ministerio de Ciencia, Innovaci\'on y Universidades, Spain.

%%%%%%%%%%%%%%%%%%%%%%%%%%%%%%%%%%%%%%%%%%%%%%%%%%%%%%%%%%%%%%%%%%%%%%%%%%%%%%%%%%%%%%

%\appendix
%\section{}\label{}

\section*{Bibliography}

\bibliographystyle{unsrt}


\bibliography{prompt_losses.bbl}

\end{document}